%% file: main-arxiv.tex
\documentclass[letterpaper,twocolumn,10pt]{article}
\usepackage{usenix}

\input{imports}

\anonymousfalse{}

\begin{document}

\date{}

\title{\Large \bf SoK: Advances and Open Problems in Web Tracking}

\ifanonymous
\author{\em Anonymous Authors}
\else
\author{
{\rm Yash Vekaria$^\star$~\ucDavis{}}, 
{\rm Yohan Beugin$^\star$~\uwMadison{}}, 
{\rm Shaoor Munir~\ucDavis{}}, 
{\rm Gunes Acar~\radboud{}}, 
{\rm Nataliia Bielova~\inria{}}, \\
{\rm Steven Englehardt~\independent{}}, 
{\rm Umar Iqbal~\washU{}},
{\rm Alexandros Kapravelos~\ncstate{}},   
{\rm Pierre Laperdrix~\cnrs{}},\\
{\rm Nick Nikiforakis~\stonybrook{}},
{\rm Jason Polakis~\uic{}},
{\rm Franziska Roesner~\uwashington{}},
{\rm Zubair Shafiq~\ucDavis{}}, 
{\rm Sebastian Zimmeck~\wu{}}\\
\ucDavis{}~\small{University of California, Davis, USA}\hfill{}
\uwMadison{}~\small{University of Wisconsin–Madison, USA}\hfill{}
\radboud{}~\small{Radboud University, Netherlands}\\
\inria{}~\small{Inria Centre at Université Côte d’Azur, France}\hfill{}
\independent{}~\small{Independent Researcher, USA}\hfill{}
\washU{}~\small{Washington University in St. Louis, USA}\\
\ncstate{}~\small{North Carolina State University, USA}\hfill{}
\cnrs{}~\small{Centre National de la Recherche Scientifique, France}\hfill{}
\stonybrook{}~\small{Stony Brook University, USA}\\
\uic{}~\small{University of Illinois Chicago, USA}\hfill{}
\uwashington{}~\small{University of Washington, USA}\hfill{}
\wu{}~\small{Wesleyan University, USA}
}
\fi

\maketitle

\begingroup\renewcommand\thefootnote{$\star$}
\footnotetext{Equal contribution.\\ An extended and living version of this document is available at \url{https://github.com/privacysandstorm/sok-advances-open-problems-web-tracking}}
\endgroup

\begin{abstract}
\input{short-version/00-abstract}
\end{abstract}

\input{short-version/01-introduction}
\input{short-version/02-background}
\input{short-version/03-methodology}
\input{short-version/04-tracking-techniques}

\input{short-version/05-tracking-mitigations}

\input{short-version/06-regulations}

\input{short-version/07-discussion}
\input{short-version/08-conclusion}


\ifanonymous
\else
\section*{Acknowledgments}
\input{acknowledgment}
\fi

\appendix
\section*{Ethical Considerations}
\input{appendix/ethics}


\section*{Open Science}
\input{appendix/open-science}


\bibliographystyle{custom-style}
\bibliography{refs}

\input{appendix/sok-methodology}

\end{document}

%% file: short-version/00-abstract.tex
Web tracking is a pervasive and opaque practice that enables personalized advertising, retargeting, and conversion tracking.
Over time, it has evolved into a sophisticated and invasive ecosystem, employing increasingly complex techniques to monitor and profile users across the web.
The research community has a long track record of analyzing new web tracking techniques, designing and evaluating the effectiveness of countermeasures, and assessing compliance with privacy regulations.
Despite a substantial body of work on web tracking, the literature remains fragmented across distinctly scoped studies, making it difficult to identify overarching trends, connect new but related techniques, and identify research gaps in the field.
Today, web tracking is undergoing a transformation, driven by fundamental shifts in the advertising industry, the adoption of anti-tracking countermeasures by browsers, and the growing enforcement of emerging privacy regulations.
This Systematization of Knowledge (SoK) aims to consolidate and synthesize this wide-ranging research, offering a comprehensive overview of the technical mechanisms, countermeasures, and regulations that shape the modern and rapidly evolving web tracking landscape.
This SoK also highlights open challenges and outlines directions for future research.

%% file: short-version/01-introduction.tex
\section{Introduction}
\label{sec:introduction}

Users access a variety of free content and services on the web, which are largely funded through online advertising.
In turn, advertising is heavily dependent on monitoring users' online activities for various purposes such as analytics, (re)targeting, and conversion tracking.
To realize these objectives, user tracking has become a pervasive part of the web---today, 92\% of webpages embed at least one third-party tracker~\cite{urban2024WebAlmanac2024}.

Over the years, researchers have conducted numerous studies to examine the evolution of web tracking mechanisms, browser developments, and regulatory compliance.
However, despite a considerable body of work, major findings remain scattered across many disparate studies, and existing survey papers have either been too high-level~\cite{ermakova2018web}, too narrowly-focused on specific forms of web tracking
such as caching and browser fingerprinting
~\cite{bujlowSurveyWebTracking2017,laperdrix2020browser}, or are not current with modern techniques anymore~\cite{mayerThirdPartyWebTracking2012}.
Furthermore, as privacy defenses improve in browsers, trackers continually adapt with new evasion techniques~\cite{narayanan2018web,localmess-usenix-sec-26}.
The result is an ever-shifting technical landscape of tracking techniques.

Regulations often govern tracking practices and although regulatory changes have had a more gradual impact than browser-based technical interventions, together they have continued to reshape the ecosystem.
Today, web tracking is undergoing a transformative change due to the introduction of privacy-enhancing protections in major web browsers, new advertising paradigms that do not rely on cross-site tracking, and evolving regulatory frameworks~\cite{hercherW3CAdPrivacy2022,chavezUpdatePlansPrivacy2025}.
These shifts call for a comprehensive study of emerging trends in the evolving tracking landscape to identify crucial research gaps.

To this end, we synthesize disparate lines of research and practices in web tracking---spanning across technical mechanisms, mitigations, and regulatory changes---to systematically provide an overview of the current state of web tracking.
We scope this SoK to how the data is \textit{collected} about users, not how that data might then be \textit{used}.
This unified perspective enables a critical reflection on how far the community has come and which research directions it should prioritize.
Our main contributions are as following:
\begin{itemize}
    \item We systematize the extensive body of research on web tracking, providing a consolidated knowledge base of advances, highlighting evolving trends, and bridging emerging but related tracking mechanisms.
    \item We systematically organize anti-tracking interventions proposed by researchers, major browser providers, and community as well as relevant regulatory and compliance actions across the EU and the US to assess how they have altered the ecosystem over the years.
    \item We identify key open challenges and promising future research directions in web tracking.
\end{itemize}

%% file: short-version/02-background.tex
\section{Background \& Prior Work}


\label{sec:threat-model}

\para{Browser's Security Model: Context-Origin Boundaries.}
The browser's context model denotes that every document runs in a container 
that maintains its state (e.g., DOM, variables, scripts) separate from others, while the origin model associates the document with an origin determining the code’s privileges.
Thus, the browser uses both the origin and context to enforce policies---it isolates contexts from each other, and when an interaction is attempted, it checks the origins involved.
Under the \textit{Same-Origin Policy (SOP)}~\cite{mdnSameoriginPolicySecurity2024}, two contexts can interact freely only if they share the same origin.

\begin{figure}[!ht]
    \centering
    \includegraphics[width=\linewidth]{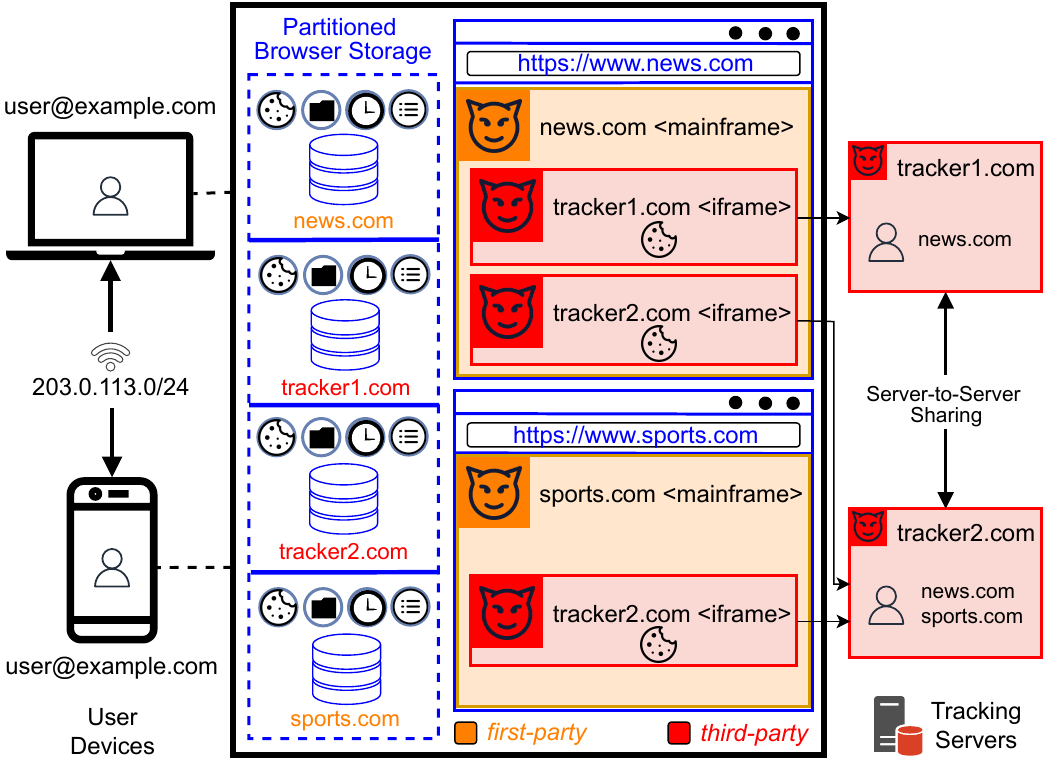}
    \caption{Conceptual overview of our threat model where \texttt{news.com} and \texttt{sports.com} are the first-party websites visited by the user from their personal device(s) and \texttt{tracker1.com} and \texttt{tracker2.com}  embedded third-parties.}
    \label{fig:threat-model}
\end{figure}

\para{Entities (\autoref{fig:threat-model}).}
%
The \textbf{\textit{user}} is an individual accessing websites on the internet through their browser (also known as user agent~\cite{mdnUserAgentMDN2025}), seeking to keep their online behavior private. 
The \textbf{\textit{browser}} mediates all interactions between the user and the web content, enforcing context-origin policies. 
The \textbf{\textit{first-party website}} is the webpage that the user directly browses for content.
It often includes resources from \textbf{\textit{third-parties}}, i.e., hosted on a different domain than the first-party website~\cite{mdnDomainMDNWeb2024}.

\para{Adversary Goals.}
A tracker is any party whose goal is to collect user’s data in order to profile or identify the user’s behavior across one site (\textit{same-site tracking}), several sites (\textit{cross-site tracking}), across contexts (\textit{cross-context tracking}) on a given device
or across devices.
%
%
Besides, a secondary goal may be to avoid detection and anti-tracking measures.

\input{tables/prior-work}

\para{Adversary Capabilities.}
%
%
%
%
\textbf{\textit{Inclusion} \symInclusion.}
A first-party tracker is assumed to directly monitor user activities within the top-level browsing context.
A third-party tracker could be embedded as an inline resource (e.g., image/script tag) within the mainframe by the first-party assuming trust delegation, an iframe with a separate context under its origin, or as a resource within a third-party iframe.
\textbf{\textit{Execution} \symExecution.}
A tracker aims to collect user data through available browser features 
within the context where it is embedded. 
A script's execution privileges are tied to its context’s origin---implying that it always ``acts as'' whatever origin its document has.
%
%
%
By including external scripts, the webpage implicitly declares that it trusts that code with its own origin’s privileges~\cite{TrackingPreventionWebKit2020}.
\textbf{\textit{Storage} \symStorage.}
Trackers may also read, write, or modify data in the user's browser storages,
which are partitioned by origin, as well as the cookie jar, which is scoped by domain (and path) rather than the full origin.
Stored data can be accessed by the tracker in subsequent visits to either the same site or a different site.
Scripts from one origin cannot read or write to another origin’s storage.
Because cookies are not origin-partitioned, third-party scripts included in the main execution context can read and write first-party cookies.
%
\textbf{\textit{Sharing} \symSharing.}
To share collected data with its own servers or a partner's, trackers can initiate network requests including the user data in one of four ways: (1)~as a URL query parameter, (2)~request payload, (3)~request header, or (4)~through HTTP cookies.
%
%
%
\textbf{\textit{Observation} \symObservation.}
A tracker present outside the browser can also passively observe and analyze network traffic from the user's browser or device.

\para{Related Systematization Work.}
Over the years, the web tracking community has produced a considerable amount of findings.
Unfortunately, these have largely remained scattered across disjoint works making it difficult to draw connections between related techniques, assess the state of the field, and identify future research directions.
Our SoK addresses these by providing the most recent, systematic, and unified view of the evolution of all known tracking techniques and their research-, industry-, and community-based defense mechanisms in parallel to major EU and US regulatory and compliance changes.
Prior survey and SoK papers in the space (see comparison overview in~\autoref{tab:prior-work}) have only partially covered such aspects; being too high-level~\cite{ermakova2018web,bujlowSurveyWebTracking2017}, too narrowly-focused on specific tracking mechanisms~\cite{laperdrix2020browser,rokickiSoKSearchLost2021}, research measurement methodologies~\cite{stafeevSoKStateKrawlers2024,riederSoKDecadesWeb2026}, or regulations exclusively~\cite{birrellSoKTechnicalImplementation2024}, or too old and outdated by newer techniques, defenses, and policies that appeared in recent years~\cite{mayerThirdPartyWebTracking2012,sanchez-rolaWebWatchingYou2017}.
%

%% file: tables/prior-work.tex
\begin{table*}[!ht]
    \small
    \centering
    \caption{Comparison overview to prior SoK and survey papers related to web tracking. \iconCookie: Cookie-based, \iconURL: Navigational,\\ \iconFingerprint: Fingerprinting, \iconJS: JavaScript-based, \iconSC: Side-channels, \iconPrivacy: Privacy proposals. \protect\pie{180}: Partial coverage compared to our SoK.
     }
    \label{tab:prior-work}
    \setlength{\tabcolsep}{2pt}
    \renewcommand{\arraystretch}{.7}
    \begin{tabular}{ccccccccccccc}
    \toprule
    & & & \multicolumn{3}{c}{\footnotesize\textbf{Tracking}} & & \multicolumn{3}{c}{\footnotesize\textbf{Defenses}}\\
    \footnotesize\textbf{Work} & \footnotesize\textbf{Year} & \footnotesize\textbf{Scope}&\rot{\footnotesize\textbf{Stateful}} &\rot{\footnotesize\textbf{Stateless}} &\rot{\footnotesize\textbf{Mixed}} &\footnotesize\textbf{Techniques} &\rot{\footnotesize\textbf{Industry}} &\rot{\footnotesize\textbf{Research}} &\rot{\footnotesize\textbf{Community}} &\footnotesize\textbf{Regulations} &\rot{\footnotesize\begin{tabular}[c]{@{}c@{}}\textbf{Future}\\ \textbf{Directions} \end{tabular}}
    &\rot{\footnotesize\begin{tabular}[c]{@{}c@{}}\textbf{Nb Papers}\\ \textbf{Studied} \end{tabular}}\\
    \midrule
         Mayer et al.~\cite{mayerThirdPartyWebTracking2012} & 2012 & Third-party tracking and policy & \pie{180} & \pie{180} & \emptypie{} & \iconCookie~\iconURL~\iconFingerprint~\iconJS & \emptypie{} & \pie{180} & \pie{180} & EU \& US & \xmark & $\approx 65$\\

         Bujlow et al.~\cite{bujlowSurveyWebTracking2017} & 2017 & Tracking techniques and defenses & \pie{180} & \pie{180} & \emptypie{} & \iconCookie~\iconURL~\iconFingerprint~\iconJS~\iconSC & \pie{180} & \pie{180} & \pie{180} & EU (briefly) & \checkmark & $\approx 70$ \\

         Sanchez-Rola et al.~\cite{sanchez-rolaWebWatchingYou2017} & 2017 & Techniques and countermeasures & \pie{180} & \pie{180} & \emptypie{} & \iconCookie~\iconFingerprint~\iconJS & \emptypie{} & \pie{180} & \pie{180} & EU \& US & \xmark &$\approx 35$\\

         Ermakova et al.~\cite{ermakova2018web} & 2018 & High-level tracking survey & \pie{180} & \pie{180} & \pie{180} & \iconCookie~\iconFingerprint~\iconJS & \emptypie{} & \pie{180} & \pie{180} & \xmark & \xmark  & $31$\\

         Laperdrix et al.~\cite{laperdrix2020browser} & 2020 & Browser fingerprinting & \emptypie{} & \pie{360} & \pie{180} & \iconFingerprint~\iconJS~\iconSC & \pie{180} & \pie{180} & \pie{360} & EU (briefly) & \checkmark &$\approx57$\\

         Rokicki et al.~\cite{rokickiSoKSearchLost2021} & 2021 & JS timers-based attacks &\emptypie{} & \pie{360} & \emptypie{} & \iconJS~\iconSC & \pie{360} & \pie{360} & \emptypie{} & \xmark & \checkmark &$\approx30$\\

         Stafeev et al.~\cite{stafeevSoKStateKrawlers2024} & 2024 & Crawlers' measurement methodology & \emptypie{} & \emptypie{} & \emptypie{} & \xmark & \emptypie{} & \emptypie{} & \emptypie{} & \xmark & \checkmark & $403 + 27$\\

         Birrell et al.~\cite{birrellSoKTechnicalImplementation2024} & 2024 & Internet privacy regulations  & \emptypie{} & \emptypie{} & \emptypie{} & \xmark & \emptypie{} & \emptypie{} & \emptypie{} & Worldwide & \checkmark & $270$\\

         Rieder et al.~\cite{riederSoKDecadesWeb2026} & 2026 & Detection measurement methodology & \pie{360} & \pie{180} & \emptypie{} & \iconCookie~\iconURL~\iconFingerprint~\iconJS & \emptypie{} & \pie{180} & \pie{180} & \xmark & \checkmark & $59 + 16$\\

         \midrule
         \textbf{Ours} & 2026 & Techniques, mitigations, regulations & \pie{360} & \pie{360} & \pie{360} & \iconCookie~\iconURL~\iconFingerprint~\iconJS~\iconSC~\iconPrivacy & \pie{360} & \pie{360} & \pie{360} & EU \& US & \checkmark & $255$\\
     \bottomrule
    \end{tabular}
\end{table*}

%% file: short-version/03-methodology.tex
\section{Methodology}
\label{sec:methodology}
\label{sec:lit-review}
\label{sec:systematization}

\begin{figure}[!ht]
    \centering
    \includegraphics[width=\columnwidth]{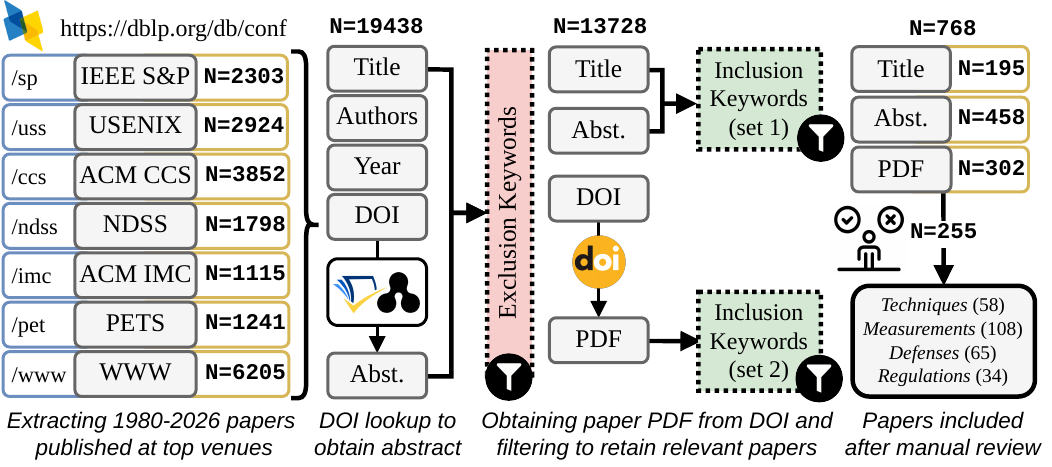}
    \caption{Overview of our literature search.}
    \label{fig:methodology}
\end{figure}

\para{Literature Review.}
To systematize the evolution of web tracking, we perform a literature search (see~\autoref{fig:methodology}) to identify all relevant papers published at any of the following seven venues---IEEE S\&P, USENIX Security, ACM CCS, NDSS, ACM IMC, PETS, and WWW---since their creation (i.e., from 1980 to 2026\footnote{Only papers available on DBLP (\url{https://dblp.org/db/conf}) by June 1st, 2026 were included. Papers were queried using conference-specific paths, e.g., for PETS, we use `conf/pet' (2000-2014) \& `journals/popets' (2015+).}).
We obtain a total of 19438 papers that we filter using keywords (see~\refappendix{sec:keywords}). 
These keywords were refined over multiple iterations by discussion among two expert authors.
The resulting list of 768 candidate papers is manually reviewed to further exclude  publications unrelated to web tracking techniques, mitigations, or regulatory compliance (e.g., IoT tracking, advertising use cases).

\para{Systematization.}
This yields a final set of 255 research papers across the following dimensions: 58 tracking techniques and 108 measurements, 65 defenses, and 34 regulatory compliance papers (see~\refappendix{sec:thematic-analysis} for details).
We systematize the evolution of web tracking along these dimensions by following Clarke and Braun’s thematic analysis~\cite{clarke2013successful} that eliminates the need for inter-rater reliability.
First, one researcher with expertise in web tracking, reviewed each included paper in detail to evaluate it based on our systematization criteria listed in~\autoref{tab:systematization-criteria}.
For each criteria, the expert assigned one or more relevant categories. 
Next, the evaluator convened with a second expert to deliberate the assigned categories, resolve any ambiguities, and reach a consensus on the final taxonomy.
We structure the SoK around this taxonomy and systematize evolution of tracking techniques, defenses, and regulatory compliance, while identifying research gaps and future directions.
When needed, we use our domain expertise to complement our explanations with other relevant publications or aspects.

\input{tables/systematization-criteria}

\para{Approach Limitations.}
%
\textbf{\textit{Selection Bias.}} We scope our SoK to the seven aforementioned venues where web tracking research has been commonly published, and so, we may have missed other relevant publications.
\textbf{\textit{Coding Bias.}} As with any qualitative analysis, our systematization is subject to a degree of subjectivity.
We mitigate this following Clarke and Braun's thematic analysis approach~\cite{clarke2013successful} coupled with our domain expertise.
%
%
\textbf{\textit{Focus on EU and US Regulations.}} We scope our approach to the EU and US contexts because they are the closest to our expertise and their regulations have often been precursors to others around the globe\footnote{See \url{https://cnil.fr/en/data-protection-around-the-world}}.
For a broader exploration of privacy regulations elsewhere and their nuances, we refer readers to the following SoK from Birrell et al. ~\cite{birrellSoKTechnicalImplementation2024}.
\textbf{\textit{Scope coverage.}} We consider adjacent topics such as advertising out-of-scope. Also, we focus on systematizing high-level advances across literature rather than detailing findings from individual papers. Due to space constraints, mixed tracking techniques and their defenses are deferred to \refappendix{sec:mixed-tracking}.

%% file: tables/systematization-criteria.tex
\begin{table*}[ht]
  \small
\centering
\caption{Summary of systematization criteria and taxonomy for tracking techniques, defenses, and regulatory compliance.}
\label{tab:systematization-criteria}
\renewcommand{\arraystretch}{1}
\begin{tabular}{p{0.004cm} p{7.8cm} p{8.66cm}}
\toprule
& \textbf{Criteria and Description} & \textbf{Taxonomy} \\
\midrule

\multirow[c]{6}{*}{\rotatebox[origin=c]{90}{\textbf{Techniques}}}
 & \textbf{Technique}: What tracking is studied in this research? & See 24 techniques in~\autoref{tab:artifact-capability-matrix}.\\
 & \textbf{Artifact(s)}: What artifact(s) is used by this technique? & See 8 artifact types listed in~\autoref{tab:artifact-taxonomy}. \\
 & \textbf{Capabilities}: What capabilities are required by this technique? & \symInclusion\ Inclusion, \symExecution\ Execution, \symObservation\ Observation, \symStorage\ Storage, \symSharing\ Sharing. \\
 & \textbf{State}: What state does this technique maintain? & Stateful, Stateless, Mixed. \\
 & \textbf{Context}: In what context does it function? & Third-party, First-party, Both. \\
 & \textbf{Goals}: What are the goals of the tracker using this technique? & Same-site, Cross-site, Cross-context. \\
\midrule

\multirow[c]{7}{*}{\rotatebox[origin=c]{90}{\textbf{Defenses}}}
 & \textbf{Technique}: What tracking does this defense protect against? & See 24 techniques in~\autoref{tab:artifact-capability-matrix}.\\
 & \textbf{Artifact(s)}: What artifact(s) does it apply protection to? & See 8 artifact types listed in~\autoref{tab:artifact-taxonomy}. \\
 & \textbf{Development}: Who developed this tracking defense? & Research-based, Industry-based, Community-based. \\
 & \textbf{Deployment}: How is this defense deployed? & \dNative\ In-browser, \dExt\ Extension, \dServer\ Infrastructural, \dTool\ Offline. \\  
 & \textbf{Support}: What is the support for this defense? & See~\autoref{tab:stateful-tracking-defenses}, \autoref{tab:stateless-tracking-defenses}, and~\autoref{fig:evolution-defenses}. \\
 & \textbf{Goals}: What are the goals of this defense mechanism? & \colorbox{cellred}{Detection \& Blocking} and \colorbox{cellyellow}{Signal Reduction \& Obfuscation}. \\
 & \textbf{Type}: How can the defense mechanism be classified? & See 8 research (\autoref{tab:defense-goal-mechanism-taxonomy}), 23 industry\&community (\autoref{tab:stateful-tracking-defenses}, \autoref{tab:stateless-tracking-defenses}).\\
\midrule

\multirow[c]{3}{*}{\rotatebox[origin=c]{90}{\small\textbf{Policies}}}
& \textbf{Policies}: What regulations and standards are studied? & See EU and US regulations, and standards in ~\autoref{tab:regulations-paper-distribution}. \\
& \textbf{Aspects}: Which aspects of these policies are covered? & See mapping of 9 policy aspects onto 6 user rights in~\autoref{tab:regulations-aspects}.\\
& \textbf{Compliance}: What are the enforcement challenges?  & Reconciling \colorbox{cellpink}{Incentives}, \colorbox{cellblue}{Expectations}, and \colorbox{cellgreen}{Responsibilities}.\\
\midrule
\multirow[c]{2}{*}{\rotatebox[origin=c]{90}{\textbf{Gaps}}}
& \textbf{Open Problems}: What are potential gaps and future directions in the evolution of these techniques, defenses, and regulations? & See the 24 open problems throughout the paper as well as 4 open problems in appendix.\\

\bottomrule
\end{tabular}
\end{table*}

%% file: short-version/04-tracking-techniques.tex
\section{Tracking Techniques}
\label{sec:tracking-techniques}

Our approach described in \autoref{sec:methodology} to systematize 58 papers on specific web tracking techniques, complemented by 108 measurement studies that characterize their real-world aspects, yields 24 distinct tracking techniques, mapped in~\autoref{tab:technique-papers}.
Fundamentally, each tracking technique relies on artifacts tied either to the user, their device, or their browsing activity.
To this end, we systematize the artifacts that trackers are observed to rely on across different papers into eight categories (\autoref{tab:artifact-taxonomy}).
\autoref{tab:artifact-capability-matrix} systematizes research to map tracker capabilities exercised over artifact categories for each tracking technique, with a paper-level breakdown represented in~\autoref{tab:paper-artifacts}. 
This section systematizes tracking techniques by state maintained by the trackers across different contexts.

%

\input{tables/techniques-artifacts-capabilities}

%

\subsection{Stateful Tracking}
\label{sec:stateful-tracking}

\textit{Stateful tracking} techniques involve maintaining \textit{state} by storing a unique identifier in the user's browser and retrieving it or updating it across visits, using either dedicated browser-provided interfaces (e.g., cookies) or as side effects of other functionality (e.g., E-Tags, HSTS upgrades)~\cite{englehardtAutomatedDiscoveryPrivacy2018,ashkansoltaniFlashCookiesPrivacy2011,solomosTalesFaviconsCaches2021}.


\subsubsection{Cookie-based Tracking}
\label{sec:stateful-techniques-cookies}

Cookie-based tracking relies on the browser's cookie jar \symStorage\ to maintain state, with cookies set and retrieved via HTTP headers \symSharing\ or JavaScript (JS) APIs such as \texttt{document.cookie} \symExecution.
Cookies were adopted for cross-site tracking almost immediately after their introduction in the 1990s~\cite{montulliHTTPStateManagement1997}, prompting early privacy concerns~\cite{kristolHTTPCookiesStandards2001,SenatorRaisesPrivacy2001}.
Despite these concerns, third-party cookie-based tracking remained dominant for nearly two decades, becoming both widespread and concentrated among a few third parties~\cite{krishnamurthyPrivacyDiffusionWeb2009,lernerInternetJonesRaiders2016,englehardtOnlineTracking1millionsite2016,yuTrackingTrackers2016,karajWhoTracksMeShedding2019}.

Browsers mediate cookie access based on the context in which they are set
~\cite{beugin2024WebAlmanac2024,singhIncoherenciesWebBrowser2010,mdnSameoriginPolicySecurity2024}.
As Safari and Firefox began blocking or partitioning third-party cookies by default~\cite{TrackingPreventionWebKit2020,mdnThirdpartyCookiesPrivacy2024}  between 2018 and 2020 (\autoref{sec:stateful-defenses}), tracking shifted to the first-party context---nearly 90\% of websites now set first-party tracking cookies, 96\% of which are set by third-party scripts, making browsing with third-party cookies disabled largely indistinguishable from the pre-blocking web~\cite{munirCookieGraphUnderstandingDetecting2023,vekaria2024WebAlmanac2024,linBrowsingThirdPartyCookies2024}.
Meanwhile, trackers continue exploiting browser- and protocol-level edge cases such as opting out of \texttt{SameSite=Lax} and instead relying on \texttt{SameSite=None} cookies~\cite{khodayariStateSameSiteStudying2022}, to retain stateful tracking capabilities~\cite{bashirHowTrackingCompanies2018}.

\subsubsection{Cookie Respawning}
\label{sec:stateful-techniques-cookie-respawning}
Once a user clears their cookies, a tracker that has stashed the same identifier somewhere else in the browser can rewrite it back into the cookie jar, ``\textit{respawning}'' or reconstructing the identifier and creating a so-called ``supercookie'' or ``evercookie''~\cite{ayensonFlashCookiesPrivacy2011,soltaniFlashCookiesPrivacy2009,acarWebNeverForgets2014} as described in~\autoref{sec:stateful-defenses-evercookie}.
Respawning was first demonstrated at scale using Flash Local Shared Objects and other legacy plugin storage~\cite{acarWebNeverForgets2014,englehardtOnlineTracking1millionsite2016}, and later shown to also draw on browser fingerprints themselves---a 2022 study measured a lower bound on fingerprint-based respawning by detecting it on 1,150 of the top 30K sites~\cite{fouadMyCookiePhoenix2022}.
More recently, respawning has expanded beyond the browser, leveraging cross-context state shared with native applications to recover identifiers~\cite{wessels2025hytrack} (further discussed in~\refappendix{sec:mixed-techniques}).

\subsubsection{Cookie Syncing}
\label{sec:stateful-techniques-cookie-syncing}
To exchange information about users across domains included on a webpage \symInclusion, third-party companies rely on \textit{cookie syncing}~\cite{acarWebNeverForgets2014}---first described by Olejnik et al.~\cite{olejnikSellingPrivacyAuction2014}---which synchronizes user identifiers maintained in each party's cookie jar \symStorage, most commonly shared via URL parameters embedded in redirects \symSharing\ between them, requiring script execution \symExecution\ to initiate the sync.
Earlier works formalized cookie syncing detection~\cite{acarWebNeverForgets2014,englehardtOnlineTracking1millionsite2016,bashirTracingInformationFlows2016}, which has since been refined~\cite{papadopoulosCookieSynchronizationEverything2019} and scaled into automated, graph- and ML-based pipelines tracing identifier flows across the ecosystem~\cite{papadopoulosCookieSynchronizationEverything2019, iqbalKhaleesiBreakerAdvertising2022}.
Cookie syncing is not just confined to the third-party context~\cite{sanchez-rolaJourneyCenterCookie2021}, as identifiers in first-party cookies have been observed to be shared with hundreds of other third-party domains~\cite{munirCookieGraphUnderstandingDetecting2023,bahramiCOOKIEGUARDCharacterizingIsolating2024}.

\subsubsection{Tracking Pixels/Tags}
\label{sec:stateful-techniques-tags}

\textit{Tracking pixels}, traditionally referred to as 1x1 image elements embedded \symInclusion\ on a webpage, shared \symSharing\ user data with a tracking endpoint over HTTP traffic artifacts, enabling same- and cross-site tracking
~\cite{Narayanan2017WebtapSpringer,englehardtOnlineTracking1millionsite2016}.
Their prevalence and privacy risks have been measured extensively, from early archaeological studies of the tracking ecosystem~\cite{lernerInternetJonesRaiders2016,englehardtOnlineTracking1millionsite2016} to recent audits of their collection of personal data~\cite{huCharacterizingPixelTracking2019,bekosPIIxelLeaksPassive2025}.
As browser restrictions rendered third-party image-based pixels increasingly ineffective, modern tracking pixels have evolved into JS-based \textit{tracking tags} that execute \symExecution\ with the embedding site's first-party privileges to collect richer behavioral data~\cite{bekosHitchhikersGuideFacebook2023,munirCookieGraphUnderstandingDetecting2023}, while trackers increasingly obfuscate pixel behavior via falsified image dimensions, redirect chains, and encoded identifiers~\cite{huCharacterizingPixelTracking2019}.
Modern tags have also expanded in scope to support tag management, bot detection, session replays, and email tracking~\cite{senolLeakyFormsStudy2022,acarNoBoundariesData2020,englehardtNoBoundariesCredentials2018,englehardtNeverSignedThis2018,fouadMissedFilterLists2020}. Recent studies have shown deployments of the same vendor's tag to vary substantially across websites in their data collection~\cite{kiesermanTrackerInstallationsAre2025,bekosPIIxelLeaksPassive2025,ghaniPixelConfigLongitudinalMeasurement2026}.

\begin{opbox}
Detecting the mere presence of a tracking tag is not enough to understand its capabilities---how do different configurations across sites and vendors determine what data is tracked, collected and shared, and how those settings evolve? 
What generalizable approaches can systematically measure tracking deployments across tags?
\end{opbox}

\subsubsection{CNAME Tracking}
\label{sec:stateful-techniques-cname}
Rather than embedding a third-party resource directly, a website can create a CNAME DNS record pointing a first-party subdomain (e.g., \textit{metrics.news.com}) to a third-party's infrastructure.
Because the browser only resolves \textit{metrics.news.com}, cookies set through this alias are treated as first-party \symStorage, allowing trackers to evade defenses that block third-party domains rather than DNS resolution~\cite{dimovaCNAMEGameLargescale2021} (a network protocol artifact). 
First documented in 2014~\cite{olejnikSellingPrivacyAuction2014,olejnikBidNotBid2014}, CNAME cloaking became widely adopted later in the decade as a response to browser anti-tracking measures. Deployed on roughly one in ten of the top-10k websites~\cite{dimovaCNAMEGameLargescale2021}, cloaked requests are indistinguishable from legitimate first-party traffic~\cite{dimovaCNAMEGameLargescale2021,chenCookieSwapParty2021}.

More broadly, CNAME cloaking supports an emerging \emph{server-side tracking} paradigm, which reuses client-side pixel/tag instrumentation to collect tracking data but routes the resulting requests through a publisher-controlled first-party endpoint---often a CNAME-aliased or A/AAAA-record subdomain~\cite{jazlanSSTGuardDetectingCharacterizing2026}---before forwarding payloads to the tracker's backend via server-to-server communication, rendering client-side defenses ineffective~\cite{fouadDevilDetailsDetection2024}.
Major platforms including Google, Meta, and TikTok have already deployed conversion APIs adopting this architecture
~\cite{fraihiClientsideServersideTracking2024}.

\begin{opbox}
As tracking logic shifts from client-side to server-side, it removes the vantage point that the research community has relied on to detect and mitigate online tracking. How can we characterize the prevalence and behavior of server-side tracking technologies, and what new methodologies (e.g., differential traffic probing, cooperation with first parties) are needed to study and defend against it?
\end{opbox}

\subsubsection{Navigational Tracking via Link Decorations}
\label{sec:stateful-techniques-link-decorations}

\textit{Link decorations} allow sharing identifiers \symSharing\ via query parameters, resource path segments, or URL fragments appended to an otherwise ordinary navigation.
The technique dates to the standardization of query parameters (RFC~1630, 1994) and the \texttt{Referer} header (RFC~1945, 1996), but became a deliberate tracking channel with the introduction of UTM parameters by Urchin/Google Analytics in 2005 and Google's click ID (\texttt{gclid}) in 2010.
As browsers began restricting third-party cookies, major ad platforms followed with their own click IDs (e.g., Meta's \texttt{fbclid}, Microsoft's \texttt{msclkid}), and over 70\% of websites now use decorations to share user data such as first-party cookies and email across navigations~\cite{munirPURLSafeEffective2024,koopInDepthEvaluationRedirect2020}.
Similarly, link \textit{shimming} involves a first-party rewriting the outbound links with decorations to route clicks through its own domain before redirecting.
While used for HTTPS upgrading or malicious-URL blocking, it equally enables the first party to log every outbound navigation~\cite{liShimShimmenyEvaluating2020}.
A recent audit of the \texttt{Referer} header further shows that inconsistent browser enforcement of Referrer-Policy still permits the cross-site identifier leakage the policy was designed to prevent~\cite{zagiReferrerPolicyImplementation2025}.

\subsubsection{Bounce Tracking}
\label{sec:stateful-techniques-bounce-tracking}

\textit{Bounce tracking} is another navigation-based mechanism that renders third-party cookie blocking ineffective~\cite{kellyBounceTrackingMitigations2022}.
Unlike link shimming, 
bounce tracking specifically involves navigating a \emph{third-party} tracker's domain in first-party context, momentarily surfacing it during a redirect chain so it can read and write identifiers (via JS APIs \symExecution\ or cookie headers \symSharing) that persist in first-party storage \symStorage, linking them with website-specific identifiers for cross-site tracking.
Bounce chains can involve two (\texttt{news.com} $\rightarrow$ \texttt{tracker.com} $\rightarrow$ \texttt{news.com}) or more unrelated sites.
%
Due to usability disruptions and defensive measures, bounce tracking is not as widely pervasive as link decorations~\cite{iqbalKhaleesiBreakerAdvertising2022}, but both mechanisms are often used together.
Measurements in 2020 found that 11.6\% of sites use one of the top 100 redirectors~\cite{koopInDepthEvaluationRedirect2020}, and in 2022 such identifiers were present in 8.1\% of crawled navigations~\cite{randallMeasuringUIDSmuggling2022}.

\begin{opbox}
Each generation of navigational tracking, from UTM parameters to click IDs to bounce tracking, emerged as its predecessor was restricted. How will trackers adapt these channels to complement server-side tracking and how can we defend against it?
\end{opbox}

\noindent Overall, every time an anti-tracking defense closes off one tracking vector, stateful trackers relocate tracking a layer further from the browser---from third-party cookies, to first-party cookies and synced identifiers, to DNS aliases, to server-side.

\begin{opbox}
As tracking logic keeps moving away from the browser, the community's tooling
needs to advance with it, or risk studying an increasingly small unrepresentative slice of how users are actually tracked.
\end{opbox}


\subsection{Stateless Tracking}
\label{sec:stateless-tracking}

Unlike stateful tracking, \textit{stateless} tracking does not rely on stored identifiers but instead collects artifacts of the user's behavior, browser, or hardware to generate a fingerprint that correlates users across sites and sessions.
Nearly all stateless techniques are realized by exercising the \symExecution\ capability, with JavaScript running in the including context's privileges \symInclusion\ .

\para{Fingerprinting.} 
Fingerprinting corresponds to the gathering of information on users’ behavioral (e.g., mouse movements, keystrokes), browser (e.g., version, screen size, installed fonts), and hardware/OS in real-time via HTTP headers and JS APIs to identify users.
%
%
Its use for tracking is hard to prevent due to the duality of fingerprinting (see~\refappendix{sec:need-for-fingerprinting}) which serves both constructive (e.g., bot and fraud detection) and destructive (e.g., tracking anonymous users) purposes. 
%
Recent work only began to differentiate intent, e.g., by using ad changes as a heuristic for fingerprinting-driven tracking~\cite{liuFirstEarlyEvidence2025}.

\begin{opbox}
Can we automatically assign the purposes of trackers to determine fingerprinting intent and block only tracking use cases while allowing the benign ones?
\end{opbox}

\subsubsection{Passive Fingerprinting} 
\label{sec:stateless-techniques-passive-fingperinting}
%
\textit{Passive fingerprinting} exploits metadata attached to every request---most commonly IP address and User-Agent string (IPUA).
Despite its instability across networks, IP alone, or combined with minimal additional signals, can identify users~\cite{mishraDontCountMe2020,klineStructureCharacteristicsUser2017}, and even IPv6 address rotations intended to improve privacy is undermined by predictable ISP schedules and stable interface-identifier components~\cite{ryeFollowScentDefeating2021}.
Recent User-Agent reduction efforts have partially shrunk this surface, but remaining passive signals and Chrome's new Client Hints API still carry substantial identifying information~\cite{eckersleyHowUniqueYour2010,senolUnveilingImpactUserAgent2023,wieflingPrivacyMeasureTurned2024}.

\subsubsection{Browser Fingerprinting} 
\label{sec:stateless-techniques-browser-fingperinting}

\textit{Browser fingerprinting} actively probes browser APIs to collect configuration information about the browser, device, and OS.
Since the first academic work on browser fingerprinting in 2009~\cite{mayerAnyPersonPamphleteer2009} and the Electronic Frontier Foundation's Panopticlick shortly after~\cite{eckersleyHowUniqueYour2010}, researchers have extensively studied its impact on privacy and diversity~\cite{laperdrixBeautyBeastDiverting2016,gomez-boixHidingCrowdAnalysis2018}, its evolution and stability over time~\cite{vastelFPSTALKERTrackingBrowser2018}, its detection~\cite{suAutomaticDiscoveryEmerging2023,nomoto2023browser}, and discovered abusable browser APIs~\cite{olejnikLeakingBatteryPrivacy2016,dasWebsSixthSense2018,linFashionFauxPas2023,huygheFPRainbowFingerprintBasedBrowser2025}.
We refer readers to Laperdrix et al.'s SoK~\cite{laperdrix2020browser} for a comprehensive understanding of foundational aspects of browser fingerprinting.
%
%
Over the years, its prevalence has grown steadily~\cite{nikiforakisCookielessMonsterExploring2013,acarFPDetectiveDustingWeb2013,englehardtOnlineTracking1millionsite2016,iqbalFingerprintingFingerprintersLearning2021}. 
%
%
Browser fingerprints are inherently unstable, with 80\% of instances changing within 10 days~\cite{vastelFPSTALKERTrackingBrowser2018,liWhoTouchedMy2020}. 
So trackers often link evolving fingerprints by combining them with stateful techniques such as cookie respawning~\cite{fouadMyCookiePhoenix2022} (\autoref{sec:stateful-techniques-cookie-respawning}). 
Meanwhile, the exploitable surface continues to widen beyond Canvas, WebGL, and Audio APIs~\cite{chaliseYourSpeakerMy2022}: permission states~\cite{nomoto2023browser}, CSS and font-rendering differences~\cite{linFashionFauxPas2023}, divergent JS execution across platforms~\cite{zafarSameScriptDifferent2025}, and, even privacy-hardening switches themselves~\cite{huygheFPRainbowFingerprintBasedBrowser2025}.
Conclusions from prior studies about fingerprint uniqueness vary with dataset scale, 
and it remains unclear whether they generalize across devices, populations~\cite{berkeHowUniqueWhose2025,pradeepNotYourAverage2023}, or real browsing conditions~\cite{muthu2025beyond,deuserBrowsingUnicityLimits2020}, rendering its purpose and integration within live systems poorly understood.
%

\begin{opbox}
What is the real impact of fingerprinting at scale, on vulnerable populations (e.g., minorities, children, marginalized), and paired with other techniques?
\end{opbox}

\subsubsection{Hardware Fingerprinting} 
\label{sec:stateless-techniques-hardware-fingperinting}

\textit{Hardware fingerprinting} exploits physical-layer variations to identify devices, first demonstrated via TCP clock-skew measurements~\cite{kohnoRemotePhysicalDevice2005} and later extended to the web through JS-accessible device~\cite{nikiforakisCookielessMonsterExploring2013} and cross-browser OS/hardware~\cite{caoCrossBrowserFingerprintingOS2017} properties.
Researchers have since demonstrated fingerprinting of specific components including CPUs~\cite{trampertBrowserBasedCPUFingerprinting2022,sanchez-rolaClockClockTimeBased2018}, GPUs via rendering variations~\cite{laorDRAWNAPARTDevice2022}, DRAM via Rowhammer-styled access patterns~\cite{venugopalanFPRowhammerDRAMBasedDevice2024}, and kernel-level PRNG state~\cite{kleinCrossLayerAttacks2021,kol2023device}.
The mobile ecosystem expands this surface through BLE frames \cite{martinHandoffAllYour2019} and motion, orientation, and acoustic sensors that are accessible without permission prompts~\cite{zhouAcousticFingerprintingRevisited2014,zhang2019sensorid,dasWebsSixthSense2018}.
%

\begin{opbox}
Hardware fingerprinting exploits manufacturing variations that cannot be mitigated via software updates. What principled abstractions or hardware-software co-designs can provably bound the fingerprintable information leaking through device APIs, while preserving functionality and security properties (e.g., hardware attestation, PUFs) that rely on the same physical-layer uniqueness?
\end{opbox}

\subsubsection{Behavioral Fingerprinting} 
\label{sec:stateless-techniques-behavioral-fingperinting}

Fine-grained interaction signals such as mouse movements, keystroke patterns, and touch gestures can reliably distinguish users and support continuous authentication~\cite{attererKnowingUsersEvery2006,zhengEfficientUserVerification2011,leivaMyMouseMy2021,fereidooniAuthentiSenseScalableBehavioral2023}.
Mobile touch dynamics alone offers roughly 20 bits of entropy without requiring any permissions~\cite{masoodTouchYoureTrappcked2018}.
%
%
Keystroke patterns link users across devices~\cite{yuanCrossdeviceTrackingIdentification2018}, engagement patterns correlate users across unrelated sites~\cite{gogaExploitingInnocuousActivity2013}, and form-filling behavior is collected at scale via session replay technologies~\cite{linFillBlanksEmpirical2020,cuiUnderstandingPrivacyNorms2025,acarNoBoundariesData2020}.
Recent work has begun assessing how well behavioral methods hold up as a general-purpose fingerprinting signal~\cite{crichtonRethinkingFingerprintingAssessment2025}.

\begin{opbox}
As AI agents automate web browsing, they will impact behavioral fingerprinting techniques, raising questions on its effectiveness, attribution, and evasion.
\end{opbox}

\subsubsection{Extension Fingerprinting} 
\label{sec:stateless-techniques-extension-fingperinting}

Browser \textit{extension fingerprinting} infers the presence of specific extensions, which can reveal sensitive user attributes (e.g., health conditions, religion)~\cite{karamiCarnusExploringPrivacy2020}.
Early approaches relied on probing web-accessible resources (WARs) that extensions expose~\cite{starovXHOUNDQuantifyingFingerprintability2017,starovExtendedTrackingPowers2017,starovUnnecessarilyIdentifiableQuantifying2019}, but because static WAR probing is comparatively easy to randomize away~\cite{trickelEveryoneDifferentClientside2019}, research shifted towards \textit{behavioral} extension fingerprinting---inferring extensions through side-effects during execution such as information leakage across components~\cite{chenMystiqueUncoveringInformation2018}, user-triggered actions~\cite{solomosDangersHumanTouch2022}, injected CSS~\cite{laperdrixFingerprintingStyleDetecting2021}, and transient DOM edits~\cite{solomosEscapingConfinesTime2022}.
Behavioral fingerprints have proven far more resilient, with 88\% remaining effective even against dedicated countermeasures \cite{karamiCarnusExploringPrivacy2020}, making defense fundamentally harder than blocking resource enumeration.
Most recently, even ad blockers have been shown to be fingerprintable through scriptless, filter-list-driven side effects, where customizing a rule set paradoxically shrinks a user's anonymity set~\cite{chehade2025double}.

\begin{opbox}
How can future browser architectures and extension frameworks be co-designed so that extensions can modify page behavior without exposing detectable side-effects, and what privacy guarantees can such designs formally provide against an adaptive adversary that combines behavioral, stylistic, and scriptless fingerprinting vectors?
\end{opbox}

\subsubsection{Side-channel Fingerprinting} 
\label{sec:stateless-techniques-side-channel-fingperinting}

\textit{Side-channel fingerprinting} exploits unintended information leakage across browser, network, and hardware layers to track users without relying on stored identifiers or active probing. These mechanisms are challenging to detect and mitigate.
%
%

\para{Kernel-based Side-channels.}
At the OS level, process-accounting interfaces leak enough about recent activity to fingerprint browsing history from kernel artifacts alone~\cite{janaMementoLearningSecrets2012}. 
Additionally, Linux's shared kernel PRNG state lets attackers correlate DNS and TCP/IP behavior into a stable device identifier that persists until reboot~\cite{kleinCrossLayerAttacks2021,kol2023device}.

\para{Network-based Side-channels.}
Encrypted traffic still leaks sensitive state through observation \symObservation\ of network-level artifacts such as packet sizes and timing~\cite{chenSideChannelLeaksWeb2010,gonzalezUserProfilingTime2016}, TLS client certificates act as persistent supercookies rarely revoked by users~\cite{foppeExploitingTLSClient2018}, and QUIC's connection-migration feature lets server-chosen connection IDs be encoded and visible to on-path observers even as clients change networks~\cite{syQUICLookWeb2019}.

\para{Storage-based Side-channels.}
Multiple browser caching layers such as favicon cache~\cite{solomosTalesFaviconsCaches2021}, HTTP cache, ETags, Alt-Svc~\cite{mishraDejaVuAbusing2021}, and DNS resolver cache~\cite{klein2019dns} were independently proven abusable as cross-context identifier stores \symStorage, since none were originally designed with cross-site isolation in mind.
More recent works have demonstrated that shared browser resource pools create cross-context timing channels in all major browsers~\cite{snyderPoolPartyExploitingBrowser2023} and automated feature-testing frameworks have surfaced previously unknown tracking vectors by probing browser features systematically~\cite{aliNavigatingMurkyWaters2023}.

\para{Timing-based Side-channels.}
Timing attacks on the web date to Felten and Schneider's 2000 demonstration that cache behavior reveals browsing history~\cite{feltenTimingAttacksWeb2000}, and have since escalated through JS-driven CPU cache attacks~\cite{orenSpySandboxPractical2015}, browser-level side-channels independent of network conditions~\cite{van2015clock}, and concurrency-based timing that eliminates the need for absolute clocks~\cite{van2020timeless}.
Most notably, Chrome's Site Isolation---introduced specifically to counter Spectre---was itself shown to introduce new cross-origin timing side-channels~\cite{jinTimingBasedBrowsingPrivacy2022}.

\begin{opbox}
Major browser features added for performance or security have often retrospectively introduced a tracking side-channel. Can browsers move from reactive patching to proactive side-channel auditing of new APIs and architectural changes before deployment? How can researchers build automated discovery frameworks directly integrable into the browser feature review process?
\end{opbox}

%% file: tables/techniques-artifacts-capabilities.tex
\begin{table*}[t]
\centering
\scriptsize
\setlength{\tabcolsep}{1pt}
\renewcommand{\arraystretch}{1}
\caption{Tracker capabilities most commonly exercised over each artifact (rows) for each tracking technique (columns). \protect\symInclusion~Inclusion, \protect\symExecution~Execution, \protect\symStorage~Storage, \protect\symSharing~Sharing, \protect\symObservation~Observation.}
\label{tab:artifact-capability-matrix}
\begin{tabular}{%
|>{\raggedright\arraybackslash}m{3.0cm}|>{\centering\arraybackslash}m{0.63cm}|>{\centering\arraybackslash}m{0.63cm}|>{\centering\arraybackslash}m{0.63cm}|>{\centering\arraybackslash}m{0.63cm}|>{\centering\arraybackslash}m{0.63cm}|>{\centering\arraybackslash}m{0.63cm}|>{\centering\arraybackslash}m{0.63cm}|>{\centering\arraybackslash}m{0.63cm}|>{\centering\arraybackslash}m{0.63cm}|>{\centering\arraybackslash}m{0.63cm}|>{\centering\arraybackslash}m{0.63cm}|>{\centering\arraybackslash}m{0.63cm}|>{\centering\arraybackslash}m{0.63cm}|>{\centering\arraybackslash}m{0.63cm}|>{\centering\arraybackslash}m{0.63cm}|>{\centering\arraybackslash}m{0.63cm}|>{\centering\arraybackslash}m{0.63cm}|>{\centering\arraybackslash}m{0.63cm}|>{\centering\arraybackslash}m{0.63cm}|>{\centering\arraybackslash}m{0.63cm}|%
}
\toprule
\multicolumn{1}{|>{\centering\arraybackslash}b{3.0cm}|}{\shortstack[c]{\textbf{Tracked}\\\textbf{Artifacts}}} & \multicolumn{1}{>{\centering\arraybackslash}b{0.63cm}|}{\rot{Third-party Cookie Tracking}} & \multicolumn{1}{>{\centering\arraybackslash}b{0.63cm}|}{\rot{Cookie Respawning}} & \multicolumn{1}{>{\centering\arraybackslash}b{0.63cm}|}{\rot{Cookie Syncing}} & \multicolumn{1}{>{\centering\arraybackslash}b{0.63cm}|}{\rot{First-party Cookie Tracking}} & \multicolumn{1}{>{\centering\arraybackslash}b{0.63cm}|}{\rot{Tracking Pixels/Tags}} & \multicolumn{1}{>{\centering\arraybackslash}b{0.63cm}|}{\rot{CNAME Tracking}} & \multicolumn{1}{>{\centering\arraybackslash}b{0.63cm}|}{\rot{Link Decorations}} & \multicolumn{1}{>{\centering\arraybackslash}b{0.63cm}|}{\rot{Bounce Tracking}} & \multicolumn{1}{>{\centering\arraybackslash}b{0.63cm}|}{\rot{Authentication-based Tracking}} & \multicolumn{1}{>{\centering\arraybackslash}b{0.63cm}|}{\rot{Cross-context Tracking}} & \multicolumn{1}{>{\centering\arraybackslash}b{0.63cm}|}{\rot{Javascript-based Tracking}} & \multicolumn{1}{>{\centering\arraybackslash}b{0.63cm}|}{\rot{Browser Fingerprinting}} & \multicolumn{1}{>{\centering\arraybackslash}b{0.63cm}|}{\rot{Extension Fingerprinting}} & \multicolumn{1}{>{\centering\arraybackslash}b{0.63cm}|}{\rot{Device Fingerprinting}} & \multicolumn{1}{>{\centering\arraybackslash}b{0.63cm}|}{\rot{Passive Fingerprinting}} & \multicolumn{1}{>{\centering\arraybackslash}b{0.63cm}|}{\rot{Kernel-based Side-channels}} & \multicolumn{1}{>{\centering\arraybackslash}b{0.63cm}|}{\rot{Network-based Side-channels}} & \multicolumn{1}{>{\centering\arraybackslash}b{0.63cm}|}{\rot{Timing-based Side-channels}} & \multicolumn{1}{>{\centering\arraybackslash}b{0.63cm}|}{\rot{Storage-based Side-channels}} & \multicolumn{1}{>{\centering\arraybackslash}b{0.63cm}|}{\rot{Behavioral Tracking}} \\
\midrule
\rh Browser Storage and Cache & \cellsym{\symStorage} & \cellsym{\symStorage} & \cellsym{\symStorage} & \cellsym{\symStorage} & \cellsym{\symStorage} & \cellsym{\symStorage} & \cellsym{\symStorage} & \cellsym{\symStorage} & \cellsym{\symStorage} & \cellsym{\symStorage} & \cellsym{\symStorage} & \ec & \ec & \ec & \ec & \ec & \ec & \ec & \cellsym{\symStorage} & \ec \\
\hline
\rh JS-Accessible Browser APIs & \cellsym{\symExecution} & \cellsym{\symExecution} & \cellsym{\symExecution} & \cellsym{\symExecution} & \cellsym{\symExecution} & \ec & \ec & \cellsym{\symExecution} & \ec & \cellsym{\symExecution} & \cellsym{\symExecution} & \cellsym{\symExecution} & \cellsym{\symExecution} & \cellsym{\symExecution} & \ec & \cellsym{\symExecution} & \ec & \cellsym{\symExecution} & \cellsym{\symExecution} & \cellsym{\symExecution} \\
\hline
\rh Webpage Artifacts & \cellsym{\symInclusion} & \cellsym{\symInclusion} & \cellsym{\symInclusion} & \cellsym{\symInclusion} & \cellsym{\symInclusion\,\symExecution} & \cellsym{\symInclusion} & \cellsym{\symInclusion} & \cellsym{\symInclusion} & \cellsym{\symInclusion} & \cellsym{\symInclusion} & \cellsym{\symInclusion\,\symExecution} & \cellsym{\symInclusion\,\symExecution} & \cellsym{\symInclusion\,\symExecution} & \cellsym{\symInclusion\,\symExecution} & \cellsym{\symInclusion} & \cellsym{\symInclusion} & \ec & \cellsym{\symInclusion} & \cellsym{\symInclusion} & \cellsym{\symInclusion\,\symExecution} \\
\hline
\rh Hardware Artifacts & \ec & \ec & \ec & \ec & \ec & \ec & \ec & \ec & \ec & \cellsym{\symExecution\,\symObservation} & \ec & \ec & \ec & \cellsym{\symExecution\,\symObservation} & \ec & \cellsym{\symExecution\,\symObservation} & \ec & \cellsym{\symExecution\,\symObservation} & \ec & \ec \\
\hline
\rh Network Protocol Artifacts & \ec & \ec & \ec & \ec & \ec & \cellsym{\symSharing} & \ec & \ec & \ec & \ec & \ec & \ec & \ec & \cellsym{\symObservation} & \cellsym{\symObservation} & \cellsym{\symObservation} & \cellsym{\symObservation} & \ec & \ec & \ec \\
\hline
\rh HTTP Traffic Artifacts & \cellsym{\symSharing} & \cellsym{\symSharing} & \cellsym{\symSharing} & \cellsym{\symSharing} & \cellsym{\symSharing} & \cellsym{\symSharing} & \cellsym{\symSharing} & \cellsym{\symSharing} & \cellsym{\symSharing} & \cellsym{\symSharing} & \cellsym{\symSharing} & \cellsym{\symSharing} & \cellsym{\symSharing} & \cellsym{\symSharing} & \cellsym{\symSharing} & \cellsym{\symSharing} & \cellsym{\symObservation} & \cellsym{\symSharing} & \cellsym{\symSharing} & \cellsym{\symSharing} \\
\hline
\rh Navigational Artifacts & \ec & \ec & \cellsym{\symSharing} & \ec & \cellsym{\symSharing} & \ec & \cellsym{\symSharing} & \cellsym{\symSharing} & \cellsym{\symSharing} & \ec & \ec & \ec & \ec & \ec & \ec & \ec & \ec & \ec & \ec & \ec \\
\hline
\rh Identifiers & \cellsym{\symStorage\,\symSharing} & \cellsym{\symStorage\,\symSharing} & \cellsym{\symStorage\,\symSharing} & \cellsym{\symStorage\,\symSharing} & \cellsym{\symStorage\,\symSharing} & \cellsym{\symStorage\,\symSharing} & \cellsym{\symSharing} & \cellsym{\symStorage\,\symSharing} & \cellsym{\symStorage\,\symSharing} & \cellsym{\symStorage\,\symSharing} & \cellsym{\symStorage\,\symSharing} & \cellsym{\symSharing} & \cellsym{\symSharing} & \cellsym{\symSharing} & \ec & \ec & \ec & \ec & \ec & \ec \\
\bottomrule
\end{tabular}
\end{table*}

%% file: short-version/05-tracking-mitigations.tex
\section{Tracking Defenses}
\label{sec:tracking-mitigations}

We systematize defenses against tracking techniques described in~\autoref{sec:tracking-techniques} across three dimensions in accordance with the criteria defined in~\autoref{tab:systematization-criteria}: \textit{industry-based}, \textit{research-based}, and \textit{community-based}.
Following the methodology described in \autoref{sec:methodology} and \refappendix{sec:thematic-analysis}, we first curate industry-based (see \autoref{tab:defenses-industrial}) and community-based (see \autoref{tab:defenses-community}) defenses and research literature comprising 65 papers on tracking defenses (see \autoref{tab:defense-mechanisms}).
Each defense technique is grouped by its goal either as \colorbox{cellred}{detection and blocking} or \colorbox{cellyellow}{signal reduction and obfuscation}. 
For industry-based and community-based defenses, we identify 12 techniques against stateful tracking (see \autoref{tab:stateful-tracking-defenses}) and 11 techniques against stateless tracking (see \autoref{tab:stateful-tracking-defenses}). 
We also systematize support of defense techniques across each browser (Edge, Firefox, Brave, Chrome, and Safari). 
For research-based defenses, we identify 8 defense techniques described in \autoref{tab:defense-goal-mechanism-taxonomy}, grouped by the two defense goals. 
\autoref{tab:defense-mechanisms} maps each paper to these defense techniques.
%
\autoref{fig:evolution-defenses} depicts the evolution of tracking defenses under a single timeline from 2000 to 2026.

\subsection{Defenses Against Stateful Tracking}
\label{sec:stateful-defenses}


\subsubsection{Industry-based Defenses (\autoref{tab:stateful-tracking-defenses})}
\label{sec:stateful-defenses-industry}

\para{Clearing State.}
\label{sec:stateful-defenses-evercookie}
A user could clear their browser's cookies at the end of session, however, ordinary cookie clearing does not remove identifiers stored in \texttt{localStorage}, \texttt{IndexedDB}, E-Tags, or the browser cache, letting trackers respawn deleted cookies from such copies~\cite{ayensonFlashCookiesPrivacy2011,acarFPDetectiveDustingWeb2013,sorensenZombiecookiesCaseStudies2013}---any API that persists even a single bit of state is a potential supercookie vector~\cite{soltaniFlashCookiesPrivacy2009,aliNavigatingMurkyWaters2023,kamkarSamyKamkarEvercookie2010,klink1334485TrackingUsing2017}.
This motivated the cache-, network-, and storage-partitioning efforts that Firefox, Chrome, and Safari each shipped between 2020 and 2021, isolating TLS sessions, DNS caches, connection pools, and the HTTP cache by top-level site~\cite{menkeStorageIsolationProject2020,privacycommunitygroupClientSideStoragePartitioning2022,mdnStatePartitioningPrivacy2024}, closing off cross-site respawning even where same-site respawning may still be possible.

\para{Restricting or Partitioning Third-party State.}
Initially, browser vendors attempted to restrict third-party cookies themselves, leaving exceptions for non-tracking uses such as single sign-on~\cite{crouchImprovingPrivacyBreaking2018}, before blocking or partitioning them outright.
Safari's ITP, introduced in 2017, used an on-device ML classifier to identify likely trackers and eventually blocked all third-party cookies by default in 2020 unless a site was explicitly visited as first-party or used the Storage Access API~\cite{TrackingPreventionWebKit2020}.
Firefox's ETP shipped off by default in 2018, on by default a year later, and added Total Cookie Protection in 2021, which partitions cookies per top-level site rather than blocking them outright~\cite{mdnThirdpartyCookiesPrivacy2024}.
Brave has blocked all third-party cookies and storage through Shields since its earliest developer preview, granting only ephemeral, per-session partitioned storage to third parties that need it~\cite{braveprivacyteamEphemeralThirdpartySite2021}.
Chrome has moved far more cautiously: it initially blocked third-party cookies only in Incognito by default, introduced CHIPS as an opt-in partitioned-cookie attribute in 2023, and after repeated delays abandoned its plan to deprecate third-party cookies altogether in 2024~\cite{chavezNewPathPrivacy2024}, leaving \texttt{SameSite} and the Storage Access API~\cite{mdnStorageAccessAPI2024} as its main developer-facing tools; Edge inherits Chromium's defaults.
%
Systematic testing of these cookie-blocking implementations has repeatedly uncovered bypass vectors via redirects, popups, and nested iframes~\cite{frankenWhoLeftOpen2018}.

\para{Restricting or Limiting First-party State.}
First-party cookies cannot be blocked without breaking site functionality, so browsers instead limit their lifetime.
Safari's ITP expires script-set first-party cookies after seven days (24 hours with link decoration) and applies CNAME and IP heuristics to unmask disguised third-party hosts~\cite{TrackingPreventionWebKit2020}; Brave applies equivalent caps and adds Forgetful Browsing to clear storage for sites the user stops visiting~\cite{bravePrivacyProtectionSecurity}; Chrome, Firefox, and Edge apply no equivalent first-party lifetime cap by default.
As third-party cookies are restricted, websites increasingly rely on a handful of \emph{identity providers} that combine first-party data with offline data (e.g., loyalty cards, point-of-sale) to construct proprietary \emph{identity graphs}, concentrating behavioral insights and eroding users' ability to maintain separate or pseudonymous personas~\cite{TargetedAdvertisingAutorite2025,munirGooglesChromeAntitrust2024}.

\begin{opbox}
How can defenses detect, audit, or constrain opaque first-party data flows within the identity provider ecosystem, which enables persistent user tracking and identity continuity across contexts and devices?
\end{opbox}

\para{Removing or Limiting Navigational Parameters.}
Browsers remove URL parameters used by trackers.
Firefox (ETP Strict), Brave, and DuckDuckGo strip known tracking query parameters from URLs, while Safari does so in Private Browsing~\cite{TrackingPreventionWebKit2020}.
All five major browsers also default to \texttt{strict-origin-when-cross-origin} Referrer-Policy, stripping parameters from cross-origin \texttt{Referer} headers.

\para{Protecting Against Bounce Tracking.}
Bounce tracking grants a tracker temporarily unrestricted first-party storage access. 
Moreover, resemblance of authentication flows to a bounce complicates the defense~\cite{kellyBounceTrackingMitigations2022}.
Firefox and Chrome mitigate it by deleting storage for domains visited only in redirects without recent user interaction~\cite{snyderNavigationalTrackingMitigations2024}, while Brave combines \textit{debouncing} (skipping known bounce-tracker redirects) with \textit{unlinkable bouncing} (routing suspected bounces through temporary, unlinkable storage)~\cite{braveprivacyteamUnlinkableBouncingMore2022}.
What counts as ``legitimate'' interaction to spare a domain varies by browser.

\para{Blocking Trackers.}
Firefox and Edge block trackers from the Disconnect list; Brave's Shields blocks via its own curated list and uses PageGraph, a page-execution graph recorder, to make decisions based on a script's runtime behavior instead of URL alone; Safari pairs its blocklist with its ML classifier.
Chrome ships no native filter-list blocking, and the Manifest V3 shift has curtailed the dynamic request-blocking APIs that extensions previously relied on.
Against CNAME-cloaked trackers, Brave resolves CNAME for every sub-resource request and checks both requested and canonical domains against its filter lists~\cite{bravePrivacyProtectionSecurity}; Safari caps cookie lifetime to seven days when it detects a differing canonical domain~\cite{TrackingPreventionWebKit2020}, extended into Advanced Tracking and Fingerprinting Protection (ATFP) in 2023; Chrome, Edge, and Firefox ship no CNAME defense.

\begin{opbox}
CNAME cloaking, server-side implementations, and edge-compute proxies erase the first-/third-party origin boundary that every major browser defense relies on. What architectural primitive could replace it without requiring visibility into server-side data flows?
\end{opbox}

\subsubsection{Research-based Defenses (\autoref{tab:defense-goal-mechanism-taxonomy})}
\label{sec:stateful-defenses-research}


\para{Detecting and Blocking Stateful Trackers.}
Most research on tracker detection represents a webpage as a graph of its HTML structure, network requests, and JS execution, and trains a classifier to separate tracking from functional resources.
AdGraph~\cite{iqbalAdGraphGraphBasedApproach2020} pioneered this approach (inspiring Brave's PageGraph), with subsequent work refining features~\cite{sibyWebGraphCapturingAdvertising2022,yangWTAGRAPHWebTracking2022}, extending to code-property graphs and real-world deployment~\cite{leeAdCPGClassifyingJavaScript2023,leeAdFlushRealWorldDeployable2024}, and fully automating filter-rule generation~\cite{leAutoFRAutomatedFilter2023}.
Complementary classifiers extended existing lists by labeling unlisted tracker domains from request features~\cite{gugelmannAutomatedApproachComplementing2015,ikramSeamlessTrackingFreeWeb2017,reitingerMLCBMachineLearning2021}, while Khaleesi classified requests from their sequential redirect context to catch cookie syncing and bounce tracking missed by per-request classifiers~\cite{iqbalKhaleesiBreakerAdvertising2022}.
A subsequent line of work moved from blocking third-party requests to classifying first-party cookies by purpose.
CookieGraph separated tracking cookies from functional ones via cookie-creation flow graphs~\cite{munirCookieGraphUnderstandingDetecting2023}, semantics-aware checkers used LLMs to compare a cookie's inferred purpose against its declared consent category~\cite{chenSemanticsAwareCookiePurpose2025}, PURL applied purpose inference to link-decoration parameters~\cite{munirPURLSafeEffective2024}, and TGNN extended LLM-based classification to tracking pixels via graph neural networks~\cite{xiongTGNNEnhancingPixel2026}.
On mobile, NoMoAds~\cite{shubaNoMoAdsEffectiveEfficient2018} and NoMoATS~\cite{shubaNoMoATSAutomaticDetection2020} intercepted app traffic through a local VPN and classified it from traffic features alone.

\begin{opbox}
Can current defenses move from binary block/allow to enforcing purpose limitations---allowing data collection for a declared purpose while preventing its repurposing for cross-site tracking---and can such enforcement be made verifiable without trusting the data collector?
\end{opbox}

\input{tables/defense-goal-mechanism-taxonomy}

\para{Reducing or Obfuscating Signals for Stateful Trackers.}
This line of research sidesteps detection by changing what state a tracker can accumulate.
Early work proposed per-origin partitioning of visited-link styling and caches~\cite{jacksonProtectingBrowserState2006}, over a decade before any browser shipped it as a default.
Subsequent systems assigned each first-party a separate non-unique browser principal~\cite{panNotKnowWhat2015}, generalized private browsing into configurable isolation levels~\cite{xuUCognitoPrivateBrowsing2015}, systematically tested cross-site leak defenses~\cite{knittelXSinatorcomFormalModel2021}, and isolated cookies set by third-party-origin scripts with first-party privileges including CNAME-cloaked ones~\cite{nikkhahbahramiCookieGuardCharacterizingIsolating2025}.
TrackMeOrNot offered per-site tracking preferences, though this does not scale to thousands of sites~\cite{mengTrackMeOrNotEnablingFlexible2016,nisenoffDefiningBrokenUser2023}.
Server-side approaches have anonymized user profiles before they reach trackers through de-identification~\cite{zhengWebUserDeidentification2008}, formal anonymization guarantees~\cite{zhuAnonymizingUserProfiles2010}, and adversarial browsing patterns via reinforcement learning~\cite{zhangHARPOLearningSubvert2022}.
Cryptographic defenses have replaced identifiers with privacy-preserving protocols, from probabilistic cookie structures that limit cross-site linkability~\cite{morBloomCookiesWeb2015} to private ad auctions~\cite{zhongIbexPrivacypreservingAd2022} and encrypted access logging~\cite{ortegaperezEncryptedAccessLogging2025}.

\subsubsection{Community-based Defenses (\autoref{tab:stateful-tracking-defenses})}
\label{sec:stateful-defenses-community}

\para{Filter Lists.}
Community-maintained blocklists predate almost every industrial defense, with Peter Lowe's list maintained since 1997, EasyList since 2005, and EasyPrivacy since 2006.
Disconnect's tracker list (2011) was later adopted as the default blocklist in Firefox, Safari, and Edge, making it the closest thing to a shared industry standard.
Despite their ubiquity, these lists face structural limitations: they are maintained by small groups of volunteers, do not capture nuanced or newly emerging techniques, accumulate stale entries---roughly 90\% of EasyList's rules are never triggered~\cite{snyderWhoFiltersFilters2020}---and being static, trackers continually evade them.

\para{Extensions and Tools.}
Adblock Plus, the earliest mainstream filter-list extension (2006), popularized browser-level ad blocking but drew criticism for its Acceptable Ads program, which whitelists certain ads in exchange for payment from advertisers.
uBlock Origin, the community's most widely deployed blocker, ships with EasyList, EasyPrivacy, and its own curated filters, and also added CNAME uncloaking on Firefox in 2019 to unmask cloaked trackers that no mainstream browser could detect natively---though Chrome's Manifest V3 transition removed it from the Chrome Web Store in 2025 by disabling the dynamic request-rule APIs it depended on.
Other extensions occupy complementary niches: AdGuard maintains its own filter lists alongside EasyList compatibility and ships as both extension and standalone network-level application, Privacy Badger learns tracker behavior from cross-site request patterns rather than static lists, Ghostery's open-sourced engine powers the public WhoTracks.me dataset, and NoScript blocks all JS by default unless explicitly permitted.
At the network layer, Pi-hole and NextDNS block tracking domains via DNS, and ClearURLs strips known tracking parameters before requests are made.

\begin{opbox}
Every defense---browser defaults, research prototypes, or extension---is evaluated in isolation, 
yet users run them in combination. 
Can we develop a reproducible framework that measures aggregate tracking resistance under realistic defense compositions?
\end{opbox}

\subsection{Defenses Against Stateless Tracking}
\label{sec:stateless-defenses}

\subsubsection{Industry-based Defenses (\autoref{tab:stateless-tracking-defenses})}
\label{sec:stateless-defenses-industry}

Browser vendors treat fingerprinting as a form of covert tracking harmful to the web, and the W3C encourages specification authors to consider how a new API expands the browser's fingerprinting surface~\cite{nottinghamUnsanctionedWebTracking2015,dotyMitigatingBrowserFingerprinting2019}.
Yet, no browser found a way to close this surface without a utility cost, and vendors disagree on whether incremental mitigation is worth pursuing without a credible path to eliminating fingerprinting entirely~\cite{snyderBraveFingerprintingPrivacy2019,rescorlaTechnicalCommentsPrivacy2021}.

\para{Reducing IPUA Exposure.}
Safari's iCloud Private Relay, Chrome's IP Protection (now discontinued), and Brave's Tor-routed private windows each route traffic through independent relays so that neither the destination nor the relay operator can link a request to a user's real IP address.
On the User-Agent side, vendors have frozen the minor version out of the string and shipped User-Agent reduction across Chrome, Edge, and Firefox.
MAC address randomization, now standard on mobile, closes off a comparable network-layer identifier.

\para{Normalizing and Randomizing API Outputs.}
The most direct defense is to make browser-exposed information less identifying.
Normalization approaches include Firefox's resist-fingerprinting mode (screen, hardware, and timing values), Safari's ATFP (fonts, plugins, and User-Agent in Private Browsing), and outright API removal where fingerprintability is disproportionate (e.g., Battery Status API) or declining to implement new ones (e.g., Network Information API in WebKit and Firefox)~\cite{TrackingPreventionWebKit2020}.
Where normalization would break legitimate site functionality, browsers instead add randomized, site-specific noise so a device's fingerprint changes across page loads---Brave introduced this as ``farbling'' in 2020~\cite{braveprivacyteamFingerprintingDefenses202020}, with Firefox~\cite{huang1816056FprandomizationMeta2023} and Safari's ATFP~\cite{wilanderPrivateBrowsing202024} later adopting it for Canvas, WebGL, and \texttt{AudioBuffer} outputs.
All five browsers additionally reduce high-resolution timer precision, closing off APIs that side-channel research (\autoref{sec:stateless-techniques-side-channel-fingperinting}) relies on.
Because normalization makes all users look alike while randomization only makes one user's fingerprint noisy over time, the two trade off differently against a patient tracker, and browsers have generally converged on randomization as the more deployable default.
Chrome's Privacy Budget proposal~\cite{lasseyMikewestPrivacybudget2019}, meant to bound per-tracker API exposure, was discontinued due to concerns about website breakage and risk of exposing additional tracking surface~\cite{snyderBraveFingerprintingPrivacy2019,rescorlaTechnicalCommentsPrivacy2021,leflerWhatHappenedPrivacy2024}.

\begin{opbox}
Can a perturbation scheme be designed with an output distribution provably indistinguishable from the natural device population?
%
%
Could a browser bound the cumulative information any page extracts about a user across its entire interaction and enforce this bound natively?
\end{opbox}

\para{Blocking JS, Tracking Scripts, or Mixed Scripts.}
Mozilla's anti-tracking policy forbids fingerprinting~\cite{mozillaSecurityTrackingPolicy2019}; Firefox blocks scripts detected to include fingerprinting code~\cite{englehardtFirefox72Blocks2020}. 
Firefox and Edge use Disconnect's fingerprinting list to block scripts, Brave's Shields uses its PageGraph execution recorder, and Safari's ATFP blocks known fingerprinting scripts in Private Browsing.
Browsers also allow disabling JS per site.

\para{Side-channel Mitigations.}
Chrome, Edge, and Firefox further isolate cross-origin content into separate OS processes (Site Isolation and Firefox's Fission) but this introduces new timing side-channels, on top of the timer-precision reductions.

\begin{opbox}
How can browsers defend against emerging side channels, including XS-Leaks and cross-site signals exposed by CDNs, that persist even when traditional fingerprinting is mitigated? 
\end{opbox}

\subsubsection{Research-based Defenses (\autoref{tab:defense-goal-mechanism-taxonomy})}
\label{sec:stateless-defenses-research}

\para{Detecting and Blocking Stateless Trackers.}
Detection has moved from heuristic-based approaches flagging known probing patterns~\cite{acarFPDetectiveDustingWeb2013} to ML classifiers over static and dynamic script behavior~\cite{iqbalFingerprintingFingerprintersLearning2021} and longitudinal models predicting which newly shipped APIs are likely to become fingerprinting vectors~\cite{bahramiFPRadarLongitudinalMeasurement2022}.
Dynamic taint tracking has been applied to follow fingerprinting-relevant values from API call to network request inside instrumented Chromium builds~\cite{boussahaFPtracerFinegrainedBrowser2024,kanyalPanoptiChromeModernInbrowser2024,hubletUserControlledPrivacyTaint2024}, while coarse server-side fingerprints have also been used defensively to flag fraudulent device configurations~\cite{kalantariBrowserPolygraphEfficient2024}.
Beyond request-level, several systems have blocked tracking at the script level.
TrackerSift~\cite{amjadTrackerSiftUntanglingMixed2021} and Duumviri~\cite{shuangDuumviriDetectingTrackers2025} separated tracking from functional code within mixed scripts, SugarCoat~\cite{smithSugarCoatProgrammaticallyGenerating2021} generated replacement scripts preserving page functionality, and finer-grained approaches have blocked at function~\cite{amjadBlockingJavaScriptBreaking2023,amjadBlockingTrackingJavaScript2024}, bytecode~\cite{ghasemisharifReadLinesDetecting2023,bahramiByteByteUnmasking2025}, graph~\cite{parkCGCContrastiveGraph2022}, or event-loop~\cite{chenDetectingFilterList2021} levels.
As blocking may cause site breakage, complementary work has also detected it from screenshots~\cite{hajjchehadeSINBADSaliencyinformedDetection2024} and studied user perception~\cite{nisenoffDefiningBrokenUser2023}.
Trackers often evade blocking: anti-adblock circumvention has been characterized through differential execution~\cite{zhuMeasuringDisruptingAntiAdblockers2018} and automated detection~\cite{leCVInspectorAutomatingDetection2021}. 
Adversarial perturbations against perceptual blockers~\cite{tramerAdVersarialPerceptualAd2019} have motivated adversarially robust classifiers~\cite{limAdVersaAdversariallyRobustPractical2026}.

\para{Reducing or Obfuscating Signals for Stateless Trackers.}
Gummy Browsers showed that a script's own fingerprint-collection routine can be turned against it, replaying a target's attribute values to impersonate a different device~\cite{liuGummyBrowsersTargeted2022}, while UNIGL rewrites WebGL shaders so GPU rendering output is identical across hardware, removing device-level variation at its source~\cite{wuRenderedPrivateMaking2019}.
On the randomization side, PriVaricator injects policy-controlled noise into fingerprintable APIs~\cite{nikiforakisPriVaricatorDeceivingFingerprinters2015}, ML-CB decides per call whether a Canvas access looks like fingerprinting before blocking~\cite{reitingerMLCBMachineLearning2021}, and FP-Sandbox degrades API output past a privacy budget~\cite{torokOnlyPayWhat2023}.
Comparative evaluations have consistently found that such defenses leave a residual detectable signature~\cite{dattaEvaluatingAntiFingerprintingPrivacy2019}.
Extension fingerprinting has been countered by Simulacrum's DOM virtualization, giving every page a consistent view regardless of installed extensions~\cite{karamiUnleashSimulacrumShifting2022}.
Behavioral tracking has been addressed by obfuscating interest profiles: Bloom Cookies encode profiles with controlled false positives~\cite{morBloomCookiesWeb2015}, HARPO poisons ad-interest profiles with reinforcement-learned decoy actions~\cite{zhangHARPOLearningSubvert2022}, and profile generalization provides $k$-anonymity for browsing history~\cite{zhuAnonymizingUserProfiles2010,zhengWebUserDeidentification2008}.
Passive IPUA fingerprinting is harder to counter in the browser alone, so proposals have acted at the network layer: unlinkable pseudonymous tokens~\cite{johnsonNymbleAnonymousIPAddress2007}, ISP-level address mixing~\cite{raghavanEnlistingISPsImprove2009}, TLS handshake metadata reduction~\cite{syEnhancedPerformancePrivacy2020}, and MAC-address randomization, whose real-world effectiveness has been found incomplete~\cite{fenskeThreeYearsLater2021}.
Side-channel defenses draw on the same partitioning techniques used against stateful supercookies~\cite{jacksonProtectingBrowserState2006}, alongside systematic cross-browser XS-Leak testing~\cite{knittelXSinatorcomFormalModel2021} and network-level tracking detection from encrypted packet metadata~\cite{leeNettrackGenericWeb2023}.


\subsubsection{Community-based Defenses (\autoref{tab:stateless-tracking-defenses})}
\label{sec:stateless-defenses-community}

\para{Filter Lists.}
Stateless tracking has no dedicated filter-list ecosystem of its own; instead, the same Disconnect, EasyList, and EasyPrivacy lists that community maintainers curate for cookie- and script-based blocking (\autoref{sec:stateful-defenses-community}) carry the fingerprinting script and known-tracker entries that Firefox, Edge, and Brave draw on for the blocking described above.

\para{Extensions and Tools.}
Tor Browser, launched in 2008, remains the reference implementation for fingerprinting resistance, providing first-party isolation and a uniform, one-fingerprint-for-all browser configuration by reporting the same values for a wide range of APIs~\cite{perryDesignImplementationTor2013}, and Mullvad Browser extended the same hardening to users who want it without routing through the Tor network itself.
NoScript blocks JavaScript per site by default, eliminating essentially every stateless technique that depends on script execution at the cost of breaking sites that need it, and AdGuard's Stealth Mode layers User-Agent overriding and referrer stripping on top of its filter-list blocking.
The EFF's Panopticlick project, and later the Markup's Blacklight and DuckDuckGo's Tracker Radar, have shifted the community's role from blocking alone to auditing, letting any user or researcher scan a site and see exactly which fingerprinting and tracking techniques it uses.


%% file: tables/defense-goal-mechanism-taxonomy.tex
\begin{table}[H]
  \small
  \centering
  \caption{Taxonomy of research-based tracking defenses along with its distribution across the literature (\# = paper count).}
  \label{tab:defense-goal-mechanism-taxonomy}
  \begin{tabular}{>{\raggedleft\arraybackslash}p{0.8cm} >{\raggedright\arraybackslash}p{6cm} >{\centering\arraybackslash}p{0.1cm}}
    \toprule
    \multicolumn{1}{c}{\textbf{Goal}} & \multicolumn{1}{c}{\textbf{Defense Technique}} & \multicolumn{1}{l}{\textbf{\#}} \\
    \midrule

    \cellcolor{cellred}
      & Machine Learning based Classification            & 28 \\
\cellcolor{cellred}
      & Program Analysis and Code Instrumentation        & 17 \\
\cellcolor{cellred}
      & Circumvention based Behavioral Classification    &  4 \\
\cellcolor{cellred}\multirow[c]{-4}{*}{\rotatebox[origin=c]{90}{\makecell{Detection \&\\Blocking}}}
      & LLM based Classification                         &  2 \\
    \midrule

    \cellcolor{cellyellow}
      & API Output Perturbation                          &  8 \\
\cellcolor{cellyellow}
      & Browser State Isolation and Partitioning         &  6 \\
\cellcolor{cellyellow}
      & Network Protocol level Modifications             &  4 \\
\cellcolor{cellyellow}\multirow[c]{-4}{*}{\rotatebox[origin=c]{90}{\makecell{Signal\\ Reduction \&\\Obfuscation}}}
      & Cryptographic Defenses                           &  6 \\
    \bottomrule
  \end{tabular}
\end{table}

%% file: short-version/06-regulations.tex
\section{Regulations \& Compliance}
\label{sec:regulations}

Here, we describe the enactment and evolution of EU and US privacy regulations.
We start by curating the 34 research papers identified in~\autoref{sec:methodology} to systematize the different regulations and standards they cover (\autoref{tab:regulations-paper-distribution}) according to~\autoref{tab:systematization-criteria}.
We map in~\autoref{tab:regulations-aspects} the mentioned policy aspects onto the six self-managed rights\footnote{For a taxonomy of rights and requirements imposed by privacy regulations, we refer to the SoK on Internet Regulations from Birrell et al.~\cite{birrellSoKTechnicalImplementation2024}.} granted by regulations and their applicability in the EU and US (California) contexts.
Then, we categorize the suggested limitations with compliance (\autoref{tab:regulations-compliance}).
Finally, we complement this taxonomy with our expertise to provide a cohesive view of the evolution of regulatory actions in the EU and US (overview in~\autoref{fig:evolution-defenses}).

\subsection{Regulatory Actions}

\para{In the EU.}
\label{sec:eu-regulations}
First data protection national laws were established in the late 1970s, followed by the EU Data Protection Directive in 1995, and the GDPR applicable to all EU members in 2018. 
Personal EU-US data transfers are currently regulated by the EU–US Data Privacy Framework.
%
The ePrivacy Directive (ePD, 2002, amended in 2009) requires
a valid user’s consent before \textit{``storing of information’’}.
This notion was redefined by GDPR by setting higher-level legal requirements~\cite{santosAreCookieBanners2020}.
EU regulators continuously update their national laws and compliance guidelines to further interpret and implement the ePD~\cite{Bielova2024-zr}, for instance to: (1)~require consent before cookies are set, read, or sent to third-parties, (2)~establish a legitimate interest exemption,
and (3)~cover various types of devices (mobile, IoT) and technologies (pixels, link decoration).
EU laws like the Digital Markets Act set up rules on valid consent for major \textit{``gatekeeper’’} companies and the Digital Services Act on dark patterns and advertising for platforms.
Currently, the Digital Omnibus proposal aims to amend, simplify, and optimize the EU's digital rulebook to reduce compliance burden and increase EU competitiveness.

\para{In the US.}
\label{sec:us-regulations}
States enact their own privacy legislation: California passed the CCPA (2020) and CPRA (2023), soon followed by other states~\cite{birrellSoKTechnicalImplementation2024,vannortwickSettingBarLow2022} as depicted in~\autoref{fig:evolution-defenses}.
Only a few narrow privacy laws exist at the federal level.
Thus, apart from mandatory provisions, the notice and choice principle, implemented via privacy policies, governs what a recipient of personal information can do with it~\cite{zimmeckInformationPrivacyLaw2013,schaubDesignSpaceEffective2015}. 
Additionally, under its jurisdiction, the FTC can consider privacy policies that misrepresent a business's data handling practices as unfair or deceptive, affecting commerce.
%
In California, recent changes to CCPA via CPRA may negatively impact the usability, scope, and visibility of the right to opt-out of sale~\cite{charatanTwoStepsForward2024}.

\begin{opbox}
Does the EU's Digital Omnibus carry similar risks to the CPRA in its proposed amendments?
\end{opbox}

\para{Policy-oriented Solutions.}
\label{sec:policy-solutions}
Several attempts were made at policy-oriented protocols through opt-out signals (see~\refappendix{sec:mixed-defenses-community}) or by the industry with the Data Rights Protocol, GDPR consent (TCF), or CCPA compliance frameworks---whose efforts overall remain in early stages~\cite{hilsMeasuringEmergenceConsent2020,matteCookieBannersRespect2020,azizJohnnyStillCant2024}.

\subsection{Regulatory Compliance}

\para{EU Regulators.} 
In the Planet49 case~\cite{bouhoulaAutomatedLargeScaleAnalysis2024}, the highest court in the EU (CJEU) established legal precedent by declaring pre-ticked boxes in consent design interfaces illegal. Similarly the French
CNIL
found that consent banners must offer a reject option on the first layer~\cite{bielovaEffectDesignPatterns2024},
and companies have been fined for setting cookies prior to consent~\cite{sunGDPRxivEstablishingState2023,marjanovDataSecurityGround2023}.
For more GDPR-related decisions, see the GDPRhub\footnote{At \url{https://gdprhub.eu/}} inventory.

\para{US Regulators.} Through its enforcement actions over the years, the FTC has effectively created a body of common law of privacy~\cite{soloveFTCNewCommon2013}.  
Similarly, state attorney generals increased their regulatory activity based on new state privacy laws~\cite{azizJohnnyStillCant2024}.

\para{In Literature.} Various studies have shown that many websites' actual behaviors are not compliant with their own privacy policies~\cite{libertAutomatedApproachAuditing2018,ouViopolicyDetectorAutomatedApproach2022,manandharAutomatedRegulationAnalysis2024}, or do not respect or register users’ 
consent correctly~\cite{carpinetoAutomaticAssessmentWebsite2016,sanchezrolaCanIOptOutYet2019,matteCookieBannersRespect2020,mehrnezhadCrossPlatformEvaluationPrivacy2020,bouhoulaAutomatedLargeScaleAnalysis2024,vannortwickSettingBarLow2022,birrellSoKTechnicalImplementation2024,Kanc-etal-25-PETs,liuOptedOutTracked2024}.
%
%
Researchers also studied the effects of GDPR, CCPA, or both on (a) privacy policies, noting improved transparency and undesired reduction of specificity in some updates~\cite{lindenPrivacyPolicyLandscape2020,hosseiniBilingualLongitudinalAnalysis2024,cuiUnderstandingPrivacyNorms2025}, (b) business practices related to incomplete data redaction~\cite{luWHOISWHOWASLargeScale2021} and deletion requests~\cite{ruppLeaveNoData2022,dimartinoRevisitingIdentificationIssues2022,mehrnezhadHowCanWould2022,jordanVICEROYGDPRCCPAcompliant2023}, and (c) non compliance of consent management platforms (CMPs) with the right to consent or opt-out~\cite{azizJohnnyStillCant2024,papadogiannakisUserTrackingPostcookie2021,bollingerAutomatingCookieConsent2022,tothDarkPatternsManipulation2022,biselliSupportingInformedChoices2024}

\begin{opbox}
While the existing focus is on privacy policy and consent, other types of compliance (e.g., EU-US data transfer law, revoking previously agreed opt-in, data minimization, consent propagation, etc.) are understudied.
\end{opbox}

\subsection{Compliance Challenges}

Multiple reasons explain such low compliance rates, we map them to the need to reconcile actors' \colorbox{cellpink}{Incentives}, \colorbox{cellblue}{Expectations}, and \colorbox{cellgreen}{Responsibilities} (see~\autoref{tab:regulations-compliance}).

\input{tables/regulations-compliance}

\para{\colorbox{cellpink}{Incentives}.} A lack of incentive or knowledge from website publishers can lead to privacy compliance being overlooked when integrating third-parties or delegating consent management to CMPs---except when legal requirements or respective guidelines exist~\cite{utzPrivacyRarelyConsidered2023,Stov-etal-23-PETs}. In turn, embedded third-parties are placed in a privileged position, giving them access to user data, and may not act as transparently as desired resulting in the use of dark patterns or server-side tracking~\cite{tothDarkPatternsManipulation2022,grayDarkPatternsLegal2021,mehrnezhadHowCanWould2022}.

\para{\colorbox{cellblue}{Expectations}.} Better enforcement power from regulators is needed~\cite{bollingerAutomatingCookieConsent2022,vannortwickSettingBarLow2022,mehrnezhadHowCanWould2022,motiTargetedTroublesomeTracking2024}. However, regulators may not have the required manpower, financial resources, and dedicated technical departments~\cite{edpbContributionEDPBReport2023} to investigate that websites are compliant not just ``at the surface''~\cite{kyiInvestigatingDeceptiveDesign2023}. The research community could step in to fill this gap, but user studies and web measurement artifacts are not always specific enough to be used by policymakers in guidelines or as evidence of violations~\cite{bielovaEffectDesignPatterns2024,Bielova2024-zr,bouhoulaAutomatedLargeScaleAnalysis2024}. 
This calls for closer collaboration to ensure compliance~\cite{saemannInvestigatingGDPRFines2022} and anticipate harms associated with future tracking techniques ~\cite{fouadDevilDetailsDetection2024,cuiUnderstandingPrivacyNorms2025,tothDarkPatternsManipulation2022,hoganMakingSensePrivate2026}.

Beyond legally-binding decisions, regulators can also perform more supportive actions such as clarifying legal aspects. For example, they should provide guidelines and endorse specific designs of compliant CMP interfaces~\cite{fouadMyCookiePhoenix2022,hosseiniBilingualLongitudinalAnalysis2024,hilsMeasuringEmergenceConsent2020,matteCookieBannersRespect2020,mehrnezhadHowCanWould2022} and consider CMPs as data controllers~\cite{papadogiannakisUserTrackingPostcookie2021,tothDarkPatternsManipulation2022}. Similarly, ambiguity should be lifted~\cite{hosseiniBilingualLongitudinalAnalysis2024,azizJohnnyStillCant2024}, such as around data deletion in the contexts of cookies respawning, legitimate interest exemption, or further data sharing with other third-parties~\cite{fouadMyCookiePhoenix2022,ruppLeaveNoData2022,ruppLeaveNoData2022}. Related is the ask to clarify the website categories to which CCPA and CPRA apply to
~\cite{vannortwickSettingBarLow2022,azizJohnnyStillCant2024,charatanTwoStepsForward2024} and the applicability of GPC in the EU context~\cite{zimmeckUsabilityEnforceabilityGlobal2023,zimmeckGeneralizableActivePrivacy2024,azizJohnnyStillCant2024}.

Finally, browsers should be expected to play a more prominent role. As an example, they could technically enforce in real time users' consent or opt-out choices~\cite{papadogiannakisUserTrackingPostcookie2021,bollingerAutomatingCookieConsent2022,bouhoulaAutomatedLargeScaleAnalysis2024}, removing the need to rely on a CMP, third-party, or web extension to fill in and respect users' preferences~\cite{azizJohnnyStillCant2024,demirLargeScaleStudyCookie2024}.

\begin{opbox}
How can we develop viable browser-based consent mechanisms, integrate automated real-time compliance verification, ensure their legal robustness, and enable effective enforcement?
\end{opbox}


\para{\colorbox{cellgreen}{Responsibilities}.} Finally, third-parties escape legal responsibility as current laws often place the main compliance obligations on website owners~\cite{santosConsentManagementPlatforms2021} even though several studies have identified problems around the default configurations of third-party services~\cite{Koch-etal-25-PETs,Rodr-etal-25-PETs,tothDarkPatternsManipulation2022}.

\begin{opbox}
What future transdisciplinary studies are needed to ensure that research and legal communities contribute to assigning responsibility to all actors fairly?
\end{opbox}

%% file: tables/regulations-compliance.tex
\begin{table}[H]
  \small
  \centering
  \footnotesize
  \caption{Taxonomy of compliance challenges along with their distribution across the literature (\# = paper count).}
  \label{tab:regulations-compliance}

  \begin{tabular}{>{\centering\arraybackslash}p{1.6cm} >{\raggedright\arraybackslash}p{5.4cm} >{\centering\arraybackslash}p{0.1cm}}
    \toprule
    \multicolumn{1}{c}{\textbf{Category}} & \multicolumn{1}{c}{\textbf{Challenge}} & \multicolumn{1}{l}{\textbf{\#}} \\
    \midrule

    \cellcolor{cellpink}
    & Third-parties: dark patterns, server-side tracking    &  12 \\
    \cellcolor{cellpink}\multirow[c]{-2}{*}{\makecell{Incentives}}
    & Publishers: overlooking compliance  &  10 \\
    \midrule

    \cellcolor{cellblue}
    & Regulators: clarification, enforcement, GPC    &  19 \\
    \cellcolor{cellblue}
    & Researchers: collaboration, evidence collection     &  12 \\
    \cellcolor{cellblue}\multirow[c]{-3}{*}{\makecell{Expectations}}
    & Browsers: acting on behalf of users    &  5 \\
    \midrule

    \cellcolor{cellgreen} \multirow[c]{-1}{*}{\makecell{Responsibilities}}& Third-parties: escaping their legal obligations &  9 \\

    \bottomrule
  \end{tabular}

\end{table}

%% file: short-version/07-discussion.tex
\section{Discussion}
\label{sec:future-outlook}

\subsection{Privacy-preserving advertising paradigms}
The Privacy Sandbox from Google was the most recent and extensive industrial attempt to deprecate cross-site identifiers, by relying more heavily on on-device computation and the browser to mediate access.
However, achieving private, but useful, ad targeting and measurement is challenging~\cite{hoganMakingSensePrivate2026}, as illustrated by the following privacy limitations uncovered in recent proposals: (a) anonymity and re-identification concerns
with FLoC~\cite{rescorlaTechnicalCommentsFLoC2021,berkePrivacyLimitationsInterestbased2022}, Topics~\cite{thomsonPrivacyAnalysisGoogles2023,jhaRobustnessTopicsAPI2023,beuginInterestdisclosingMechanismsAdvertising2024,beuginPublicReproducibleAssessment2024}, Protected Audience~\cite{thomsonProtectedAudiencePrivacy2024,longEvaluatingGooglesProtected2024}, or Apple's Private Click Measurement~\cite{thomsonAnalysisApplesPrivate2022} and Private Relay~\cite{zohaibInvestigatingTrafficAnalysis2023},
(b) flawed implementations or side-channel leakage~\cite{turatiLocalitySensitiveHashingDoes2023,aliNavigatingMurkyWaters2023,calderonioFledgingWillContinue2024,nisenoffExploitingSharedStorage2025}, and (c)  abuse~\cite{senolUnveilingImpactUserAgent2023,wieflingPrivacyMeasureTurned2024}.

\begin{opbox}
What new privacy-preserving technologies can be built by learning from issues in prior proposals? 
\end{opbox}

By late 2025, Google deprecated most Privacy Sandbox APIs, leaving third-party cookies largely opt-in in Chrome~\cite{chavezUpdatePlansPrivacy2025,gribRiseFallGoogle2026}, while such restrictions already exist by default in Brave, Firefox, or Safari.
Overall, Privacy Sandbox demonstrated that besides technical challenges, economic and other (misaligned) incentives often dominate with advertising~\cite{beuginTechnicalReportNeed2025,beuginLessonsAdoptionDeprecation2026}.
Ideally, browsers would coordinate work on privacy-preserving advertising through the W3C's Private Advertising Technology Community Group (PATCG), chartered in 2021, rather than shipping unilateral, mutually incompatible APIs.
Meta and Mozilla's Interoperable Private Attribution proposal is currently under such W3C incubation.

\begin{opbox}
What would a standardized, cross-browser on-device computation primitive look like that advertisers could build against regardless of which browser a user runs?
\end{opbox}

\subsection{Tracking in other ecosystems}

While our focus was on web tracking, similar tracking mechanisms also exist in mobile apps and IoT ecosystems---using often a richer set of sensor information available via OS APIs rather than web APIs. Some key differences and similarities exist: reliance on cookies on the web and on device-level identifiers such as the Ad ID (known as \texttt{IDFA} on iOS, \texttt{AAID} on Android, or \texttt{TIFA} on Samsung Smart TVs), and vendors making different design choices across ecosystems. For instance, while Safari blocks third-party cookies by default, iOS instead asks permission to give access to device-level identifier.

\begin{opbox}
If cross-device tracking is understood theoretically, characterizing its occurrence in practice and defending against it requires more efforts. How do techniques and protections combine and diverge across ecosystems?
\end{opbox}

\subsection{AI as an emerging tracking surface}

AI is beginning to reshape web tracking on multiple fronts.
Browsing behaviors are shifting as AI assistants navigate the web on behalf of users. This introduces a new context where account-linked identity meets rich conversational and behavioral signals otherwise not shared on the web~\cite{vekaria2025big}.
As AI platforms already embed trackers~\cite{jazlan2026tracking}, introduce ads in chatbots (e.g., ChatGPT), or use LLM to infer sensitive user attributes from ordinary browsing data~\cite{chenWhenAdsBecome2026}, harms and privacy concerns may amplify.
Browsers have a key role to play in ensuring security and privacy of novel AI integrations.

\begin{opbox}
Will advertising and tracking shift to AI platforms, creating new tracking surfaces or amplifying the existing ones? Conversely, can real-time adaptive defenses be built with LLMs to surpass today's static heuristics?
\end{opbox}

%% file: short-version/08-conclusion.tex
\section{Conclusion}

Decades after its introduction, web tracking still remains an \textit{archetypal cat-and-mouse game}.
%
%
This adversarial dynamic shows that purely reactive approaches cannot deliver privacy guarantees for online users.
\textit{Regulations alone are insufficient} as enforcement lags the speed of technical changes in evolving tracking mechanisms, and trackers often find tolerated gray zones to bypass regulations.
Regulators and the measurement community need to collaborate to incorporate agile, evidence-driven auditing methods to avoid any oversight and ensure that regulations evolve competitively with the technical reality.
On the other hand, while \textit{browsers can be powerful gatekeepers, they provide an unreliable line of defense}.
Default protections vary widely across browser vendors, mitigations ship years after the issues are identified, and commercial incentives often result in more permissive designs.
Future research must therefore look beyond ``\textit{fix it in the browser}'' remedies and explore complementary approaches that truly safeguard users' privacy.
The current tracking landscape demands a \textbf{default privacy-first solution} where users can control their privacy.
By systematizing evolution of web tracking, its defenses, and regulatory compliance, this SoK calls for a privacy-first web in which tracking is systematically measured and proactively audited, while users remain in control of their data by default.

%% file: acknowledgment.tex
This material is based upon work supported by the National Science Foundation
under Grant Numbers 
2041894, 
2138138, 
2047260, 
2343611, 
and
2143363. 
%
Any opinions, findings, and conclusions or
recommendations expressed in this material are those of the author(s) and do not
necessarily reflect the views of the National Science Foundation.
This work was also supported in part by the Semiconductor Research Corporation (SRC) and DARPA, 
the Agence Nationale de la Recherche through the ANR-21-CE39-0019 FACADES and the ProjetIA-22-PECY-0002 iPoP projects,  
and ANR MRSEI TULIP (ANR-24-MRS0-0004), Inria  DATA4US Exploratory Action project. 

%% file: appendix/ethics.tex
\label{sec:ethics-considerations}

This work is a Systematization of Knowledge (SoK) on web tracking.
As such, it does not involve any data collection, user studies, or measurement and deployment of tracking technologies.
Instead, it systematizes developments in the field on online tracking through research and industrial advancements to identify gaps and open problems.
Nevertheless, we discuss the ethical implications of our work in shaping the future research directions and influencing the involved stakeholders.\\
%

\noindent \textbf{Stakeholder Analysis.}
Our work discusses six stakeholders:
(1) \textit{Online users}:
are individuals whose web activity may be observed, inferred, or profiled using different tracking mechanisms.
(2) \textit{Browser vendors}:
are the developers of browsers that mediate access to the web, while designing and deploying technical countermeasures to balance privacy, usability, performance, and compatibility.
(3) \textit{Website publishers}:
operate websites and monetize their content by relying on analytics, personalization, and advertising, while maintaining functionality under evolving privacy constraints.
(4) \textit{Advertising and tracking platforms}:
that design and operate tracking infrastructures and are affected by regulatory changes, browser policies, and public scrutiny.
(5) \textit{Regulators and policymakers}:
are responsible for enforcing data protection and competition laws. They rely on accurate technical understanding to craft effective and enforceable regulation.
(6) \textit{Research community}:
academic and industrial researchers studying web tracking and privacy to understand, measure, and defend against evolving tracking techniques.\\

\noindent \textbf{Impact of the Research.}
Next, we list the positive and negative impacts of our research on different stakeholders.

\noindent \textbf{\textit{Positive Impacts.}}
(1) \textit{Improved Understanding (Impact on Users, Regulators, and Researchers)}:
By systematizing advances in tracking techniques, defenses, and measurement methodologies, this work benefits users and regulators by clarifying what forms of tracking are known, which ones are mitigated, and which ones remain unresolved.
This SoK also serves as a unified source of knowledge for early career researchers, users, and regulators to understand technicalities of different tracking mechanisms.
(2) \textit{Future Research Directions (Impact on Researchers and Browser Vendors)}:
Our work identifies underexplored areas and open problems based on advancements in tracking landscape allowing new as well as established researchers to focus on important directions of research to improve the state of web tracking rather than making marginal improvements or working well-explored problems.
Similarly, it can also inform browser vendors to anticipate emerging forms of web tracking and work towards in-browser countermeasures.
(3) \textit{Informing Platform and Policy Design (Impact on Browser Vendors and Regulators)}:
By synthesizing years of developments in web tracking, this SoK informs browser vendors and policymakers about unintended consequences and gaps in current approaches, while guiding future policy changes---both browser-level as well as legal.

\noindent \textbf{\textit{Negative Impacts.}}
(1) \textit{Potential for Adversarial Use (Impact on Users)}:
This paper can inform motivated adversaries (i.e., trackers) to exploit the identified gaps in online tracking research by deploying them within their tracking technologies to circumvent existing protections.
Identified open problems may further motivate to make these mechanisms stealthier.
(2) \textit{Misinterpretation or Selective Use of Findings (Impact on Policy)}:
Industry or policy actors may selectively cite parts of this SoK to justify particular technical or regulatory choices, potentially oversimplifying nuanced trade-offs discussed in the specific academic work in the literature.\\

\noindent \textbf{Justification for Research.}
Despite these risks, we argue that publishing this SoK is ethically justified and necessary.
Web tracking is already widely deployed at scale, often with limited transparency to users and regulators.
The primary ethical risk lies not in documenting this ecosystem, but in allowing knowledge fragmentation or obscurity to persist.
Several factors mitigate the highlighted potential negative impacts.
First, this work does not introduce new tracking mechanisms, datasets, or attack surfaces; it systematizes and critically evaluates information that is already publicly available on the internet or published in the research literature.
Second, by explicitly framing open problems, we aim to encourage research that prioritizes user protection, accountability, and robustness rather than exploitation.
Finally, increased visibility into the limitations of existing defenses and measurements enables more informed decision-making by browser vendors, regulators, and researchers, reducing the likelihood of misleading claims about privacy guarantees.
Overall, this SoK contributes to an ethically grounded understanding of web tracking.
By establishing what is known, what remains unsolved, and where future research efforts are needed the most, this work supports long-term improvements in web privacy that outweigh the potential risks from systematizing this knowledge.

%% file: appendix/open-science.tex
\label{sec:open-science}

The main artifact produced by our work includes the thematic organization and categorization of the literature on web tracking identified as relevant in~\autoref{sec:methodology}.
As a result, we release the code to obtain papers, keywords used to filter them as well as the list of publications organized under tracking techniques, defenses, and regulatory compliance along with the assigned codes, annotations, and identified themes. Our artifacts can be obtained at: 
\url{https://osf.io/zsy7e/overview?view_only=a346b98bf89244b5b17d842f6fb3fb13}

\ifanonymous
Additionally, we maintain an extended and live version of this SoK at \textit{URL omitted for anonymous submission} to provide more details on the concepts covered and enable regular updates based on ecosystem changes, new findings, and community suggestions on our \LaTeX{} code repository.
\else
Additionally, we maintain an extended and live version of this SoK at \url{https://github.com/privacysandstorm/sok-advances-open-problems-web-tracking} to provide more details on the concepts covered and enable regular updates based on ecosystem changes, new findings, and community suggestions on our \LaTeX{} code repository.
\fi
%

%% file: appendix/sok-methodology.tex
\section{SoK Methodology}
\label{sec:appendix-methodology}

\subsection{Search Criteria for Literature Review}
\label{sec:keywords}

Web tracking comprises a vast body of literature published over the last few decades---spanning across technical mechanisms, defenses, and regulatory compliance. Through our literature search (see~\autoref{sec:methodology}), we obtained a list of 19438 papers along with their metadata (title, authors, year, and DOI), missing abstracts were completed by performing a DOI-based query on Semantic Scholar\footnote{\url{https://www.semanticscholar.org/}} and OpenAlex\footnote{\url{https://openalex.org/}}.
To further filter that list, we first use exclusion keywords to remove potential papers that could be incorrectly retained under web tracking.

\begin{exclusionkeywordbox}
\textbf{Exclusion Keywords:}
\noindent
NOT [\textbf{Title} OR \textbf{abstract}: crypto* OR
blockchain* OR
website fingerprint* OR
``tor'' OR
``poster:'' OR
differential* priva* OR
``federated learning'' OR
``membership inference'' OR
``trusted execution'' OR
``unlearning'' OR
``inversion'' OR
multi-party computation* OR
airtag* OR
privacy label* OR
sdk* OR
``app store'' OR
``google play'' OR
``social media'' OR
smart tv* OR
smart home* OR
``iot'' OR
``virtual reality'' OR
``3g'' OR
``4g'' OR
``5g'' OR
``wireless'' OR
``censorship'' OR
``voice'' OR
``gaze'' OR
``eye'' OR
``genome'' OR
scam* OR
``bitcoin'' OR
fuzz* OR
cve* OR
certificate authorit* OR
public key* OR
vehicle* OR
robot*]
\end{exclusionkeywordbox}

Second, two sets of inclusion keywords are used to narrow down potential candidates for manual review.
The first set is used to filter publications that are related to web tracking techniques and defenses based on their title and abstract text.

\begin{keywordbox}
\textbf{Inclusion Keywords for Techniques and Defenses:}
\noindent
[\textbf{Title} OR \textbf{abstract}: tracking OR
web track* OR
online track* OR
user track* OR
track* user* OR
web privacy OR
online privacy OR
user* privacy OR
user *identifi* OR
track* circumvent* OR
track* eva* OR
track* defen* OR
track* block* OR
ad*block* OR
ad* block*]
\end{keywordbox}

The second set of keywords is used to identify papers discussing privacy regulations and compliance in the context of web tracking in their full text.
We rely on full text search to identify papers that may not primarily be regulation-focused (hence lower probability of a match in title or abstract), but may still discuss regulatory implications of the studied tracking technique or defense in the paper.

\begin{keywordbox}
\textbf{Inclusion Keywords for Regulations and Compliance:}
\noindent
[\textbf{Full paper}: regulat* compliance OR
regulation* OR
track* violat* OR
privacy polic* OR
privacy law* OR
privacy regulation* OR
``privacy compliance'' OR
``gdpr'' OR
``ccpa'' OR
``cpra'' OR
``cookie banner'' OR
``cookie compliance'' OR
compliance with cookie* OR
``cookie consent'' OR
track* consent OR
consent management* OR
opt*out OR
opt* out]
\end{keywordbox}

\subsection{Thematic Literature Analysis}
\label{sec:thematic-analysis}

We consider a paper to be a \textit{tracking technique} paper if it provides an in-depth or foundational understanding of a specific tracking mechanism (e.g., first or reference paper on a technique).
Similarly, a \textit{tracking defense} paper comprises research that builds a system or an approach to block, prevent, or defend against a tracking technique, while \textit{regulatory compliance} research discusses different aspects of public policies (i.e., privacy regulation, standard, or guideline), their enforcement, and perhaps limitations.
Other papers focusing on measuring the prevalence or impact of different tracking techniques are classified as \textit{measurement} research and used throughout this SoK to support systematization of advances in web tracking.
This thematic organization is available as part of our artifact (see our~\refopenscience{sec:open-science} statement).

We complement our systematization of research-based tracking defenses with industry-based ones by performing a systematic review of changelogs and release notes for each version of five popular browsers---Brave, Google Chrome, Microsoft Edge (including former Internet Explorer), Mozilla Firefox, and Apple Safari---since their creation.
%
Specifically, we extract tracking defense related features across stable versions of browsers based on the description of changes and apply our defense-specific systematization criteria from~\autoref{tab:systematization-criteria} to each defense technique.
For each tracking technique and defense technique identified, we also find and associate community-built defenses (if any).

\input{tables/artifact-systematization}

\newpage
\input{appendix/need-for-fp.tex}

\input{appendix/mixed-tracking}

\input{tables/appendix-tracking-technique-papers-techniques}

\input{tables/appendix-tracking-techniques-papers-artifacts-capabilities}
\input{tables/defense-deployment-mechanism-type}

\begin{figure*}[!ht]
    \centering
    \includegraphics[width=\textwidth]{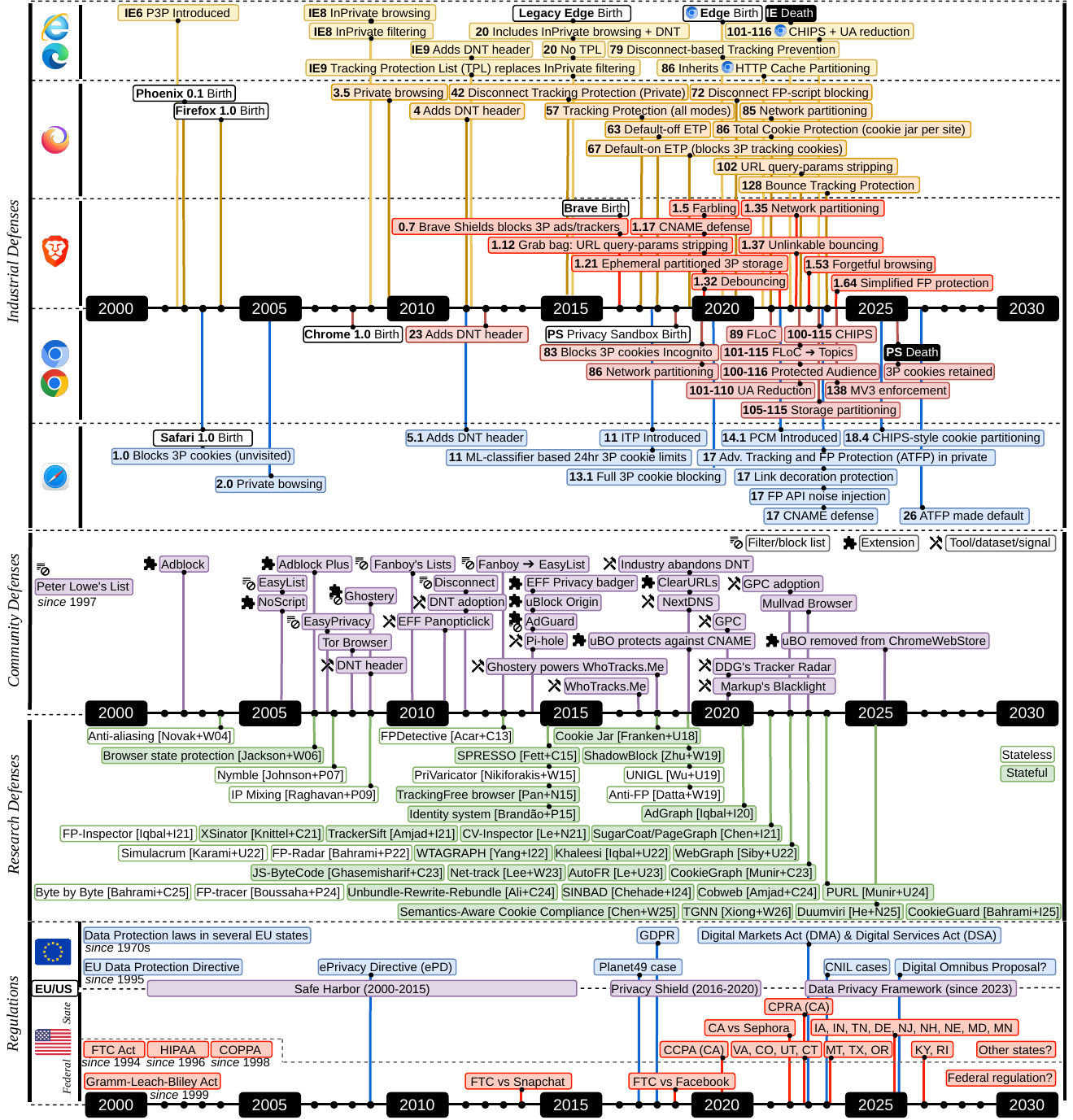}
    \caption{Evolution of \emph{industry-}, \emph{community-}, and \emph{research-based} defenses from 2000--2026 with EU and US \textit{regulations}.}
    \label{fig:evolution-defenses}
\end{figure*}

\input{tables/stateful-tracking-defenses}
\input{tables/stateless-tracking-defenses}

\input{tables/defenses-industrial}
\input{tables/defenses-community}

\input{tables/regulations-paper-distribution}
\input{tables/regulations-aspects}

%% file: tables/artifact-systematization.tex
\begin{table*}[ht]
  \small
\centering
\caption{Web tracking artifact taxonomy: mapping of tracked artifacts to artifact labels.}
\label{tab:artifact-taxonomy}
\renewcommand{\arraystretch}{1.2}
\begin{tabular}{@{} p{3.4cm} p{13.8cm} @{}}
\toprule
\textbf{Artifact Label} & \textbf{Artifacts Observed in the Literature Review} \\
\midrule

Browser Storage \& Cache &
  \textit{Cookies};
  \textit{Web Storage} (localStorage, sessionStorage, IndexedDB);
  \textit{Browser Caches} (HTTP, HSTS, favicon, Alt-Svc, Service Worker, and Cache API) \\

\midrule

JavaScript-Accessible Browser APIs &
  \textit{Fingerprinting APIs} (e.g., Navigator, Screen, Canvas, WebGL, AudioContext);
  \textit{Device APIs} (Battery Status, Media Devices, Gamepad, Speech Synthesis voices); 
  \textit{Browser Permissions} (e.g., camera, microphone) \\

\midrule

Hardware Artifacts &
  \textit{GPU Rendering Behavior} (timing variations, memory allocation, performance counters);
  \textit{Sensor Calibration Data} (accelerometer or gyroscope imperfections, speaker frequency response);
  \textit{Clock Drift Patterns} (system time offset, TCP timestamp skew);
  \textit{CPU Microarchitectural State} (scheduling statistics, memory footprint) \\

\midrule

Network Protocol Artifacts &
  \textit{IP Address};
  \textit{TLS/SSL Handshake Parameters} (cipher suites, SNI, client certificates);
  \textit{TCP/IP Stack Characteristics} (source port selection, IP ID fields, IPv6 flow labels);
  \textit{DNS Records} (CNAME records, reverse hostnames, TTL values);
  \textit{QUIC/HTTP2+ Transport Metadata} (connection IDs, retry tokens, transport parameters) \\

\midrule

HTTP Traffic Artifacts &
  Request Metadata, Response Content, Headers, Payload \\

\midrule

Navigational Artifacts &
  \textit{URL Components} (host, domain, path, query parameters);
  \textit{Redirect Chains} \\

\midrule

Webpage Artifacts &
  \textit{Resources} (DOM elements and their modifications, stylesheets, scripts);
  \textit{Interaction Patterns} (click traces, touch gesture characteristics, keystrokes, mouse movements, scroll depth, click coordinates, interaction timing);
  \textit{Rendering Characteristics} (font rendering, CSS feature detection, element bounding dimensions) \\

\midrule

Identifiers &
  \textit{Authentication IDs} (OAuth tokens, SSO redirect parameters, session IDs);
  \textit{PII} (email addresses, phone, PII hashes);
  \textit{Device-specific IDs} (IDFA, EMEI, MAC address); 
  \textit{Advertiser-specific IDs} (e.g., UUIDs)\\

\bottomrule
\end{tabular}
\end{table*}

%% file: appendix/need-for-fp.tex
\section{Need for Fingerprinting}
\label{sec:need-for-fingerprinting}

Constructively, fingerprinting can be thought of as a form of intrusion detection.
Web applications can learn about the browsing environment or behavior of their first-party users and associate it with specific user identities~\cite{LinPhishSheepsClothing2022}.
For example, a website can learn that the user is using a smartphone with specific dimensions or a desktop browser with a specific kind of GPU.
If the user's credentials are ever stolen and an attacker attempts to login to that service, the service can extract the attacker's fingerprint, observe a major difference against fingerprints from prior sessions, and request additional authentication data from the attacker (such as a one-time password).
The same techniques can be constructively used to differentiate real users from malicious bots or attackers engaging in ad fraud.
In practice, large online services commonly incorporate fingerprinting-derived signals into broader abuse-mitigation pipelines to detect automated account creation, credential stuffing, and spam. 
These signals are typically combined with behavioral features and rate-limiting mechanisms, enabling services to adapt authentication requirements based on assessed risk and maintaining service integrity in adversarial environments where traditional identifiers are unreliable.

Destructively, the same techniques that can identify user-impersonating attackers and bots, are turned against users who wish to remain anonymous.
With fingerprinting, a web app may be able to determine that a certain anonymous user is in fact eponymous, since their browser fingerprint matches that of a known user on the same platform.
This undesired re-identification occurs despite user's attempt to hide by deleting cookies or using the browser's private mode.
In a cross-site context, fingerprinting can be abused to link unrelated website visits, even when third-party cookies are disabled.

%% file: appendix/mixed-tracking.tex
\section{Mixed Tracking}
\label{sec:mixed-tracking}

\subsection{Mixed Tracking Techniques}
\label{sec:mixed-techniques}

Prior stateful and stateless approaches can also be combined to track users across contexts---bridging identity across login boundaries, devices, browsers, and native apps. 

\subsubsection{JavaScript-based Tracking}
\label{sec:mixed-techniques-js-tracking}

%
JavaScript can be used to read and write browser cookies and storage for persistence by stateful techniques while also be used to probe browser APIs to construct a fingerprint by stateless techniques.
Although JS-based approaches don't constitute a different tracking technique, it is included here as a canonical example of mixed tracking in practice.

\subsubsection{Authentication-based Tracking}
\label{sec:mixed-techniques-auth-tracking}

Authentication creates a deterministic, cross-context identifier (an account or email address) far more durable than any cookie or fingerprint.
Logged-in users are served more third-party trackers and personalized content~\cite{kaizerCharacterizingWebsiteBehaviors2016}, and logged-in pages exhibit more fingerprinting than other pages on the same site, suggesting trackers use authentication to link a stateless fingerprint to stable account identifiers to extend identification into anonymous sessions~\cite{senolDoubleEdgedSword2024}.
Moreover, OAuth-based single sign-on itself routinely leaks identifying information to identity providers and relying parties~\cite{dimovaEverybodysLookingSSOmething2023}.
As email addresses survive cookie clears and device changes, first parties increasingly hash and share them with advertising partners to build cross-site identity graphs that restore the linkage that third-party cookie restrictions were designed to prevent.

\subsubsection{Cross-device Tracking}
\label{sec:mixed-techniques-cross-device}

When deterministic identifiers are unavailable (e.g., logged out state), trackers fall back to probabilistic signals to link activity across devices---shared IP addresses, browsing patterns~\cite{caoCrossBrowserFingerprintingOS2017}, and behavioral features such as typing dynamics~\cite{yuanCrossdeviceTrackingIdentification2018}.
These signals are combined into proprietary \textit{cross-device graphs}, a practice found to be largely opaque to both users and regulators~\cite{zimmeckPrivacyAnalysisCrossdevice2017,brookmanCrossDeviceTrackingMeasurement2017}.
Because probabilistic identifiers are inherently noisy (e.g., ISPs dynamically rotate and share public IP addresses across households), trackers apply heuristics such as ignoring commercial, private, and proxied IP ranges or setting temporal thresholds to filter false matches.

\begin{opbox}
Given the lack of ground truth and diversity of proprietary linking algorithms forming cross-device identity graphs, how can cross-device tracking be reliably measured and systematically attributed to uncover specific signals used by different trackers?
\end{opbox}

\subsubsection{Other Cross-Context Tracking}
\label{sec:mixed-techniques-cross-context}

Trackers also bridge contexts within the same device.
Cross-browser fingerprinting uses OS- and hardware-level features stable across browsers to link users across browser boundaries~\cite{caoCrossBrowserFingerprintingOS2017}.
On mobile, Android Custom Tabs and WebViews share the browser's cookie jar with native apps, letting a tracker plants an identifier in one context that resurfaces in the other, surviving app reinstalls and device backups~\cite{wessels2025hytrack}.
%
Recently, some native apps and in-browser scripts have been found to silently link identities by communicating over \texttt{localhost}, a channel neither the browser's permission model nor the OS sandbox was designed to police~\cite{localmess-usenix-sec-26}.
These vectors share a common root cause---OS- and browser-level integration points enabling legitimate app-to-web handoffs that are invisible to in-browser tracking defenses.

\begin{opbox}
Cross-context tracking exploits integration boundaries between browsers, between apps and the web, and between user profiles that no single platform controls end to end. What architectural principles can unify isolation across these boundaries without fragmenting the seamless cross-context experiences users and developers expect?
\end{opbox}

\subsection{Defenses Against Mixed Tracking}
\label{sec:mixed-defenses}

\subsubsection{Industry-based Defenses}
\label{sec:mixed-defenses-industry}

\para{Against Cross-device and Cross-context Tracking.}
Deterministic cross-device tracking through shared login accounts is inherently difficult to defend against at the browser level, since the user has explicitly linked their devices by authenticating.
Browsers instead address the authentication pathway with purpose-built APIs: the Storage Access API, now shipped by all five major browsers, lets a third party request first-party-equivalent storage explicitly. Chrome and Edge additionally support Federated Credential Management (FedCM), which preserves single sign-on while preventing the identity provider from learning which relying party the user logs into. Brave blocks unrecognized identity providers by default.
On mobile, Apple's ATT framework has required explicit opt-in before an app can access the device's advertising identifier since iOS 14.5, and Google's Privacy Sandbox for Android proposed similar restrictions before being abandoned.
However, WebViews and in-app browsers share few of the standalone browser's tracking defenses---no extension support, no storage partitioning, no filter-list blocking---leaving users exposed in these embedded contexts.

\begin{opbox}
In cases such as WebViews and in-app browsers, can browser-level tracking protections be extended to embedded rendering engines, or does defense in these contexts require OS-level enforcement that applies regardless of which engine serves the page?
\end{opbox}

\subsubsection{Research-based Defenses}
\label{sec:mixed-defenses-research}

\para{Hardening the Identity Layer.}
Because cross-device and authentication-based tracking often runs through a shared identity provider, several proposals harden that layer directly.
SPRESSO inserts a forwarder between the relying party and the identity provider so the provider cannot learn which site a user logs into~\cite{fettSPRESSOSecurePrivacyRespecting2015}, encrypted access logging lets a service audit device access without retaining a persistent identifier~\cite{ortegaperezEncryptedAccessLogging2025}, and identity-system reform proposals re-architect national- and platform-level infrastructure to decouple shared identity from cross-service activity correlation~\cite{brandaoMendingTwoNationScale2015}.

\para{Cryptographic Advertising Proposals.}
A parallel line of work shows that measurements advertisers need such as conversion attribution and bid optimization, can be computed without either party learning individual user data.
Adnostic, Privad, and ObliviAd demonstrated this for on-device ad selection~\cite{toubianaAdnosticPrivacyPreserving2010,guhaPrivadPracticalPrivacy2011,backesObliviAdProvablySecure2012}, and Ibex later extended the guarantee to conversion attribution and bid optimization jointly using cryptographic aggregation~\cite{zhongIbexPrivacypreservingAd2022}.
Few of these designs have been adopted by browser vendors, largely due to computation and coordination costs, but they remain the clearest evidence that interest-based advertising does not inherently require disclosing anything about an individual user.

\subsubsection{Community-based Defenses}
\label{sec:mixed-defenses-community}

\para{Universal Opt-out Signals.}
Several attempts were made at implementing opt out and consent signals for users to communicate their privacy preferences to services. However, the adoption and enforcement of such signals by both senders and recipients is an unresolved coordination problem~\cite{hilsPrivacyPreferenceSignals2021}.

\noindent
\textit{Platform for Privacy Preferences Project (P3P)}~\cite{reaglePlatformPrivacyPreferences1999,cranorPlatformPrivacyPreferences2002,byersSearchingPrivacyDesign2005,cranorPlatformPrivacyPreferences2006} enabled websites to communicate their privacy practices to users in a standardized and fine-grained format.
The P3P specification was published by the W3C in 2002, but was rapidly criticized for being too complex to implement or understand.
Its utility was limited by the low number of sites that were adopting it~\cite{cranorUseP3PUser2002} and those implementing it correctly and transparently~\cite{cranorAnalysisP3PDeployment2003, leonTokenAttemptMisrepresentation2010}.
Ultimately, the specification was officially obsoleted by the W3C in 2018.

\noindent
\textit{Do Not Track (DNT)}~\cite{fieldingTrackingPreferenceExpression2019}, proposed in 2009 as a binary opt out signal, was adopted by every major browser but was never given legal force.
Indeed, CalOPPA, which influenced the design of DNT, only requires online services to say whether or not they respect it~\cite{CaliforniaCodeBPC2003}.
The W3C working group for DNT closed in 2019 without industry consensus.

\noindent\textit{Global Privacy Control (GPC)}~\cite{zimmeckGlobalPrivacyControl2024} can be viewed as a successor to DNT, but succeeds where DNT failed by being legally enforceable: California's 2022 settlement with Sephora~\cite{bontaAttorneyGeneralBonta2022} established that ignoring GPC violates the CCPA~\cite{attorneygeneralbecerraCCPARequiresBusinesses2021}, and Colorado has since imposed a similar requirement in 2024~\cite{coloradodepartmentoflawUniversalOptOutShortlist2024}.
Additional recognition of universal opt outs by states like Connecticut, Montana, Oregon, and Texas is driving broader adoption than DNT ever achieved, although it is still slow despite people finding the signal useful and usable~\cite{zimmeckUsabilityEnforceabilityGlobal2023,zimmeckGeneralizableActivePrivacy2024,azizJohnnyStillCant2024}.
DuckDuckGo, Brave, and Firefox send GPC by default; Chrome added support only in 2026~\cite{googleGlobalPrivacyControl2025}.
As GPC is not tied to any single storage or fingerprinting mechanism, it acts as a cross-cutting signal that covers every technique discussed in this SoK, given a site honors it.

\begin{opbox}
The applicability of GPC in the EU within ePD and GDPR context is an open discussion~\cite{berjonGPCGDPR2021,bielovaEffectDesignPatterns2024
}. Will EU regulators incorporate GPC as an opt-out signal? 
\end{opbox}

%% file: tables/appendix-tracking-technique-papers-techniques.tex
\newlength{\artfirstcol}
\newlength{\artdatacol}

\setlength{\artfirstcol}{2.6cm} 
\setlength{\artdatacol}{0.58cm}  

\newcolumntype{T}{>{\centering\arraybackslash}b{\artdatacol}}
\newcolumntype{Z}{>{\centering\arraybackslash}b{\artdatacol}}

\newcommand{\gtick}{\textcolor{green!50!black}{\checkmark}}

\begin{table*}[ht] 
\centering
\scriptsize
\setlength{\tabcolsep}{0.6pt}
\resizebox{\textwidth}{!}{%
\begin{tabular}{%
>{\raggedright\arraybackslash}m{\artfirstcol}|%
>{\centering\arraybackslash}m{1.4cm}|
*{24}{T}%
}
 
\toprule
\multicolumn{1}{>{\centering\arraybackslash}b{\artfirstcol}|}{\shortstack[c]{\textbf{Tracking Technique}\\\textbf{Literature}}} & \multicolumn{1}{>{\centering\arraybackslash}b{1.4cm}|}{\shortstack[c]{\textbf{Venue}\\\textbf{Year}}} & \multicolumn{1}{Z}{\rot{Third-party Cookie Tracking}} & \multicolumn{1}{Z}{\rot{Cookie Respawning}} & \multicolumn{1}{Z}{\rot{Cookie Syncing}} & \multicolumn{1}{Z}{\rot{First-party Cookie Tracking}} & \multicolumn{1}{Z}{\rot{Tracking Pixels/Tags}} & \multicolumn{1}{Z}{\rot{CNAME Tracking}} & \multicolumn{1}{Z}{\rot{Server-side Tracking}} & \multicolumn{1}{Z}{\rot{Link Decorations}} & \multicolumn{1}{Z}{\rot{Bounce Tracking}} & \multicolumn{1}{Z}{\rot{Authentication-based Tracking}} & \multicolumn{1}{Z}{\rot{Javascript-based Tracking}} & \multicolumn{1}{Z}{\rot{Browser Fingerprinting}} & \multicolumn{1}{Z}{\rot{Browser Extension Fingerprinting}} & \multicolumn{1}{Z}{\rot{Device Fingerprinting}} & \multicolumn{1}{Z}{\rot{Passive Fingerprinting}} & \multicolumn{1}{Z}{\rot{Kernel-based Side-channels}} & \multicolumn{1}{Z}{\rot{Network-based Side-channels}} & \multicolumn{1}{Z}{\rot{Storage-based Side-channels}} & \multicolumn{1}{Z}{\rot{Timing-based Side-channels}} & \multicolumn{1}{Z}{\rot{Behavioral Tracking}} & \multicolumn{1}{Z}{\rot{Cross-context Tracking}} & \multicolumn{1}{Z}{\rot{FLoC / Topics API}} & \multicolumn{1}{Z}{\rot{Protected Audience}} & \multicolumn{1}{Z}{\rot{Ad-Blocking Circumvention}} \\
\midrule
Felten et al.~\cite{feltenTimingAttacksWeb2000} &  CCS'00 &   &   &   &   &   &   &   &   &   &   &   &   &   &   &   &   &   &   &  \scalebox{1.15}{\textbf{\gtick}} &   &   &   &   &   \\
Kohno et al.~\cite{kohnoRemotePhysicalDevice2005} &  S\&P'05 &   &   &   &   &   &   &   &   &   &   &   &   &   &  \scalebox{1.15}{\textbf{\gtick}} &   &   &   &   &   &   &   &   &   &   \\
Chen et al.~\cite{chenSideChannelLeaksWeb2010} &  S\&P'10 &   &   &   &   &   &   &   &   &   &   &   &   &   &   &   &   &  \scalebox{1.15}{\textbf{\gtick}} &   &   &   &   &   &   &   \\
Yen et al.~\cite{yenHostFingerprintingTracking2012} &  NDSS'12 &   &   &   &   &   &   &   &   &   &   &   &  \scalebox{1.15}{\textbf{\gtick}} &   &  \scalebox{1.15}{\textbf{\gtick}} &   &   &   &   &   &   &   &   &   &   \\
Jana et al.~\cite{janaMementoLearningSecrets2012} &  S\&P'12 &   &   &   &   &   &   &   &   &   &   &   &   &   &   &   &  \scalebox{1.15}{\textbf{\gtick}} &   &   &   &   &   &   &   &   \\
Nikiforakis et al.~\cite{nikiforakisCookielessMonsterExploring2013} &  S\&P'13 &   &   &   &   &   &   &  \scalebox{1.15}{\textbf{\gtick}} &   &   &   &   &   &   &  \scalebox{1.15}{\textbf{\gtick}} &   &   &   &   &   &   &   &   &   &   \\
Acar et al.~\cite{acarWebNeverForgets2014} &  CCS'14 &   &  \scalebox{1.15}{\textbf{\gtick}} &  \scalebox{1.15}{\textbf{\gtick}} &   &   &   &   &   &   &   &  \scalebox{1.15}{\textbf{\gtick}} &  \scalebox{1.15}{\textbf{\gtick}} &   &   &   &   &   &   &   &   &   &   &   &   \\
Zhou et al.~\cite{zhouAcousticFingerprintingRevisited2014} &  CCS'14 &   &   &   &   &   &   &   &   &   &   &   &   &   &  \scalebox{1.15}{\textbf{\gtick}} &   &   &   &   &   &   &   &   &   &   \\
Oren et al.~\cite{orenSpySandboxPractical2015} &  CCS'15 &   &   &   &   &   &   &   &   &   &   &   &   &   &   &   &   &   &   &  \scalebox{1.15}{\textbf{\gtick}} &   &   &   &   &   \\
Das et al.~\cite{dasWebsSixthSense2018, dasTrackingMobileWeb2016, dasEveryMoveYou2018} &  NDSS'16 &   &   &   &   &   &   &   &   &   &   &  \scalebox{1.15}{\textbf{\gtick}} &   &   &  \scalebox{1.15}{\textbf{\gtick}} &   &   &   &   &   &   &   &   &   &   \\
Laperdrix et al.~\cite{laperdrixBeautyBeastDiverting2016} &  S\&P'16 &   &   &   &   &   &   &  \scalebox{1.15}{\textbf{\gtick}} &   &   &   &   &  \scalebox{1.15}{\textbf{\gtick}} &   &   &   &   &   &   &   &   &   &   &   &   \\
Jain et al.~\cite{jainMiningLatentClient2016} &  PETS'16 &   &   &   &   &   &   &   &   &   &   &   &   &   &   &   &   &  \scalebox{1.15}{\textbf{\gtick}} &   &   &   &   &   &   &   \\
Gonzalez et al.~\cite{gonzalezUserProfilingTime2016} &  IMC'16 &   &   &   &   &   &   &   &   &   &   &   &   &   &   &   &   &  \scalebox{1.15}{\textbf{\gtick}} &   &   &   &   &   &   &   \\
Kaizer et al.~\cite{kaizerCharacterizingWebsiteBehaviors2016} &  IMC'16 &   &   &   &  \scalebox{1.15}{\textbf{\gtick}} &   &   &   &   &   &  \scalebox{1.15}{\textbf{\gtick}} &  \scalebox{1.15}{\textbf{\gtick}} &   &   &   &   &   &   &   &   &  \scalebox{1.15}{\textbf{\gtick}} &   &   &   &   \\
Cao et al.~\cite{caoCrossBrowserFingerprintingOS2017} &  NDSS'17 &   &   &   &   &   &   &   &   &   &   &  \scalebox{1.15}{\textbf{\gtick}} &   &   &  \scalebox{1.15}{\textbf{\gtick}} &   &   &   &   &   &   &   &   &   &   \\
Starov et al.~\cite{starovXHOUNDQuantifyingFingerprintability2017} &  S\&P'17 &   &   &   &   &   &   &  \scalebox{1.15}{\textbf{\gtick}} &   &   &   &   &   &  \scalebox{1.15}{\textbf{\gtick}} &   &   &   &   &   &   &   &   &   &   &   \\
Vastel et al.~\cite{vastelFPSTALKERTrackingBrowser2018} &  S\&P'18 &   &   &   &   &   &   &  \scalebox{1.15}{\textbf{\gtick}} &   &   &   &   &  \scalebox{1.15}{\textbf{\gtick}} &   &   &   &   &   &   &   &   &   &   &   &   \\
Foppe et al.~\cite{foppeExploitingTLSClient2018} &  PETS'18 &   &   &   &   &   &   &   &   &   &   &   &   &   &   &   &   &  \scalebox{1.15}{\textbf{\gtick}} &   &   &   &   &   &   &   \\
Masood et al.~\cite{masoodTouchYoureTrappcked2018} &  PETS'18 &   &   &   &   &   &   &   &   &   &   &  \scalebox{1.15}{\textbf{\gtick}} &   &   &   &   &   &   &   &   &  \scalebox{1.15}{\textbf{\gtick}} &   &   &   &   \\
Olejnik et al.~\cite{olejnikLeakingBatteryPrivacy2016} &  USENIX'18 &   &   &   &   &   &   &   &   &   &   &  \scalebox{1.15}{\textbf{\gtick}} &  \scalebox{1.15}{\textbf{\gtick}} &   &   &   &   &   &   &   &   &   &   &   &   \\
Bashir et al.~\cite{bashirHowTrackingCompanies2018} &  IMC'18 &   &   &   &   &   &   &   &   &   &   &  \scalebox{1.15}{\textbf{\gtick}} &   &   &   &   &   &   &   &   &   &   &   &   &  \scalebox{1.15}{\textbf{\gtick}} \\
Klein et al.~\cite{klein2019dns} &  NDSS'19 &   &   &   &   &   &   &   &   &   &   &  \scalebox{1.15}{\textbf{\gtick}} &   &   &   &   &   &   &  \scalebox{1.15}{\textbf{\gtick}} &   &   &   &   &   &   \\
Hu et al.~\cite{huCharacterizingPixelTracking2019} &  S\&P'19 &   &   &   &   &   &  \scalebox{1.15}{\textbf{\gtick}} &   &   &   &   &   &   &   &   &   &   &   &   &   &   &   &   &   &   \\
Martin et al.~\cite{martinHandoffAllYour2019} &  PETS'19 &   &   &   &   &   &   &   &   &   &   &   &   &   &  \scalebox{1.15}{\textbf{\gtick}} &   &   &   &   &   &   &   &   &   &   \\
Sy et al.~\cite{syQUICLookWeb2019} &  PETS'19 &   &   &   &   &   &   &   &   &   &   &   &   &   &   &   &   &  \scalebox{1.15}{\textbf{\gtick}} &  \scalebox{1.15}{\textbf{\gtick}} &   &   &   &   &   &   \\
Karami et al.~\cite{karamiCarnusExploringPrivacy2020} &  NDSS'20 &   &   &   &   &   &   &   &   &   &   &  \scalebox{1.15}{\textbf{\gtick}} &   &  \scalebox{1.15}{\textbf{\gtick}} &   &   &   &   &   &   &   &   &   &   &   \\
Koop et al.~\cite{koopInDepthEvaluationRedirect2020} &  PETS'20 &   &   &   &   &   &   &   &  \scalebox{1.15}{\textbf{\gtick}} &   &   &   &   &   &   &   &   &   &   &   &   &   &   &   &   \\
Li et al.~\cite{liShimShimmenyEvaluating2020} &  USENIX'20 &   &   &   &   &   &   &   &  \scalebox{1.15}{\textbf{\gtick}} &   &   &   &   &   &   &   &   &   &   &   &   &   &   &   &   \\
Mishra et al.~\cite{mishraDontCountMe2020} &  WWW'20 &   &   &   &   &   &   &   &   &   &   &   &   &   &   &  \scalebox{1.15}{\textbf{\gtick}} &   &   &   &   &   &   &   &   &   \\
Solomos et al.~\cite{solomosTalesFaviconsCaches2021} &  NDSS'21 &   &   &   &   &   &   &   &   &   &   &   &   &   &   &   &   &   &  \scalebox{1.15}{\textbf{\gtick}} &   &   &   &   &   &   \\
Klein et al.~\cite{kleinCrossLayerAttacks2021} &  S\&P'21 &   &   &   &   &   &   &   &   &   &   &   &   &   &  \scalebox{1.15}{\textbf{\gtick}} &   &  \scalebox{1.15}{\textbf{\gtick}} &   &   &   &   &   &   &   &   \\
Dimova et al.~\cite{dimovaCNAMEGameLargescale2021} &  PETS'21 &   &   &   &   &   &  \scalebox{1.15}{\textbf{\gtick}} &   &   &   &   &  \scalebox{1.15}{\textbf{\gtick}} &   &   &   &   &   &   &   &   &   &   &   &   &   \\
Mishra et al.~\cite{mishraDejaVuAbusing2021} &  PETS'21 &   &   &   &   &   &   &   &   &   &   &   &   &   &   &   &   &   &  \scalebox{1.15}{\textbf{\gtick}} &   &   &   &   &   &   \\
Laperdrix et al.~\cite{laperdrixFingerprintingStyleDetecting2021} &  USENIX'21 &   &   &   &   &   &   &   &   &   &   &  \scalebox{1.15}{\textbf{\gtick}} &   &  \scalebox{1.15}{\textbf{\gtick}} &   &   &   &   &   &   &   &   &   &   &   \\
Chen et al.~\cite{chenCookieSwapParty2021} &  WWW'21 &   &   &   &  \scalebox{1.15}{\textbf{\gtick}} &   &   &   &   &   &   &  \scalebox{1.15}{\textbf{\gtick}} &   &   &   &   &   &   &   &   &   &   &   &   &   \\
Rye et al.~\cite{ryeFollowScentDefeating2021} &  IMC'21 &   &   &   &   &   &   &   &   &   &   &   &   &   &   &  \scalebox{1.15}{\textbf{\gtick}} &   &   &   &   &   &   &   &   &   \\
Laor et al.~\cite{laorDRAWNAPARTDevice2022} &  NDSS'22 &   &   &   &   &   &   &   &   &   &   &  \scalebox{1.15}{\textbf{\gtick}} &   &   &  \scalebox{1.15}{\textbf{\gtick}} &   &   &   &   &   &   &   &   &   &   \\
Jin et al.~\cite{jinTimingBasedBrowsingPrivacy2022} &  S\&P'22 &   &   &   &   &   &   &  \scalebox{1.15}{\textbf{\gtick}} &   &   &   &   &   &   &   &   &   &   &   &  \scalebox{1.15}{\textbf{\gtick}} &   &   &   &   &   \\
Khodayari et al.~\cite{khodayariStateSameSiteStudying2022} &  S\&P'22 &  \scalebox{1.15}{\textbf{\gtick}} &   &   &  \scalebox{1.15}{\textbf{\gtick}} &   &   &   &   &   &   &   &   &   &   &   &   &   &   &   &   &   &   &   &   \\
Randall et al.~\cite{randallMeasuringUIDSmuggling2022} &  IMC'22 &   &   &   &   &   &   &   &   &  \scalebox{1.15}{\textbf{\gtick}} &   &  \scalebox{1.15}{\textbf{\gtick}} &   &   &   &   &   &   &   &   &   &   &   &   &   \\
Berke et al.~\cite{berkePrivacyLimitationsInterestbased2022} &  CCS'22 &   &   &   &   &   &   &   &   &   &   &  \scalebox{1.15}{\textbf{\gtick}} &   &   &   &   &   &   &   &   &   &   &  \scalebox{1.15}{\textbf{\gtick}} &   &   \\
Solomos et al.~\cite{solomosEscapingConfinesTime2022} &  CCS'22 &   &   &   &   &   &   &   &   &   &   &  \scalebox{1.15}{\textbf{\gtick}} &   &  \scalebox{1.15}{\textbf{\gtick}} &   &   &   &   &   &   &   &   &   &   &   \\
Ali et al.~\cite{aliNavigatingMurkyWaters2023} &  NDSS'23 &   &   &   &   &   &   &   &   &   &   &  \scalebox{1.15}{\textbf{\gtick}} &   &   &   &   &   &   &  \scalebox{1.15}{\textbf{\gtick}} &   &   &   &   &  \scalebox{1.15}{\textbf{\gtick}} &   \\
Nomoto et al.~\cite{nomoto2023browser} &  NDSS'23 &   &   &   &   &   &   &   &   &   &   &  \scalebox{1.15}{\textbf{\gtick}} &  \scalebox{1.15}{\textbf{\gtick}} &   &   &   &   &   &   &   &   &   &   &   &   \\
Lin et al.~\cite{linFashionFauxPas2023} &  S\&P'23 &   &   &   &   &   &   &  \scalebox{1.15}{\textbf{\gtick}} &   &   &   &   &  \scalebox{1.15}{\textbf{\gtick}} &   &   &   &   &   &   &   &   &   &   &   &   \\
Jha et al.~\cite{jhaRobustnessTopicsAPI2023} &  PETS'23 &   &   &   &   &   &   &   &   &   &   &  \scalebox{1.15}{\textbf{\gtick}} &   &   &   &   &   &   &   &   &   &   &  \scalebox{1.15}{\textbf{\gtick}} &   &   \\
Kol et al.~\cite{kol2023device} &  USENIX'23 &   &   &   &   &   &   &   &   &   &   &   &   &   &  \scalebox{1.15}{\textbf{\gtick}} &   &  \scalebox{1.15}{\textbf{\gtick}} &   &   &   &   &   &   &   &   \\
Snyder et al.~\cite{snyderPoolPartyExploitingBrowser2023} &  USENIX'23 &   &   &   &   &   &   &   &   &   &   &  \scalebox{1.15}{\textbf{\gtick}} &   &   &   &   &   &   &  \scalebox{1.15}{\textbf{\gtick}} &   &   &  \scalebox{1.15}{\textbf{\gtick}} &   &   &   \\
Su et al.~\cite{suAutomaticDiscoveryEmerging2023} &  WWW'23 &   &   &   &   &   &   &   &   &   &   &  \scalebox{1.15}{\textbf{\gtick}} &  \scalebox{1.15}{\textbf{\gtick}} &   &   &   &   &   &   &   &   &   &   &   &   \\
Beugin et al.~\cite{beuginInterestdisclosingMechanismsAdvertising2024} &  PETS'24 &   &   &   &   &   &   &   &   &   &   &  \scalebox{1.15}{\textbf{\gtick}} &   &   &   &   &   &   &   &   &   &   &  \scalebox{1.15}{\textbf{\gtick}} &   &   \\
Fouad et al.~\cite{fouadDevilDetailsDetection2024} &  PETS'24 &   &   &   &   &   &   &  \scalebox{1.15}{\textbf{\gtick}} &   &   &   &   &   &   &   &   &   &   &   &   &   &   &   &   &   \\
Chehade et al.~\cite{chehade2025double} &  USENIX'25 &   &   &   &   &   &   &   &   &   &   &  \scalebox{1.15}{\textbf{\gtick}} &   &  \scalebox{1.15}{\textbf{\gtick}} &   &   &   &   &   &   &   &   &   &   &   \\
Wessels et al.~\cite{wessels2025hytrack} &  USENIX'25 &   &  \scalebox{1.15}{\textbf{\gtick}} &   &   &   &   &   &   &   &   &   &   &   &   &   &   &   &   &   &   &  \scalebox{1.15}{\textbf{\gtick}} &   &   &   \\
Huyghe et al.~\cite{huygheFPRainbowFingerprintBasedBrowser2025} &  WWW'25 &   &   &   &   &   &   &   &   &   &   &  \scalebox{1.15}{\textbf{\gtick}} &  \scalebox{1.15}{\textbf{\gtick}} &   &   &   &   &   &   &   &   &   &   &   &   \\
Bekos et al.~\cite{bekosPIIxelLeaksPassive2025} &  CCS'25 &   &   &   &   &  \scalebox{1.15}{\textbf{\gtick}} &   &   &   &   &   &  \scalebox{1.15}{\textbf{\gtick}} &   &   &   &   &   &   &   &   &   &   &   &   &   \\
Ukani et al.~\cite{ukaniLocalFramesExploiting2025} &  CCS'25 &  \scalebox{1.15}{\textbf{\gtick}} &   &   &   &   &   &   &   &   &   &  \scalebox{1.15}{\textbf{\gtick}} &  \scalebox{1.15}{\textbf{\gtick}} &   &   &   &   &   &   &   &   &   &   &   &  \scalebox{1.15}{\textbf{\gtick}} \\
 
\bottomrule
\end{tabular}%
}
\caption{Literature on papers studying web tracking techniques over the years.}
\label{tab:technique-papers}
\end{table*}

%% file: tables/appendix-tracking-techniques-papers-artifacts-capabilities.tex
\begin{table}[!t]
\centering
\scriptsize
\setlength{\tabcolsep}{0.6pt}
\renewcommand{\arraystretch}{1.0}
\begin{tabular}{%
>{\raggedright\arraybackslash}m{2.6cm}|%
*{8}{>{\centering\arraybackslash}m{0.7cm}}%
}

\toprule
\multicolumn{1}{>{\centering\arraybackslash}b{2.6cm}|}{\shortstack[c]{\textbf{Tracking Technique}\\\textbf{Literature}}} & \multicolumn{1}{>{\centering\arraybackslash}b{0.7cm}}{\rot{Browser Storage and Cache}} & \multicolumn{1}{>{\centering\arraybackslash}b{0.7cm}}{\rot{JavaScript-Accessible Browser APIs}} & \multicolumn{1}{>{\centering\arraybackslash}b{0.7cm}}{\rot{Webpage Artifacts}} & \multicolumn{1}{>{\centering\arraybackslash}b{0.7cm}}{\rot{Hardware Artifacts}} & \multicolumn{1}{>{\centering\arraybackslash}b{0.7cm}}{\rot{Network Protocol Artifacts}} & \multicolumn{1}{>{\centering\arraybackslash}b{0.7cm}}{\rot{HTTP Traffic Artifacts}} & \multicolumn{1}{>{\centering\arraybackslash}b{0.7cm}}{\rot{Navigational Artifacts}} & \multicolumn{1}{>{\centering\arraybackslash}b{0.7cm}}{\rot{Identifiers}} \\
\midrule
Felten et al.~\cite{feltenTimingAttacksWeb2000} & {\symStorage} & {\symExecution} & {\symInclusion} &  &  &  &  & \\
Kohno et al.~\cite{kohnoRemotePhysicalDevice2005} &  &  &  & {\symObservation} & {\symObservation} &  &  & \\
Chen et al.~\cite{chenSideChannelLeaksWeb2010} &  &  &  &  & {\symObservation} &  &  & \\
Yen et al.~\cite{yenHostFingerprintingTracking2012} & {\symStorage} & {\symExecution} & {\symInclusion} &  & {\symObservation} & {\symInclusion\,\symSharing} &  & {\symObservation} \\
Jana et al.~\cite{janaMementoLearningSecrets2012} &  &  &  & {\symObservation} & &  &  & \\
Nikiforakis et al.~\cite{nikiforakisCookielessMonsterExploring2013} &  & {\symExecution} & {\symInclusion\,\symExecution} &  &  & {\symInclusion\,\symSharing} &  & \\
Zhou et al.~\cite{zhouAcousticFingerprintingRevisited2014} &  &  &  & {\symExecution} &  &  &  & \\
Acar et al.~\cite{acarWebNeverForgets2014} & {\symStorage} & {\symExecution} & {\symInclusion\,\symExecution} &  &  & {\symInclusion\,\symSharing} & {\symSharing} & {\symStorage\,\symSharing}\\
Oren et al.~\cite{orenSpySandboxPractical2015} &  & {\symExecution} &  & {\symExecution} &  &  &  & \\
Das et al.~\cite{dasWebsSixthSense2018, dasTrackingMobileWeb2016, dasEveryMoveYou2018} &  & {\symExecution} &  & {\symExecution} &  &  &  & \\
Laperdrix et al.~\cite{laperdrixBeautyBeastDiverting2016} &  & {\symExecution} & {\symInclusion\,\symExecution} &  &  & {\symInclusion\,\symSharing} &  & \\
Jain et al.~\cite{jainMiningLatentClient2016} &  &  &  &  & {\symObservation} &  &  & {\symObservation}\\
Gonzalez et al.~\cite{gonzalezUserProfilingTime2016} &  &  &  &  & {\symObservation} & {\symObservation} &  & \\
Kaizer et al.~\cite{kaizerCharacterizingWebsiteBehaviors2016} & {\symStorage} &  & {\symInclusion\,\symExecution} &  &  & {\symInclusion\,\symSharing} &  & {\symStorage}\\
Cao et al.~\cite{caoCrossBrowserFingerprintingOS2017} &  & {\symExecution} & {\symInclusion\,\symExecution} & {\symExecution} &  &  &  & \\
Starov et al.~\cite{starovXHOUNDQuantifyingFingerprintability2017} &  &  & {\symInclusion\,\symExecution} &  &  &  &  & \\
Vastel et al.~\cite{vastelFPSTALKERTrackingBrowser2018} &  & {\symExecution} & {\symInclusion\,\symExecution} &  &  & {\symInclusion\,\symSharing} &  & \\
Foppe et al.~\cite{foppeExploitingTLSClient2018} &  &  &  &  & {\symObservation} &  &  & {\symObservation}\\
Masood et al.~\cite{masoodTouchYoureTrappcked2018} &  & {\symExecution} & {\symInclusion\,\symExecution} &  &  &  &  & \\
Olejnik et al.~\cite{olejnikLeakingBatteryPrivacy2016} &  & {\symExecution} &  &  &  &  &  & \\
Bashir et al.~\cite{bashirHowTrackingCompanies2018} & {\symStorage} & {\symExecution} & {\symInclusion} &  &  & {\symInclusion\,\symSharing} &  & \\
Klein et al.~\cite{klein2019dns} &  & {\symExecution} & {\symInclusion} &  & {\symObservation} &  &  & \\
Hu et al.~\cite{huCharacterizingPixelTracking2019} & {\symStorage} &  & {\symInclusion} &  &  & {\symInclusion\,\symSharing} & {\symSharing} & {\symStorage\,\symSharing}\\
Sy et al.~\cite{syQUICLookWeb2019} &  &  &  &  & {\symObservation} &  &  & {\symObservation}\\
Martin et al.~\cite{martinHandoffAllYour2019} &  & &  & {\symObservation} &  &  &  & {\symObservation}\\
Karami et al.~\cite{karamiCarnusExploringPrivacy2020} &  &  & {\symInclusion\,\symExecution} &  &  & {\symInclusion\,\symSharing} &  & \\
Koop et al.~\cite{koopInDepthEvaluationRedirect2020} & {\symStorage} &  & {\symInclusion} &  &  &  & {\symSharing} & \\
Li et al.~\cite{liShimShimmenyEvaluating2020} &  &  & {\symInclusion} &  &  & {\symInclusion\,\symSharing} & {\symSharing} & \\
Mishra et al.~\cite{mishraDontCountMe2020} & {\symStorage} &  &  &  & {\symObservation} & {\symInclusion\,\symSharing} &  & \\
Solomos et al.~\cite{solomosTalesFaviconsCaches2021} & {\symStorage} &  &  &  &  &  &  & \\
Klein et al.~\cite{kleinCrossLayerAttacks2021} &  &  &  & {\symObservation} & {\symObservation} &  &  & \\
Mishra et al.~\cite{mishraDejaVuAbusing2021} & {\symStorage} &  & {\symInclusion} &  &  & {\symInclusion\,\symSharing} &  & \\
Dimova et al.~\cite{dimovaCNAMEGameLargescale2021} & {\symStorage} &  & {\symInclusion} &  & {\symSharing} & {\symInclusion\,\symSharing} &  & \\
Laperdrix et al.~\cite{laperdrixFingerprintingStyleDetecting2021} &  & {\symExecution} & {\symInclusion\,\symExecution} &  &  &  &  & \\
Chen et al.~\cite{chenCookieSwapParty2021} & {\symStorage} &  & {\symInclusion\,\symExecution} &  &  & {\symInclusion\,\symSharing} &  & \\
Rye et al.~\cite{ryeFollowScentDefeating2021} &  &  &  &  & {\symObservation} &  &  & \\
Laor et al.~\cite{laorDRAWNAPARTDevice2022} &  & {\symExecution} & {\symInclusion\,\symExecution} & {\symExecution} &  &  &  & \\
Khodayari et al.~\cite{khodayariStateSameSiteStudying2022} & {\symStorage} &  &  &  &  & {\symSharing} &  & \\
Jin et al.~\cite{jinTimingBasedBrowsingPrivacy2022} &  & {\symExecution} &  & {\symExecution} &  &  &  & \\
Randall et al.~\cite{randallMeasuringUIDSmuggling2022} & {\symStorage} & {\symExecution} & {\symInclusion\,\symExecution} &  &  & {\symInclusion\,\symSharing} & {\symSharing} & {\symStorage\,\symSharing}\\
Berke et al.~\cite{berkePrivacyLimitationsInterestbased2022} &  & {\symExecution} &  &  &  &  &  & \\
Solomos et al.~\cite{solomosEscapingConfinesTime2022} &  & {\symExecution} & {\symInclusion\,\symExecution} &  &  &  &  & \\
Nomoto et al.~\cite{nomoto2023browser} &  & {\symExecution} &  &  &  &  &  & \\
Ali et al.~\cite{aliNavigatingMurkyWaters2023} & {\symStorage} &  &  &  &  &  &  & \\
Lin et al.~\cite{linFashionFauxPas2023} &  & {\symExecution} & {\symInclusion\,\symExecution} &  &  &  &  & \\
Jha et al.~\cite{jhaRobustnessTopicsAPI2023} &  & {\symExecution} &  &  &  &  &  & \\
Kol et al.~\cite{kol2023device} &  &  &  & {\symObservation} & {\symObservation} &  &  & \\
Snyder et al.~\cite{snyderPoolPartyExploitingBrowser2023} & {\symStorage} & {\symExecution} &  &  &  &  &  & \\
Su et al.~\cite{suAutomaticDiscoveryEmerging2023} &  & {\symExecution} & {\symInclusion\,\symExecution} &  &  &  &  & \\
Beugin et al.~\cite{beuginInterestdisclosingMechanismsAdvertising2024} &  & {\symExecution} &  &  &  &  &  & \\
Fouad et al.~\cite{fouadDevilDetailsDetection2024} & {\symStorage} &  & {\symInclusion} &  & {\symSharing} & {\symInclusion\,\symSharing} &  & \\
Chehade et al.~\cite{chehade2025double} &  &  & {\symInclusion\,\symExecution} &  &  & {\symInclusion\,\symSharing} &  & \\
Wessels et al.~\cite{wessels2025hytrack} & {\symStorage} &  &  &  &  &  &  & {\symStorage}\\
Huyghe et al.~\cite{huygheFPRainbowFingerprintBasedBrowser2025} &  & {\symExecution} & {\symInclusion\,\symExecution} &  &  &  &  & \\
Ukani et al.~\cite{ukaniLocalFramesExploiting2025} & {\symStorage} & {\symExecution} & {\symInclusion\,\symExecution} &  &  &  &  & \\
Bekos et al.~\cite{bekosPIIxelLeaksPassive2025} & {\symStorage} & {\symExecution} & {\symInclusion\,\symExecution} &  &  & {\symInclusion\,\symSharing} & {\symSharing} & \\

\bottomrule
\end{tabular}
\caption{Tracker capabilities exercised over artifacts as studied in each tracking techniques related paper.}
\label{tab:paper-artifacts}
\end{table}

%% file: tables/defense-deployment-mechanism-type.tex
\newcommand{\cmark}{\scalebox{1.15}{\textbf{\gtick}}}

\begin{table}[!t]
\centering
\scriptsize
\setlength{\tabcolsep}{0.5pt}
\renewcommand{\arraystretch}{0.94}
\begin{tabular}{%
>{\raggedright\arraybackslash}m{3.4cm}|%
>{\centering\arraybackslash}m{1.5cm}|%
>{\centering\arraybackslash}m{0.7cm}|%
*{8}{>{\centering\arraybackslash}m{0.3cm}}%
}

\toprule
\multicolumn{1}{>{\centering\arraybackslash}b{3.4cm}|}{\shortstack[c]{\textbf{Tracking Defense}\\\textbf{Literature}\\\textbf{(Defense System Name)}}} &
\multicolumn{1}{>{\centering\arraybackslash}b{1.5cm}|}{\shortstack[c]{\textbf{Venue}\\\textbf{Year}}} &
\multicolumn{1}{>{\centering\arraybackslash}b{0.7cm}|}{\rot{Deployment Type}} &
\multicolumn{1}{>{\cellcolor{cellred}\centering\arraybackslash}b{0.32cm}}{\rot{ML-based}} &
\multicolumn{1}{>{\cellcolor{cellred}\centering\arraybackslash}b{0.32cm}}{\rot{Program Instrum.}} &
\multicolumn{1}{>{\cellcolor{cellred}\centering\arraybackslash}b{0.32cm}}{\rot{Circumvention-based}} &
\multicolumn{1}{>{\cellcolor{cellred}\centering\arraybackslash}b{0.32cm}}{\rot{LLM-based}} &
\multicolumn{1}{>{\cellcolor{cellyellow}\centering\arraybackslash}b{0.32cm}}{\rot{API Perturbation}} &
\multicolumn{1}{>{\cellcolor{cellyellow}\centering\arraybackslash}b{0.32cm}}{\rot{State Partitioning}} &
\multicolumn{1}{>{\cellcolor{cellyellow}\centering\arraybackslash}b{0.32cm}}{\rot{Network Modif.}} &
\multicolumn{1}{>{\cellcolor{cellyellow}\centering\arraybackslash}b{0.32cm}}{\rot{Cryptographic}} \\
\midrule

Anti-aliasing~\cite{novakAntialiasingWeb2004} & WWW'04 & \dNative &  &  &  &  & \cmark &  &  &  \\
Browser state protection~\cite{jacksonProtectingBrowserState2006} & WWW'06 & \dNative &  &  &  &  & \cmark & \cmark &  &  \\
XSS Prevention with Taint~\cite{vogtCrossSiteScripting2007} & NDSS'07 & \dNative &  & \cmark &  &  &  &  &  &  \\
Nymble~\cite{johnsonNymbleAnonymousIPAddress2007} & PETS'07 & \dNative\,\dServer &  &  &  &  &  &  &  & \cmark \\
De-identification~\cite{zhengWebUserDeidentification2008} & WWW'08 & \dServer &  &  &  &  & \cmark &  &  &  \\
IP Address Mixing~\cite{raghavanEnlistingISPsImprove2009} & PETS'09 & \dServer &  &  &  &  &  &  & \cmark &  \\
UPS anonymization~\cite{zhuAnonymizingUserProfiles2010} & WWW'10 & \dExt\,\dServer &  &  &  &  & \cmark &  &  &  \\
FPDetective~\cite{acarFPDetectiveDustingWeb2013} & CCS'13 & \dTool &  & \cmark &  &  &  &  &  &  \\
Privee~\cite{zimmeckPriveeArchitectureAutomatically2014} & USENIX'14 & \dTool & \cmark &  &  &  &  &  &  &  \\
SPRESSO~\cite{fettSPRESSOSecurePrivacyRespecting2015} & CCS'15 & \dNative\,\dServer &  &  &  &  &  &  &  & \cmark \\
UCognito~\cite{xuUCognitoPrivateBrowsing2015} & CCS'15 & \dNative &  &  &  &  &  & \cmark &  &  \\
Bloom Cookies~\cite{morBloomCookiesWeb2015} & NDSS'15 & \dExt\,\dServer &  &  &  &  &  &  &  & \cmark \\
TrackingFree Browser~\cite{panNotKnowWhat2015} & NDSS'15 & \dNative &  &  &  &  &  & \cmark &  &  \\
Auto-blacklist~\cite{gugelmannAutomatedApproachComplementing2015} & PETS'15 & \dExt\,\dTool & \cmark &  &  &  &  &  &  &  \\
Identity System Reform~\cite{brandaoMendingTwoNationScale2015} & PETS'15 & \dServer &  &  &  &  &  &  &  & \cmark \\
PriVaricator~\cite{nikiforakisPriVaricatorDeceivingFingerprinters2015} & WWW'15 & \dNative &  &  &  &  & \cmark &  &  &  \\
TrackMeOrNot~\cite{mengTrackMeOrNotEnablingFlexible2016} & WWW'16 & \dExt &  &  &  &  &  & \cmark &  &  \\

One-class tracker classifier~\cite{ikramSeamlessTrackingFreeWeb2017} & PETS'17 & \dTool & \cmark &  &  &  &  &  &  &  \\
Anti-adblock diff. exec.~\cite{zhuMeasuringDisruptingAntiAdblockers2018} & NDSS'18 & \dNative\,\dTool &  & \cmark & \cmark &  &  &  &  &  \\
NoMoAds~\cite{shubaNoMoAdsEffectiveEfficient2018} & PETS'18 & \dServer & \cmark &  &  &  &  &  & \cmark &  \\
AdVersarial~\cite{tramerAdVersarialPerceptualAd2019} & CCS'19 & \dServer & \cmark &  &  &  &  &  &  &  \\
UNIGL~\cite{wuRenderedPrivateMaking2019} & USENIX'19 & \dNative &  & \cmark &  &  & \cmark &  &  &  \\
ShadowBlock~\cite{zhuShadowBlockLightweightStealthy2019} & WWW'19 & \dNative &  &  & \cmark &  &  &  &  &  \\
AdGraph~\cite{iqbalAdGraphGraphBasedApproach2020} & S\&P'20 & \dNative\,\dTool & \cmark &  &  &  &  &  &  &  \\
NoMoATS~\cite{shubaNoMoATSAutomaticDetection2020} & PETS'20 & \dServer & \cmark &  &  &  &  &  &  &  \\
TFO-TLS Enhancement~\cite{syEnhancedPerformancePrivacy2020} & PETS'20 & \dServer &  &  &  &  &  &  & \cmark &  \\
XSinator~\cite{knittelXSinatorcomFormalModel2021} & CCS'21 & \dTool &  &  &  &  &  & \cmark &  &  \\
FP-Inspector~\cite{iqbalFingerprintingFingerprintersLearning2021} & S\&P'21 & \dNative\,\dTool & \cmark &  &  &  &  &  &  &  \\
SugarCoat/PageGraph-based~\cite{chenDetectingFilterList2021} & S\&P'21 & \dNative\,\dTool &  & \cmark & \cmark &  &  &  &  &  \\
TrackerSift~\cite{amjadTrackerSiftUntanglingMixed2021} & IMC'21 & \dTool &  & \cmark &  &  &  &  &  &  \\
CV-Inspector~\cite{leCVInspectorAutomatingDetection2021} & NDSS'21 & \dTool & \cmark &  & \cmark &  &  &  &  &  \\
MAC Randomization~\cite{fenskeThreeYearsLater2021} & PETS'21 & \dServer &  &  &  &  &  &  & \cmark &  \\
ML-CB~\cite{reitingerMLCBMachineLearning2021} & PETS'21 & \dExt & \cmark &  &  &  &  &  &  &  \\
Ibex~\cite{zhongIbexPrivacypreservingAd2022} & CCS'22 & \dNative\,\dServer &  &  &  &  &  &  &  & \cmark \\
WTAGRAPH~\cite{yangWTAGRAPHWebTracking2022} & S\&P'22 & \dNative\,\dExt & \cmark &  &  &  &  &  &  &  \\
HARPO~\cite{zhangHARPOLearningSubvert2022} & NDSS'22 & \dExt & \cmark &  &  &  &  &  &  &  \\
FP-Radar~\cite{bahramiFPRadarLongitudinalMeasurement2022} & PETS'22 & \dTool & \cmark &  &  &  &  &  &  &  \\
Khaleesi~\cite{iqbalKhaleesiBreakerAdvertising2022} & USENIX'22 & \dExt & \cmark &  &  &  &  &  &  &  \\
Simulacrum~\cite{karamiUnleashSimulacrumShifting2022} & USENIX'22 & \dNative &  &  &  &  & \cmark &  &  &  \\
WebGraph~\cite{sibyWebGraphCapturingAdvertising2022} & USENIX'22 & \dNative\,\dTool & \cmark &  &  &  &  &  &  &  \\
CGC~\cite{parkCGCContrastiveGraph2022} & WWW'22 & \dTool & \cmark &  &  &  &  &  &  &  \\
AdCPG~\cite{leeAdCPGClassifyingJavaScript2023} & CCS'23 & \dTool & \cmark & \cmark &  &  &  &  &  &  \\
CookieGraph~\cite{munirCookieGraphUnderstandingDetecting2023} & CCS'23 & \dNative\,\dTool & \cmark &  &  &  &  &  &  &  \\
GDPR Runtime~\cite{kleinGeneralDataProtection2023} & CCS'23 & \dServer &  & \cmark &  &  &  &  &  &  \\
JS-ByteCode~\cite{ghasemisharifReadLinesDetecting2023} & CCS'23 & \dTool & \cmark & \cmark &  &  &  &  &  &  \\
FP-Sandbox~\cite{torokOnlyPayWhat2023} & S\&P'23 & \dNative &  &  &  &  & \cmark &  &  &  \\
Tracking JS blocking~\cite{amjadBlockingJavaScriptBreaking2023} & PETS'23 & \dNative &  & \cmark &  &  &  &  &  &  \\
AutoFR~\cite{leAutoFRAutomatedFilter2023} & USENIX'23 & \dExt\,\dTool & \cmark &  &  &  &  &  &  &  \\
Net-track~\cite{leeNettrackGenericWeb2023} & WWW'23 & \dServer & \cmark &  &  &  &  &  &  &  \\
Cobweb~\cite{amjadBlockingTrackingJavaScript2024} & CCS'24 & \dNative\,\dExt &  & \cmark &  &  &  &  &  &  \\
Unbundle-Rewrite-Rebundle~\cite{aliUnbundleRewriteRebundleRuntimeDetection2024} & CCS'24 & \dExt &  & \cmark &  &  &  &  &  &  \\
SINBAD~\cite{hajjchehadeSINBADSaliencyinformedDetection2024} & S\&P'24 & \dTool & \cmark &  &  &  &  &  &  &  \\
Browser Polygraph~\cite{kalantariBrowserPolygraphEfficient2024} & IMC'24 & \dServer & \cmark &  &  &  &  &  &  &  \\
FP-tracer~\cite{boussahaFPtracerFinegrainedBrowser2024} & PETS'24 & \dNative &  & \cmark &  &  &  &  &  &  \\
User-Controlled Privacy~\cite{hubletUserControlledPrivacyTaint2024} & PETS'24 & \dNative &  & \cmark &  &  &  &  &  &  \\
PURL~\cite{munirPURLSafeEffective2024} & USENIX'24 & \dNative\,\dTool & \cmark &  &  &  &  &  &  &  \\
AdFlush~\cite{leeAdFlushRealWorldDeployable2024} & WWW'24 & \dExt & \cmark &  &  &  &  &  &  &  \\
PanoptiChrome~\cite{kanyalPanoptiChromeModernInbrowser2024} & WWW'24 & \dNative &  & \cmark &  &  &  &  &  &  \\
Byte-level FP Detection~\cite{bahramiByteByteUnmasking2025} & CCS'25 & \dTool & \cmark & \cmark &  &  &  &  &  &  \\
CookieGuard~\cite{nikkhahbahramiCookieGuardCharacterizingIsolating2025} & IMC'25 & \dNative\,\dExt &  &  &  &  &  & \cmark &  &  \\
Duumviri~\cite{shuangDuumviriDetectingTrackers2025} & NDSS'25 & \dExt\,\dTool & \cmark & \cmark &  &  &  &  &  &  \\
CSAL~\cite{ortegaperezEncryptedAccessLogging2025} & USENIX'25 & \dNative\,\dServer &  &  &  &  &  &  &  & \cmark \\
Semantics-aware Cookies~\cite{chenSemanticsAwareCookiePurpose2025} & WWW'25 & \dTool &  &  &  & \cmark &  &  &  &  \\
AdVersa~\cite{limAdVersaAdversariallyRobustPractical2026} & WWW'26 & \dExt & \cmark &  &  &  &  &  &  &  \\
TGNN~\cite{xiongTGNNEnhancingPixel2026} & WWW'26 & \dTool &  &  &  & \cmark &  &  &  &  \\

\bottomrule
\end{tabular}
\caption{Defense deployment and mechanism types for tracking defense literature. Deployment: \dNative\ in-browser; \dExt\ browser extension; \dServer\ infrastructure-level; \dTool\ offline.}
\label{tab:defense-mechanisms}
\end{table}

%% file: tables/stateful-tracking-defenses.tex
\begin{table*}[t]
  \scriptsize
\centering
\caption{\textbf{Industrial and community defenses against stateful tracking} across two defense goals: \colorbox{cellred}{detection \& blocking} and \colorbox{cellyellow}{signal reduction \& obfuscation}. Protects against: \iconCookie: Cookie-based tracking, \iconURL: Navigational tracking, \iconFingerprint: Fingerprinting, \iconJS: JavaScript-based tracking, \iconSC: Side-channels. ``(D)'' is by-default, ``(P)'' is private-mode only, and ``(O)'' is opt-in based.}
\label{tab:stateful-tracking-defenses}
\footnotesize
\renewcommand{\arraystretch}{1.1}
\scalebox{0.74}{%
\begin{tabular}{
  m{3.15cm}
  m{3.0cm}
  m{3.0cm}
  m{3.0cm}
  m{3.0cm}
  m{3.0cm}
  m{3.0cm}
}
\toprule
\textbf{Defense Technique}\newline (\textbf{Protects Against}) & \textbf{Brave} & \textbf{Chrome} & \textbf{Edge} & \textbf{Firefox} & \textbf{Safari} & \textbf{Community Defense} \\
\toprule

\cellcolor{cellred} Blocking\newline3P cookies/storage\newline\newline \iconCookie
  & all 3P cookies and storage blocked via Shields (D)
  & 3P cookies and storage blocked in Incognito (P)
  & known tracker 3P cookies blocked (D)
  & known tracker 3P cookies blocked in ETP stnd. (D);\newline all 3P cookies blocked in ETP Strict (D)
  & all 3P cookies and storage blocked (D)
  & Tor blocks 3P cookies and storage\newline uBO, Privacy Badger, AdGuard, Ghostery blocks known tracker cookies via filter lists \\
\midrule

\cellcolor{cellyellow} Partitioning\newline3P cookies/storage\newline\newline \iconCookie\,\iconSC
  & uses Chromium's storage partitioning;\newline ephemeral 3P cookies/storage cleared on tab close (D)
  & opt-in partitioned cookies via \texttt{Partitioned} attribute: CHIPS;\newline storage partitioned by top-level site (D)
  & uses Chromium's CHIPS and its storage partitioning (D)
  & Total Cookie Protection: separate cookie jar per top-level site for all embedded 3P origins;\newline all storage APIs isolated per top-level site (D)
  & \texttt{Partitioned} cookie attribute supported;\newline Access to the 3P storage restricted under ITP;\newline Storage API provides a controlled opt-in (D)
  & ---\\
\midrule

\cellcolor{cellred} Blocking\newline1P tracking cookies/storage\newline\newline \iconCookie
  & Brave Shields blocks 1P-disguised trackers via filter lists (D)
  & NA
  & NA
  & NA
  & ITP purges 1P cookies from classified trackers after inactivity (D)
  & uBO and AdGuard supports targeted deletion of known 1P tracking cookies via filter rules and scriptlets \\
\midrule

\cellcolor{cellyellow} Limiting\newline1P cookie/storage lifetime\newline\newline \iconCookie
  & Caps lifetime of script-set cookies to 7 days;\newline Forgetful Browsing auto-clears 1P storage per-site (O)
  & NA
  & NA
  & NA
  & all script-set 1P cookies/storage expire after 7 days without interaction;\newline Expiry reduced to 24h when link decoration is detected (D)
  & ---\\
\midrule

\cellcolor{cellyellow} HTTP cache partitioning\newline\newline \iconCookie\,\iconSC
  & Inherits Chromium cache partitioning (D)
  & Cache keyed by top-level site (D)
  & Inherits Chromium cache partitioning (D)
  & Network partitioning includes cache isolation per top-level site (D)
  & ITP's isolation addresses cache-based tracking vectors (D)
  & ---\\
\midrule

\cellcolor{cellyellow} Network state partitioning\newline\newline \iconCookie\,\iconURL\,\iconJS\,\iconFingerprint
  & Network state partitioned (D)
  & Included with broader partitioning effort (D)
  & Inherits Chromium network partitioning (D)
  & TLS sessions, DNS cache, HSTS, OCSP, connections all isolated per top-level site (D)
  & Isolation at network-level under ITP's tracking prevention framework (D)
  & ---\\
\midrule

\cellcolor{cellred} URL parameter stripping\newline\newline \iconURL
  & Strips the known tracking query parameters (D)
  & NA
  & NA
  & Query-parameter stripping for known tracking parameters in ETP Strict and Private Browsing (O)
  & ITP detects link decora- tion and reduces script-writable storage lifetime from 7 to 24 hours (D);\newline ATFP's link tracking pro- tection strips params (P)
  & ClearURLs strips utm\_*, gclid, fbclid\newline DuckDuckGo strips by default \\
\midrule

\cellcolor{cellyellow} Referrer policy restrictions\newline\newline \iconURL
  & Strict referrer policy strips cross-origin referrer detail (D)
  & Defaults to strict-origin-when-cross-origin (D)
  & Inherits Chromium referrer policy (D)
  & Defaults to strict-origin-when-cross-origin (D)
  & Defaults to strict-origin-when-cross-origin (D)
  & AdGuard stealth strips referrers;\newline uBO can modify referrer headers via dynamic filtering \\
\midrule

\cellcolor{cellred} Blocking\newline CNAME-cloaked trackers\newline\newline \iconURL\,\iconCookie
  & Blocks CNAME-cloaked trackers (D)
  & NA
  & NA
  & NA;\newline uBO performs CNAME uncloaking through the \texttt{browser.dns} API
  & ITP applies CNAME/IP heuristics for detecting cloaked 3P hosts;\newline ATFP blocks CNAME-cloaked trackers (D)
  & uBO performs CNAME uncloaking (only Firefox);\newline AdGuard does CNAME resolution;\newline Extension-based defenses lack full DNS visibility \\
\midrule

\cellcolor{cellred} Protecting\newline against bounce tracking\newline\newline \iconURL\,\iconCookie
  & Debouncing skips known bounce-tracker redirects;\newline Unlinkable bouncing rou- tes visits through temporary unlinkable storage (D)
  & Heuristics-based deletion of site storage for bounce domains without any recent user interaction (D)
  & NA
  & ETP Strict uses blocklists; deletes storage without recent interaction (D)
  & Behavioral heuristics for deletion of storage for suspected trackers lacking legitimate interaction (D)
  & ---\\
\midrule

\cellcolor{cellred} Blocking known trackers\newline via filter/block lists\newline\newline \iconCookie\,\iconURL\,\iconJS\,\iconFingerprint
  & Shields blocks known 3P ads, trackers, crypto (D)
  & NA
  & Balanced and Strict mode blocks trackers from Disconnect list (D)
  & ETP blocks trackers from Disconnect list (D)
  & ITP and ATFP blocks kn- own and CNAME-cloaked trackers (D)
  & \textit{List}: easylist, easyprivacy, disconnect, Fanboy's lists, Peter Lowe's list, AdGuard lists, DuckDuckGo Tracker Radar;\newline \textit{Extension}: Ghostery, Ad- Guard, Privacy Badger, uBO;\newline \textit{DNS}: Pi-hole, NextDNS \\
\midrule

\cellcolor{cellred} ML-based\newline ad/tracker blocking\newline\newline \iconCookie\,\iconURL\,\iconJS\,\iconFingerprint
  & PageGraph detects trackers via page-level graph analysis (D)
  & NA
  & NA
  & NA
  & ITP uses ML classifier to identify tracking domains from cross-site loading patterns (D)
  & Privacy Badger learns fr- om observed cross-site behavior;\newline Blacklight scans sites for trackers and fingerprinting \\

\bottomrule
\end{tabular}%
}
\end{table*}

%% file: tables/stateless-tracking-defenses.tex
\begin{table*}[t]
  \scriptsize
\centering
\caption{\textbf{Industrial and community defenses against stateless tracking} across defense goals: \colorbox{cellred}{detection \& blocking} and \colorbox{cellyellow}{signal reduction \& obfuscation}. Protects against: \iconCookie: Cookie-based tracking, \iconURL: Navigational tracking, \iconFingerprint: Fingerprinting, \iconJS: JavaScript-based tracking, \iconSC: Side-channels. ``(D)'' is by-default, ``(P)'' is private-mode only, and ``(O)'' is opt-in based.}
\label{tab:stateless-tracking-defenses}
\footnotesize
\renewcommand{\arraystretch}{1.1}
\scalebox{0.74}{%
\begin{tabular}{
  m{3.15cm}
  m{3.0cm}
  m{3.0cm}
  m{3.0cm}
  m{3.0cm}
  m{3.0cm}
  m{3.0cm}
}
\toprule
\textbf{Defense Technique}\newline (\textbf{Protects Against}) & \textbf{Brave} & \textbf{Chrome} & \textbf{Edge} & \textbf{Firefox} & \textbf{Safari} & \textbf{Community Defense} \\
\toprule

\cellcolor{cellyellow} Masking or hiding\newline IP address\newline\newline \iconFingerprint
  & Private Window with Tor routes traffic through Tor network (O)
  & IP Protection hides IP via two-hop proxy system (P) (discontinued)
  & NA
  & NA;\newline Mozilla VPN available
  & iCloud's Private Relay routes Safari traffic via two relays to hide IP (O)
  & Tor Browser routes all traffic via onion network \\
\midrule

\cellcolor{cellyellow} Reducing\newline User-Agent string\newline\newline \iconFingerprint
  & Inherits Chromium UA reduction (D)
  & UA reduction and UA Client Hints (D)
  & Inherits Chromium UA reduction and Client Hints (D)
  & Uses the reduced form of Gecko UA string; resist-fingerprinting feature reports generic UA (D)
  & Freezes Safari version in UA string; limits exposed platform detail (D)
  & AdGuard's Stealth Mode can override UA \\
\midrule

\cellcolor{cellred} Blocking\newline fingerprinting scripts\newline\newline \iconJS\,\iconFingerprint
  & PageGraph records page execution for fingerprinting detection;\newline Shields blocks known fingerprinting scripts (D)
  & NA
  & Fingerprinting tracker blo- cking via Disconnect lists (D)
  & Fingerprinting script blo- cking via Disconnect lists (D)
  & ATFP blocks known fingerprinting scripts (P)
  & NoScript blocks JS (D);\newline uBO, AdGuard block fingerprinting domains;\newline Privacy Badger detects fi- ngerprinting-like behavior\\
\midrule

\cellcolor{cellyellow} Normalizing\newline browser API outputs\newline\newline \iconFingerprint
  & NA
  & NA
  & NA
  & resist-fingerprinting standardizes API outputs to uniform values for system specs like hardware and screen metrics (O)
  & ATFP normalizes fonts, plugins, UA (D);\newline ATFP standardizes some outputs such as screen, window, hardware metrics by default in private mode (O)
  & Tor normalizes API outputs for a uniform fingerprint\newline Mullvad Browser applies same hardening for non-Tor use (D) \\
\midrule

\cellcolor{cellyellow} Randomizing\newline browser API outputs\newline\newline \iconFingerprint
  & Farbling does site-specific randomization of API outputs via noise injection so that fingerprint changes per page load (D)
  & NA
  & NA
  & Randomizes API outputs of Canvas, JS floating po- int math operations, and peripheral touch support (D)
  & ATFP injects noise to au-\newline dio, canvas, WebGL by default in private mode (O)
  & ---\\
\midrule

\cellcolor{cellyellow} Reducing precision\newline of browser API outputs\newline\newline \iconFingerprint\,\iconSC
  & Reduces timer's precision (D)
  & Reduces \texttt{performance.now} resolution (D)
  & Inherits Chromium timer precision reductions (D)
  & Reduces event time and \texttt{performance.now}  precision\newline resist-fingerprinting mode applies broader reductions by injecting noise into dynamic hardware hooks like Canvas and timer (D)
  & Reduces timer's precision (D)
  & ---\\
\midrule

\cellcolor{cellred} Removing or declining\newline fingerprintable APIs\newline\newline \iconFingerprint
  & Restricts or disables the fingerprinting-prone APIs where feasible (D)
  & Removed Battery Status API (D)
  & Inherits Chromium API removals (D)
  & Removed Battery Status API; declined Network Information API (D)
  & Declined Network Information API, Web Bluetooth; WebKit policy considers fingerprinting impact for new APIs (D)
  & ---\\
\midrule

\cellcolor{cellred} Blocking JavaScript\newline\newline \iconJS\,\iconFingerprint
  & Brave Shields allows per-site script blocking; aggressive mode blocks all scripts (O)
  & Available via site settings (O)
  & Available via site settings (O)
  & Available via about:config and site permissions (O)
  & Available via site settings (O)
  & NoScript blocks JS by default per-site; highly effective but significant usability burden \\
\midrule

\cellcolor{cellred} Neutralization of\newline mixed scripts\newline\newline \iconJS\,\iconFingerprint\,\iconCookie
  & SugarCoat generates rep-\newline lacement scripts preserv- ing the functionality while neutralizing tracking (D)
  & NA
  & NA
  & NA;\newline uBO scriptlets can override specific functions
  & NA
  & ---\\
\midrule

\cellcolor{cellyellow} Defending against\newline Extension fingerprinting\newline\newline \iconFingerprint
  & Randomizes artifacts detectable by extensions (D)
  & NA
  & NA
  & Extension APIs somewhat isolated but not fully hardened
  & Extensions are run within sandboxed context with limited web-accessible resources (D)
  & ---\\
\midrule

\cellcolor{cellyellow} Side-channel defenses\newline\newline \iconSC
  & Inherits Chromium miti- gations;\newline Shields provides additio-\newline nal isolation (D)
  & Site Isolation COOP/\newline COEP headers;\newline State partitioning (D)
  & Inherits Chromium Site\newline Isolation \& COOP/COEP (D)
  & Site Isolation (Fission);\newline Network partitioning;\newline Reduced timer precision (D)
  & Process-per-tab isolation\newline ITP state partitioning (D)
  & ---\\

\bottomrule
\end{tabular}%
}
\end{table*}

%% file: tables/defenses-industrial.tex
\newcommand{\tablescale}{0.8}        
\newcommand{\rowheight}{1.0}          

\newlength{\colBrowser}   \setlength{\colBrowser}{0.7cm}
\newlength{\colDate}      \setlength{\colDate}{2cm}
\newlength{\colVersion}   \setlength{\colVersion}{2.2cm}
\newlength{\colProtects}  \setlength{\colProtects}{2cm}
\newlength{\colDesc}      \setlength{\colDesc}{13.2cm}

\begin{table*}[t]
  \scriptsize
\centering
\caption{\textbf{Evolution of industry-driven tracking defenses across different browsers.} \iconCookie: Cookie-based tracking, \iconURL: Navigational tracking, \iconFingerprint: Fingerprinting, \iconJS: JavaScript-based tracking, \iconSC: Side-channels, \iconPrivacy: Privacy proposals.}
\label{tab:defenses-industrial}
\footnotesize
\renewcommand{\arraystretch}{\rowheight}
\scalebox{\tablescale}{%
\begin{tabular}{
  >{\centering\arraybackslash}m{\colBrowser}
  >{\centering\arraybackslash}m{\colDate}
  >{\centering\arraybackslash}m{\colVersion}
  >{\centering\arraybackslash}m{\colProtects}
  m{\colDesc}
}
\toprule
\textbf{Browser} & \textbf{Release Date} & \textbf{Browser Version} & \textbf{Protects Against} & \textbf{Description of the Introduced Tracking Defenses} \\
\toprule

\multirow{4}{*}[0pt]{\iconIE}
  & Aug 2001 & 6  & \iconCookie                                & P3P support; 3P cookies restricted by default unless an acceptable P3P compact policy is present \\
  & Mar 2009 & 8  & \iconCookie\,\iconSC       & Introduced InPrivate Browsing and InPrivate Filtering (blocks 3P content loaded across $>$10 sites) \\
  & Mar 2011 & 9  & \iconCookie\,\iconURL\,\iconFingerprint  & Tracking Protection Lists (TPL) replaces InPrivate Filtering; Do Not Track (DNT) support added \\
  & Jun 2022 & 11 &       ---       & Last IE~11 desktop app retired and disabled, redirecting to Microsoft Edge \\
\midrule

\multirow{7}{*}[-5pt]{\iconEdge}
  & Jul 2015    & 20      & --- & Edge launches with Windows~10, InPrivate browsing, and DNT; drops IE's TPL \\
  & Jan 2020    & 79      & \iconCookie\,\iconURL\,\iconFingerprint & Chromium Edge with Disconnect-based blocking; blocks cryptomining+fingerprinting trackers \\
  & Oct 2020    & 86      & \iconSC                                  & Inherits Chromium HTTP cache partitioning \\
  & 2022--2023  & 101--116 & \iconFingerprint\, \iconPrivacy                           & Inherits Chromium User-Agent reduction (reduced UA string; UA Client Hints) \\
  & May 2023    & 114     & \iconCookie\,\iconSC\, \iconPrivacy                     & Inherits Chromium CHIPS (partitioned 3P cookies) \\
  & Aug 2023    & 115     & \iconSC                                  & Inherits Chromium Storage Partitioning \\
  & Jun 2024    & 139     & ---                                & Inherits Chromium Manifest~V3 enforcement; MV2 ad/tracker blockers phased out \\
\midrule

\multirow{14}{*}[-15pt]{\iconFirefox}
  & Sep 2002 & 0.1       & \iconCookie                                 & Launch of Phoenix~0.1 with basic cookie management \\
  & Nov 2004 & 1         & \iconCookie                                 & Launch of Firefox~1.0: cookie manager with exceptions; option to block 3P cookies \\
  & Jun 2009 & 3.5       & ---                                  & Private Browsing mode introduced \\
  & Mar 2011 & 4         & ---                             & Do Not Track (DNT) header adopted; first browser to implement DNT \\
  & Nov 2015 & 42        & \iconCookie\,\iconURL\,\iconFingerprint          & Tracking Protection blocks known trackers in Private Browsing using Disconnect list \\
  & Nov 2017 & 57        & \iconCookie\,\iconURL\,\iconFingerprint                   & Tracking Protection available outside Private Browsing (optional) \\
  & Oct 2018 & 63        & \iconCookie\,\iconJS          & Enhanced Tracking Protection (ETP) introduced (off by default) \\
  & Jun 2019 & 67        & \iconCookie\,\iconJS                                 & ETP turned on by default for new installs (Standard mode) \\
  & Jan 2020 & 72        & \iconFingerprint\,\iconJS                   & Fingerprinting-script blocking enabled by default via Disconnect's fingerprinting list \\
  & Jul 2020 & 79        & \iconURL                                    & Basic redirect (bounce) tracking protection introduced \\
  & Jan 2021 & 85        & \iconCookie\,\iconSC                     & Supercookie protection via network partitioning (caches/connections isolated per top-level site) \\
  & Feb 2021 & 86        & \iconCookie                     & Total Cookie Protection introduced (separate cookie jar per website under ETP Strict) \\
  & Jun 2022 & 102       & \iconURL                                    & URL query-parameter stripping for known tracking parameters (ETP Strict and Private Browsing) \\
  & Aug 2024     & 128       & \iconURL                                    & Bounce Tracking Protection introduced to ETP Strict \\
\midrule

\multirow{10}{*}[-15pt]{\iconSafari}
  & Jun 2003 & 1    & \iconCookie                                          & Launch of Safari~1.0: blocks 3P cookies from unvisited sites by default \\
  & Apr 2005 & 2    & ---                                           & Private Browsing mode introduced; first browser to offer it \\
  & Jul 2011     & 5.1  & ---                                      & DNT header support added \\
  & Sep 2017 & 11   & \iconCookie                         & ITP~1.0: ML identifies tracking cookies; 24h 3P cookie window; purges 30d+ unused cookies \\
  & Sep 2019 & 13   & \iconURL\,\iconJS\,\iconSC                        & ITP~2.3: 7-day cap on all script-writable storage to counter storage and link-decoration tracking \\
  & Mar 2020 & 13.1 & \iconCookie                                          & Complete 3P cookie blocking enabled by default; first mainstream browser to block all 3P cookies \\
  & Apr 2021 & 14.1 & \iconPrivacy                        & Private Click Measurement (PCM) enabled by default for privacy-preserving ad click attribution \\
  & Sep 2023 & 17   & \iconCookie\,\iconURL\,\iconFingerprint\ & Advanced Tracking and Fingerprinting Protection (ATFP) in Private Browsing: Link Tracking Protection, API noise injection, blocks known/CNAME-cloaked trackers \\
  & Mar 2025 & 18.4 & \iconCookie\,\iconSC                              & Support for the CHIPS-styled partitioned cookie attribute introduced \\
  & Sep 2025 & 26   & \iconFingerprint                                     & Advanced Fingerprinting Protection extended to default (non-Private) browsing \\
\midrule

\multirow{14}{*}[-15pt]{\iconChrome}
  & Dec 2008 & 1              & ---                    & Launch of Chrome~1.0 stable: Incognito mode \\
  & Nov 2012 & 23             & ---               & DNT header support added \\
  & Aug 2019 & ---            & \iconPrivacy                  & Privacy Sandbox initiative announced \\
  & May 2020 & 83             & \iconCookie\       & 3P cookies blocked by default in Incognito mode; new cookie controls \\
  & Oct 2020 & 86             & \iconSC                    & HTTP cache partitioning (cache keyed by top-level site) \\
  & Apr 2021 & 89             & \iconPrivacy & FLoC origin trial begins (interest cohorts as a 3P cookie alternative) \\
  & Jan 2022 & 101--115 (OT)  
  & \iconPrivacy                  & FLoC abandoned; Topics API proposed as replacement \\
  & Apr 2022 & 100--116 (OT)  
  & \iconPrivacy & FLEDGE in-browser ad-auction origin trial begins \\
  & Jun 2022 & 101--110 (OT)  
  & \iconFingerprint\,\iconPrivacy              & UA Reduction introduced: browser minor version reduced to 0.0.0 in the UA string \\
  & May 2023 & 100--115 (OT)  
  & \iconCookie\,\iconSC,\iconPrivacy & CHIPS (Cookies Having Independent Partitioned State) introduced \\
  & Aug 2023 & 105--115 (OT)  
  & \iconSC                    & Storage Partitioning (localStorage, IndexedDB, CacheStorage partitioned by top-level site) \\
  & Apr 2025 & ---            & \iconCookie\,\iconPrivacy     & Google drops planned 3P cookie prompt; cookies remain; deprecation effort ended \\
  & Sep 2025 & 138            & ---                  & Manifest~V3 enforcement; MV2 ad/tracker blockers phased out \\
  & Oct 2025 & ---            & \iconPrivacy                  & Privacy Sandbox APIs (Topics, Protected Audience, Attribution Reporting, etc.) retired \\
\midrule

\multirow{11}{*}[-10pt]{\iconBrave}
  & Jan 2016 & 0.7  & \iconCookie\,\iconURL\,\iconFingerprint\,\iconSC & Developer preview: blocks 3P ads/trackers and 3P storage by default; HTTPS Everywhere \\
  & Nov 2019 & 1    & \iconCookie\,\iconURL\,\iconFingerprint\,\iconSC           & Launch of Brave~1.0: Shields blocks 3P ads/trackers, cross-site cookies, and fingerprinting by default \\
  & Mar 2020 & 1.5  & \iconFingerprint                                        & Fingerprint randomization (farbling) introduced \\
  & Jul 2020 & 1.12 & \iconURL                                                & URL query-parameter stripping, stricter referrer policy, Reporting API limits \\
  & Jul 2020 & 1.17 & \iconCookie\,\iconURL\,\iconFingerprint                                & Defense against CNAME-cloaked trackers \\
  & Feb 2021 & 1.21 & \iconCookie\,\iconSC                                 & Ephemeral 3P site storage (partitioned and cleared when site closes) \\
  & Oct 2021 & 1.32 & \iconURL                                                & Debouncing introduced; skips known bounce-tracker redirects \\
  & Dec 2021 & 1.35 & \iconCookie\,\iconSC                                 & Partitioning of network state \\
  & Mar 2022 & 1.37 & \iconURL                                                & Unlinkable Bouncing: routes suspected-site visits through temporary, unlinkable storage \\
  & May 2023 & 1.53 & \iconCookie\,\iconSC                                 & Forgetful Browsing: auto-clear first-party storage per-site \\
  & Jan 2024 & 1.64 & \iconFingerprint                                        & Simplified fingerprinting protections \\

\bottomrule
\end{tabular}%
}
\end{table*}

%% file: tables/defenses-community.tex
\newlength{\colCYear}    \setlength{\colCYear}{0.5cm}
\newlength{\colCName}    \setlength{\colCName}{3.8cm}
\newlength{\colCType}    \setlength{\colCType}{0.6cm}
\newlength{\colCDesc}    \setlength{\colCDesc}{15.5cm}

\begin{table*}[t]
  \scriptsize
\centering
\caption{\textbf{Timeline of community-driven tracking defenses.} \iconBlocklist: Filter/block list, \iconExtension: Browser extension, \iconTool: represents tools, standards, dataset, or browser.}
\label{tab:defenses-community}
\footnotesize
\renewcommand{\arraystretch}{\rowheight}
\scalebox{\tablescale}{%
\begin{tabular}{
  >{\centering\arraybackslash}m{\colCYear}
  m{\colCName}
  >{\centering\arraybackslash}m{\colCType}
  m{\colCDesc}
}
\toprule
\textbf{Year} & \textbf{Community Defense} & \textbf{Type} & \textbf{Description} \\
\toprule

1997 & Peter Lowe's list (pgl.yoyo.org) & \iconBlocklist & Initially created for personal local DNS use, this crowdsourced adblock list is widely consumed by hosts files, Pi-hole, AdGuard, uBO \\
2002 & Adblock & \iconExtension & Ad blocking extension created for Firefox; later replaced by its successors \\
2005 & EasyList & \iconBlocklist & Created as a maintainable ad-blocking subscription for the Adblock extension; has become the de-facto core ad filter list \\
2005 & NoScript & \iconExtension & Blocks JavaScript content by default per-site, cutting JS-based tracking and fingerprinting \\
2006 & Adblock Plus (Eyeo) & \iconExtension & Adblock extension re-written as Adblock Plus with subscribable filter lists (e.g.\ EasyList) \\
2006 & EasyPrivacy & \iconBlocklist & Companion list to EasyList targeting trackers, analytics, tracking pixels/beacons, and web bugs (not just ads) \\
2008 & Tor Browser (Tor Project) & \iconTool & Reference anti-fingerprinting and onion-routing browser; provides first-party isolation and uniform fingerprint (one-fingerprint-for-all) \\
2009 & Ghostery & \iconExtension & Detects and (optionally) blocks 3P trackers; also exposes who is tracking on each page \\
2009 & Do Not Track (DNT) & \iconTool & Proposed by researchers as a browser header to opt out of cross-site tracking \\
2010 & Fanboy's Annoyance List & \iconBlocklist & List targeting cookie-consent notices, social-media widgets, and in-page annoyances (basis for EasyList Cookie List) \\
2010 & EFF Panopticlick & \iconTool & EFF tool demonstrating browser fingerprint uniqueness and tracker exposure \\
2011 & Disconnect & \iconBlocklist & List blocks 3P trackers; its open tracker list got later on adopted by Firefox, Safari, and Edge \\
2011 & Do Not Track (DNT) & \iconTool & Implemented by Firefox, IE, Safari, Opera, then Chrome; W3C Tracking Protection WG chartered to standardize it \\
2013 & Fanboy $\rightarrow$ EasyList & \iconBlocklist & Fanboy's List merged into EasyList, consolidating the dominant English ad-filter lists \\
2014 & Privacy Badger (EFF) & \iconExtension & Heuristic tracker blocker that learns about trackers from observed cross-site behavior rather than a static list \\
2014 & uBlock Origin & \iconExtension & uBlock released for Chrome/Opera \\
2014 & AdGuard & \iconBlocklist\,\iconExtension & Extension blocks ad/tracker domains based on various lists; Stealth Mode strips referrers, 3P cookies, and reduces fingerprinting surface \\
2014 & Pi-hole & \iconTool & Open-source network-wide DNS sinkhole (Raspberry Pi); blocks ad/tracker domains for an entire network via DNS \\
2017 & WhoTracks.me & \iconTool & Open dataset and reporting of trackers' prevalence and behavior across the web \\
2018 & Ghostery powers WhoTracks.Me & \iconTool & Ghostery goes open source; powers the WhoTracks.me tracker dataset \\
2019 & Do Not Track (DNT) & \iconTool & W3C Tracking Protection Working Group closed without DNT adoption; DNT effectively abandoned by industry \\
2019 & ClearURLs & \iconExtension & Strips known tracking parameters (e.g.\ utm\_*, gclid, fbclid) from URLs before requests and on copy \\
2019 & NextDNS & \iconTool & Cloud DNS resolver with configurable ad/tracker blocklists, applied across devices/browsers \\
2019 & uBO protects against CNAME & \iconExtension & uBO (Firefox) adds CNAME-uncloaking to defeat first-party-disguised (CNAME-cloaked) trackers; advanced dynamic filtering/scriptlets \\
2020 & Global Privacy Control (GPC) & \iconTool & EFF-led universal opt-out signal; legally enforceable under CCPA/CPRA; adopted by DuckDuckGo, Brave, Abine, Firefox \\
2020 & DuckDuckGo Tracker Radar & \iconTool & Open, auto-generated tracker dataset (domains $\rightarrow$ owners $\rightarrow$ behaviors); powers DDG protections, Vivaldi, and Blacklight \\
2020 & Blacklight & \iconTool & Markup's web tool that scans any site and reports trackers, cookies, session recorders, pixels, and fingerprinting \\
2022 & GPC Adoption & \iconTool & California Sephora settlement establishes that ignoring GPC violates CCPA, driving site adoption \\
2023 & Mullvad Browser & \iconTool & Tor Project + Mullvad: Tor Browser's anti-fingerprinting hardening for everyday (non-Tor-network) use \\
2025 & uBO removed from Chrome & \iconExtension & uBO removed from the Chrome Web Store as Chrome disables MV2 extensions by July 2025; remains full-featured on Firefox/Brave \\

\bottomrule
\end{tabular}%
}
\end{table*}

%% file: tables/regulations-paper-distribution.tex
\begin{table*}[!ht]
  \small
  \centering
  \caption{Taxonomy of policies along with their distribution across the literature (\# = paper count).}
  \label{tab:regulations-paper-distribution}
\begin{subtable}[t]{0.29\linewidth}
\centering
  \begin{tabular}{>{\centering\arraybackslash}p{4cm} >{\centering\arraybackslash}p{0.1cm}}
    \toprule
    \multicolumn{1}{c}{\textbf{EU Regulation}} & \multicolumn{1}{l}{\textbf{\#}} \\
    \midrule
    GDPR & 27 \\
    ePrivacy Directive (ePD) & 9 \\
    Digital Services Act (DSA) & 3 \\
    Digital Markets Act (DMA) & 1 \\
    \bottomrule
  \end{tabular}
  \caption{EU Regulations}
\end{subtable}
\begin{subtable}[t]{0.29\linewidth}
\centering
  \begin{tabular}{>{\centering\arraybackslash}p{4cm} >{\centering\arraybackslash}p{0.1cm}}
    \toprule
    \multicolumn{1}{c}{\textbf{US Regulation}} & \multicolumn{1}{l}{\textbf{\#}} \\
    \midrule
    CCPA (California) & 11 \\
    CPRA (California) & 3 \\
    Other states & 2 \\
    FTC Act or COPPA (Federal) & 2\\
    \bottomrule
  \end{tabular}
  \caption{US Regulations}
\end{subtable}
\begin{subtable}[t]{0.4\linewidth}
\centering
  \begin{tabular}{>{\centering\arraybackslash}p{6cm} >{\centering\arraybackslash}p{0.1cm}}
    \toprule
    \multicolumn{1}{c}{\textbf{Standard}} & \multicolumn{1}{l}{\textbf{\#}} \\
    \midrule
    GPC & 5 \\
    P3P & 3\\
    DNT & 3 \\
    IAB TCF or CCPA Compliance Framework & 3 \\
    \bottomrule
  \end{tabular}
  \caption{Opt-Out Signals and Industry Frameworks}
\end{subtable}
\end{table*}

%% file: tables/regulations-aspects.tex
\begin{table*}[!ht]
    \small
    \centering
    \caption{Taxonomy of policy aspects (along with its distribution across literature) mapped onto rights and context applicability.}
    \label{tab:regulations-aspects}
    \begin{tabular}{cccccccccc}
    \toprule
    && \multicolumn{6}{c}{\textbf{Right to}} \\
    \textbf{Policy Aspect} & \rot{\textbf{\# Paper count}} &\rot{\textbf{Access}} &\rot{\textbf{Portability}} &\rot{\textbf{Correct}} &\rot{\textbf{Delete}} &\rot{\textbf{Opt-out}} &\rot{\textbf{Consent}} & \textbf{GDPR/ePD Context} & \textbf{CCPA/CPRA Context}\\
    \midrule

    User Consent & 18 & \pie{0} & \pie{0} & \pie{0} & \pie{0} & \pie{0} & \pie{360} & \checkmark\ Opt-in design & \xmark\ Opt-out design\\

    Notice & 8 & \pie{360} & \pie{0} & \pie{0} & \pie{0} & \pie{0} & \pie{360} & \checkmark\ Prelude for consent & \checkmark\ Disclosure requirements \\

    Sharing \& Consent Propagation & 7 & \pie{0} & \pie{0} & \pie{0} & \pie{360} & \pie{360} & \pie{360} & \checkmark\ Processor obligations & \xmark\ Third-party notices\\

    Opt-out Sale & 6 & \pie{0} & \pie{0} & \pie{0} & \pie{360} & \pie{360} & \pie{0} & \xmark\ Opt-in design & \checkmark\ Opt-out design\\

    Data Access Request & 5 & \pie{360} & \pie{360} & \pie{360} & \pie{360} & \pie{0} & \pie{0} & \checkmark\ Controller must provide info. & \checkmark\ Requests to know \\

    Data Minimization & 4 & \pie{360} & \pie{360} & \pie{360} & \pie{360} & \pie{360} & \pie{360} & \checkmark\ Processing principle & \checkmark\ CPRA amendment \\

    Legitimate Interest Exemption & 5 & \pie{0} & \pie{0} & \pie{0} & \pie{0} & \pie{360} & \pie{360} & \checkmark\ Specific purposes only& \xmark\ No equivalent\\

    Data Storage (persistence) & 3 & \pie{360} & \pie{360} & \pie{360} & \pie{360} & \pie{0} & \pie{0} & \checkmark\ Storage limitations & \checkmark\ Retention schedules\\

    Privacy by Design & 1 & \pie{0} & \pie{360} & \pie{360} & \pie{0} & \pie{0} & \pie{360} & \checkmark\ By design and by default & \checkmark\ CPRA amendment\\

     \bottomrule
    \end{tabular}
\end{table*}